\documentclass[a4paper,11pt]{JHEP3}
\usepackage[OT4]{fontenc}
\usepackage{amsmath}
\usepackage{amssymb}

\usepackage[cp1250]{inputenc} 

\def\bra#1{\left\langle\hskip 1pt #1\hskip .5pt\right|}
\def\ket#1{\left|\hskip .5pt #1\hskip 1pt\right\rangle}
\def\drangle{\rangle\!\rangle}
\def\dlangle{\langle\!\langle}

\newcommand{\A}{{\rm\scriptscriptstyle A}}
\newcommand{\R}{{\rm\scriptscriptstyle R}}
\newcommand{\NS}{{\rm\scriptscriptstyle NS}}
\newcommand{\li}{{\rm\scriptscriptstyle L}}
\newcommand{\gi}{{\scriptscriptstyle \Gamma}}
\newcommand{\pmp}{{\scriptstyle \pm} {\sf  p}}

\newcommand{\p}{{\sf  p}}
\newcommand{\bli}{{b^{\rm\scriptscriptstyle L}}}
\newcommand{\bgi}{{b^{\scriptscriptstyle \Gamma}}}
\newcommand{\ali}{{\alpha^{\rm\scriptscriptstyle L}}}
\newcommand{\agi}{{\alpha^{\scriptscriptstyle \Gamma}}}

\newcommand{\SL}{{\scriptscriptstyle\rm SL}}
\newcommand{\LL}{{\scriptscriptstyle\rm LL}}

\renewcommand{\theequation}{\arabic{section}.\arabic{equation}}
\author{
     Leszek Hadasz\footnote{e-mail: hadasz@th.if.uj.edu.pl} \\
	 M. Smoluchowski Institute of Physics,
	 Jagiellonian University,\\
	 W.~Reymonta 4,
	 30-059~Krak\'ow, Poland

}
\author{
     Zbigniew Jask\'{o}lski \footnote{e-mail: jask@ift.uni.wroc.pl}\\
	 Institute of Theoretical Physics,
	 University of Wroc{\l}aw,\\
	 pl. M. Borna 1, 95-204~Wroc{\l}aw, Poland
}

\abstract{ The AGT motivated relation between  the tensor product of the ${\cal N}=1$ super-Liouville field theory with the imaginary free fermion (SL) and
a certain projected tensor product  of the real and the imaginary Liouville field theories (LL) is analyzed. Using conformal field theory techniques we give a complete proof of the equivalence in the NS sector.
It is shown that the SL-LL correspondence is based on the equivalence of chiral objects including
suitably chosen chiral structure constants of all the three Liouville theories involved.
\vspace*{1cm}}

\title{ Super-Liouville - Double Liouville correspondence }

\preprint{}

\begin{document}
\section{Introduction}

It was conjectured several years ago  that
partition functions of ${\cal N} = 2$ superconformal ${\rm SU}(N)$ gauge theories in four dimensions are directly related to
correlation functions of the two-dimensional Liouville/Toda field theories \cite{Alday:2009aq,Wyllard:2009hg}.
This by now well established AGT relation has been extended
and generalized in various directions leading to many new developments on both sides of the correspondence.
One of the essential generalizations was the  proposal that ${\cal N}=2$ ${\rm SU}(N)$ gauge theories on $\mathbb{R}^4/\mathbb{Z}_p$ should be related to
certain coset conformal fields theories. This conjecture was first formulated in \cite{Belavin:2011pp} where also the first checks in the case of $N=p=2$ corresponding to the
${\cal N}=1$ super-Liouville theory were presented. It was soon clarified that the  general case $ p>2$ should correspond
to para-Liouville/Toda theories \cite{Nishioka:2011jk}.  Some further checks of the AGT relation for $N=p=2$ where done in the NS sector in \cite{Belavin:2011tb,Bonelli:2011jx,Bonelli:2011kv}
and in the R sector in \cite{Ito:2011mw,Belavin:2012aa}. It was in particular observed in \cite{Bonelli:2011jx,Bonelli:2011kv} that the blow-up formula
for the Nekrasov partition function suggests a precise relation between ${\cal N}=1$  super-Liouville and Liouville conformal blocks.
An interesting explanation of this phenomenon on the CFT side was given in \cite{Wyllard:2011mn}.
It was motivated by old results \cite{Crnkovic:1989gy,Crnkovic:1989ug,Lashkevich:1992sb,Lashkevich:1993fb} relating various rational models
realized as quotients,
$$
V(p,m) \sim \frac{\widehat{\rm SU}(2)_p\times \widehat{\rm SU}(2)_m}{\widehat{\rm SU}(2)_{p+m}}.
$$
The case relevant for the present paper is the relation between the Virasoro minimal models $V(m)=V(1,m)$ and the ${\cal N}=1$  superconformal models $SV(m)=V(2,m)$:
$$
V(1)\otimes SV(m) \sim V(m)\otimes_P V(m+1), \;\;\;\;m=1,2,\dots,
$$
where the symbol $\otimes_P$ denotes a projected tensor product in which only selected pairs of conformal families are present.
The nonrational counterpart
of this relation  proposed in \cite{Wyllard:2011mn} takes the schematic form
\begin{equation}
\label{mrel}
\hbox{free fermion}\,\otimes\,{\cal N} =1\ \hbox{super-Liouville}
\hskip 8pt \sim \hskip 8pt
\hbox{Liouville} \ \otimes_P\, \hbox{Liouville}.
\end{equation}
In the NS sector this relation  has been made much more precise in \cite{Belavin:2011sw} where it was used
as an essential element of the proof of the AGT correspondence in the case of  $N=p=2$.
The considerations of \cite{Belavin:2011sw} are based on the instanton partition function computed on the resolved space \cite{Bonelli:2011jx,Bonelli:2011kv}.
It is interesting to note that a different instanton
counting scheme for the ALE spaces, the orbifolded instanton counting, yields in general a slightly different partition function \cite{Ito:2013kpa} and it is still an open question whether
its CFT counterpart exists.
The extension of (\ref{mrel}) to the Ramond sector along with some nontrivial checks were presented in  \cite{Schomerus:2012se}.

Although most of the ingredients and constructions were already discussed in \cite{Belavin:2011sw} and \cite{Schomerus:2012se}
a precise formulation of (\ref{mrel})  as
an exact equivalence of CFT models is still an open problem.
The product of the free fermion and the ${\cal N}=1$ super-Liouville theory (SL) is a perfectly well defined model of CFT but
an exact definition of  the double Liouville theory (LL) on the r.h.s.\ of (\ref{mrel}) has not yet been clarified.
There are also some essential elements of the proof missing.
The aim of the present paper is to fill these gaps.

Let us briefly describe  our proposal for the precise meaning of equivalence (\ref{mrel}).
The chiral symmetry algebra ${\sf A}_{\NS}$ of the SL theory
is the product of the ${\cal N}=1$ super-Virasoro algebra ${\sf SVir}_{\NS}$ and the Heisenberg algebra ${\sf H}_{\NS}$ of  fermion oscillators.
One has therefore a $1-1$ correspondence between
${\sf SVir}_{\NS}$- and ${\sf A}_{\NS}$-primaries.
Any highest weight representation ${\cal A}_{\Delta_p}$ of ${\sf A}_{\NS}$  with the central charge $c={3\over 2} +3Q^2, Q=b+b^{-1}$
and the highest weight
$
\Delta_p= {Q^2\over 8}+{p^2\over 2}
$
carries a representation of two mutually commuting Virsoro algebras $\{L^\li_n\}$, $\{L^\gi_n\}$
with the corresponding central charges given by
\cite{Wyllard:2011mn,Crnkovic:1989gy,Crnkovic:1989ug,Lashkevich:1992sb,Lashkevich:1993fb,Belavin:2011sw}:
\begin{equation}
\label{defblibgi}
\begin{array}{rlllllllllllllll}
c^{\li}&=&1+6\,\left(Q^{\li}\right)^2,&& Q^{\li}&=& \displaystyle b^{\li} +{1\over b^{\li}},&&b^{\li}&=&\displaystyle \frac{2b}{\sqrt{2-2b^2}},\\
c^{\gi}&=&1-6\,\left( Q^{\gi} \right)^2,&& Q^{\gi}&=& \displaystyle {1\over b^{\gi}} -b^{\gi},&&\displaystyle {1\over b^{\gi}}&=&\displaystyle \frac{2}{\sqrt{2-2b^2}}.
\end{array}
\end{equation}
For real $b$ they are on the opposite sides of the
$c=1$ barrier. We assume $b<1$ throughout the paper.
Then
$$
c^{\li}>25,\;\;\;1>c^{\gi}
$$
and the parameterizations above coincide with the  standard ones of the Liouville theory \cite{Dorn:1994xn,Zamolodchikov:1995aa,Teschner:1995yf,Teschner:2001rv} and of the
generalized minimal models (GMM, the time-like Liouvile theory) \cite{Zamolodchikov:2005fy,Zamolodchikov:2005sj}.

It was shown in \cite{Belavin:2011sw} that
the ${\sf Vir} \oplus {\sf Vir}$ representation
on ${\cal A}_{\Delta_p}$ decomposes into irreducible components
\begin{equation}
\label{decva}
{\cal A}_{\Delta_p}= \bigoplus\limits_{j\in \mathbb{Z}} {\sf V}_{\Delta^{\li}(p, j)}\otimes {\sf V}_{\Delta^{\gi}(p, j)}
\end{equation}
where ${\sf V}_{\Delta^{\li}(p, j)}$, ${\sf V}_{\Delta^{\gi}(p, j)}$ are Verma modules of the corresponding Virasoro algebras with the highest weights
$$
\Delta^{\li}(p, j) =
 \frac{1}{1-b^2}\left(\frac{Q^2}{8} + \frac{\left(p + ij\,b\right)^2}{2}\right),
\;\;\;\;
\Delta^{\gi}(p, j)=
\frac{1}{1-b^{-2}}\left(\frac{Q^2}{8} + \frac{\left(p + ij\, b^{-1}\right)^2}{2}\right).
$$
Decomposition (\ref{decva}) implies that the spectrum of the double Liouville theory, although
diagonal in the continuous parameter $p,$ is
non-diagonal in the discrete index $j$. This important novelty
requires  appropriate off diagonal extensions of
the DOZZ \cite{Dorn:1994xn,Zamolodchikov:1995aa,Teschner:1995yf} and the
GMM \cite{Zamolodchikov:2005fy,Zamolodchikov:2005sj} structure constants. A simple idea is  to
split the diagonal constants into chiral parts.
As we shall see this receives strong support from our calculations.
Indeed it turns out that the SL-LL equivalence is to large extend based on the relations between
chiral structure constants.

In the standard normalization  of the  Liouville theory the reflection amplitude and the two-point function are equal \cite{Dorn:1994xn,
Zamolodchikov:1995aa,Teschner:1995yf}. This is suitable for the analytic continuation arguments and natural from the path integral point of view \cite{Harlow:2011ny}.
Other normalizations were discussed in the context of the GMM  \cite{Zamolodchikov:2005fy,McElgin:2007ak}.
In the present discussion it is convenient to choose the symmetric normalization
$$
\Phi_\alpha = \Phi_{Q-\alpha}
$$
with no restrictions on the two-point functions. This simplifies the relation between fields on both sides of the correspondence.

In the symmetric normalization the DOZZ structure constant for primary fields $\Phi_{\alpha}$ can be written as
\begin{eqnarray}
\nonumber
C^{\rm\scriptscriptstyle DOZZ }_b(\alpha_3,\alpha_2,\alpha_1)& \equiv & \Big\langle
\Phi_{\alpha_3}(\infty,\infty)
\Phi_{\alpha_2}(1,1)
\Phi_{\alpha_1}(0,0)\Big\rangle_{\!\!\li}   \\
&=&\label{DOZZ:threepoint}
M^\li_b
{\sf C}^\li_b(\alpha_3,\alpha_2,\alpha_1)
\bar {\sf C}^\li_b(\alpha_3,\alpha_2,\alpha_1)
\\[6pt]
\nonumber
{\sf C}^\li_b(\alpha_3,\alpha_2,\alpha_1)
& = &
\frac{
\Gamma_{b}\left( \alpha_{123}- Q\right)\,
\prod\limits_{i<j}
\Gamma_{b}\left( \alpha_{ij}\right) }
{
\prod\limits_{i}\sqrt{
\Gamma_{b}\left( 2 \alpha_i\right) \Gamma_{b}\left(2Q- 2 \alpha_i\right)}}
\ ,
\\
\nonumber
\bar {\sf C}^\li_b(\alpha_3,\alpha_2,\alpha_1)
&=&
\frac{
\Gamma_{b}\left( 2Q-\alpha_{123}\right)\,
\prod\limits_{i<j}
\Gamma_{b}\left(Q- \alpha_{ij}\right) }
{
\prod\limits_{i}\sqrt{
\Gamma_{b}\left(Q- 2 \alpha_i\right) \Gamma_{b}\left( 2 \alpha_i-Q\right)}}
\ ,
\end{eqnarray}
where $\alpha_{123}=\alpha_1+\alpha_2+\alpha_3,\ \alpha_{12}=\alpha_1+\alpha_2-\alpha_3,$ etc\footnote{A definition of the Barnes double gamma
function along with some of its properties is given in Appendix C.}.

The same can be done for the GMM structure constant
\begin{eqnarray}
\nonumber
C^{\rm\scriptscriptstyle GMM }_b(\alpha_3,\alpha_2,\alpha_1)
& \equiv & \Big\langle
\Phi_{\alpha_3}(\infty,\infty)
\Phi_{\alpha_2}(1,1)
\Phi_{\alpha_1}(0,0)\Big\rangle_{\!\!\gi}\\
\label{GMM:threepoint}
&=&
M^\gi_b
{\sf C}^\gi_b(\alpha_3,\alpha_2,\alpha_1)
\bar {\sf C}^\gi_b(\alpha_3,\alpha_2,\alpha_1),
\\[6pt]
\nonumber
{\sf C}^\gi_b(\alpha_3,\alpha_2,\alpha_1)
& = &
\frac
{
\prod\limits_{i}\sqrt{
\Gamma_{b}\left( b+ 2 \alpha_i\right) \Gamma_{b}\left(2b^{-1}-b- 2 \alpha_i\right)}
}
{
\Gamma_{b}\left( \alpha_{123}- b^{-1}+2b\right)\,
\prod\limits_{i<j}
\Gamma_{b}\left( \alpha_{ij}+b\right) }
\ ,
\\
\nonumber
\bar {\sf C}^\gi_b(\alpha_3,\alpha_2,\alpha_1)
&=&
\frac
{
\prod\limits_{i}\sqrt{
\Gamma_{b}\left( b^{-1}- 2 \alpha_i\right) \Gamma_{b}\left(2b -b^{-1}+ 2 \alpha_i\right)}
}
{
\Gamma_{b}\left(2b^{-1} -b- \alpha_{123}\right)\,
\prod\limits_{i<j}
\Gamma_{b}\left( b^{-1}-\alpha_{ij}\right) }
\ .
\end{eqnarray}
Decompositions (\ref{DOZZ:threepoint}) and (\ref{GMM:threepoint}) are
based on the splitting  of
$$
\Upsilon_b(x)={1\over \Gamma_b(x)\Gamma_b(b^{-1}+b-x)}
$$
into the Barnes gamma factors and are  to some extend arbitrary.
What is important for further calculations is that one uses the same splitting of $\Upsilon_b$
for terms with the same arguments in the DOZZ and the GMM structure constants. Once   splittings are chosen
one can introduce the off-diagonal extensions of the Liouville structure constants:
\begin{eqnarray}
\label{gistru}
C^{\sigma }_b(\alpha_3,\alpha_2,\alpha_1,\overline\alpha_3,\overline\alpha_2,\overline\alpha_1)
&=&
M^\sigma_b
{\sf C}^\sigma_{b}(\alpha_3,\alpha_2,\alpha_1)
\bar {\sf C}^\sigma_b(\overline\alpha_3,\overline\alpha_2,\overline\alpha_1),\;\;\;\sigma=\li, \gi.
\end{eqnarray}
We shall split in a similar manner
the NS super-Liouville structure constants \cite{Rashkov:1996jx,Poghosian:1996dw} (see also \cite{Fukuda:2002bv,Belavin:2007gz}).
In the symmetric normalization the NS structure constants (which coincide with the SL structure constants) can be written as
\begin{eqnarray}
\label{constant1}
C^\NS_b(\alpha_3,\alpha_2,\alpha_1) & = & \Big\langle
\Phi_{\alpha_3}(\infty,\infty)
\Phi_{\alpha_2}(1,1)
\Phi_{\alpha_1}(0,0)\Big\rangle_{\!\!\SL}\\
\nonumber
& =& M^\NS_b
{\sf C}^\NS_b(\alpha_3,\alpha_2,\alpha_1)
\bar {\sf C}^\NS_b(\alpha_3,\alpha_2,\alpha_1),
\\[6pt]
\label{constant2}
\widetilde C_b^\NS(\alpha_3,\alpha_2,\alpha_1)& = & \Big\langle
\Phi_{\alpha_3}(\infty,\infty)
{\cal G}_{-{1\over 2}}\overline {\cal G}_{-{1\over 2}}
\Phi_{\alpha_2}(1,1)
\Phi_{\alpha_1}(0,0)
\Big\rangle_{\!\!\SL}
\\
& =& 2iM^\NS_b
{\sf D}^\NS_b(\alpha_3,\alpha_2,\alpha_1)
\bar {\sf D}^\NS_b(\alpha_3,\alpha_2,\alpha_1) \nonumber,
\end{eqnarray}
where
\begin{eqnarray*}
{\sf C}^\NS_b(\alpha_3,\alpha_2,\alpha_1)
& = &
\frac{
\Gamma^\NS_{b}\left( \alpha_{123}- Q\right)\,
\prod\limits_{i<j}
\Gamma^\NS_{b}\left( \alpha_{ij}\right) }
{
\prod\limits_{i}\sqrt{
\Gamma^\NS_{b}\left( 2 \alpha_i\right) \Gamma^\NS_{b}\left(2Q- 2 \alpha_i\right)}}
\ ,
\\
\bar {\sf C}^\NS_b(\alpha_3,\alpha_2,\alpha_1)
&=&
\frac{
\Gamma^\NS_{b}\left( 2Q-\alpha_{123}\right)\,
\prod\limits_{i<j}
\Gamma^\NS_{b}\left(Q- \alpha_{ij}\right) }
{
\prod\limits_{i}\sqrt{
\Gamma^\NS_{b}\left(Q- 2 \alpha_i\right) \Gamma^\NS_{b}\left( 2 \alpha_i-Q\right)}}
\ ,
\\
{\sf D}^\NS_b(\alpha_3,\alpha_2,\alpha_1)
& = &
\frac{
\Gamma^\R_{b}\left( \alpha_{123}- Q\right)\,
\prod\limits_{i<j}
\Gamma^\R_{b}\left( \alpha_{ij}\right) }
{
\prod\limits_{i}\sqrt{
\Gamma^\NS_{b}\left( 2 \alpha_i\right) \Gamma^\NS_{b}\left(2Q- 2 \alpha_i\right)}}
\ ,
\\
\bar {\sf D}^\NS_b(\alpha_3,\alpha_2,\alpha_1)
&=&
\frac{
\Gamma^\R_{b}\left( 2Q-\alpha_{123}\right)\,
\prod\limits_{i<j}
\Gamma^\R_{b}\left(Q- \alpha_{ij}\right) }
{
\prod\limits_{i}\sqrt{
\Gamma^\NS_{b}\left(Q- 2 \alpha_i\right) \Gamma^\NS_{b}\left( 2 \alpha_i-Q\right)}}
\ ,
\end{eqnarray*}
and
\begin{eqnarray*}
\Gamma^\NS_b(x)
&=&\textstyle
\Gamma_b\left(\frac{x}{2}\right)\Gamma_b\left(\frac{x+Q}{2}\right),\;\;\;\;\;\;
\Gamma^\R_b(x)
\;=\;
\Gamma_b\left(\frac{x+b}{2}\right)\Gamma_b\left(\frac{x+b^{-1}}{2}\right).
\end{eqnarray*}

The modular properties of the toric partition function of the SL theory
imply that the GSO projection on the even with respect
to the common left and right parity subspace is a necessary consistency condition. This yields
\begin{equation}
\label{hilbertevenNS}
{\cal H}^{\rm\scriptscriptstyle SL} = \int\limits_{\mathbb{R}_+} dp\,
\left[
\left({\cal A}_{\Delta_p} \right)_{\!\!\rm\scriptscriptstyle even}
\!\!\otimes
\left( \overline{{\cal A}}_{\Delta_p}\right)_{\!\!\rm\scriptscriptstyle even}
\oplus
\left({\cal A}_{\Delta_p}\right)_{\!\!\rm\scriptscriptstyle odd}
\!\!\otimes
\left( \overline{{\cal A}}_{\Delta_p}\right)_{\!\!\rm\scriptscriptstyle odd}\,
\right].
\end{equation}
On the r.h.s.\ of (\ref{mrel}) it corresponds to
\begin{equation}
\label{hilberteven}
{\cal H}^{\rm\scriptscriptstyle LL} = \int\limits_{\mathbb{R}_+} dp\,
\begin{array}[t]{c}
{\displaystyle\bigoplus} \\[-5pt]
{\scriptscriptstyle
j,\bar j \,\in\, \mathbb{Z}}
\\[-7pt]
{\scriptscriptstyle
j+\bar j \,\in\, 2\mathbb{Z}}
\end{array}
\left({\sf V}_{\Delta^{\li}(p, j)}\otimes \overline{\sf V}_{\Delta^{\li}(p, \bar j)}\right)
\otimes
\left({\sf V}_{\Delta^{\gi}(p, j)}\otimes \overline{\sf V}_{\Delta^{\gi}(p, \bar j)}\right).
\end{equation}
The spaces of states ${\cal H}^{\rm\scriptscriptstyle SL}$, ${\cal H}^{\rm\scriptscriptstyle LL}$ and structure constants (\ref{gistru}), (\ref{constant1}),
(\ref{constant2})  unambiguously   define both sides of the SL-LL correspondence in the NS sector.
The main statement we are concerned with in this paper can be formulated as follows.

{\it
If the relative normalization condition
\begin{equation}
\label{normcond}
{ M^\NS_b\over M^\li_\bli M^\gi_\bgi}= {\Upsilon_\bgi (\bgi)
{\textstyle\Upsilon_b\left(b\right)\Upsilon_b\left({Q\over 2}\right)}
\over
\Upsilon_\bli (\bli)
}
b^{-{b^2\over 1-b^2} {Q^2\over 2}}
\left({1-b^2\over 2}\right)^{-{Q^2\over 4}+{1\over 2}}
\end{equation}
is satisfied then there exists a map ${\cal I}$ from the algebra of local fields of the NS sector of the GSO projected SL model to the algebra of local fields
 of the GSO projected  LL model preserving all correlation functions on the sphere:
 $$
 \Big\langle
\Phi_{\alpha_n}\dots
\Phi_{\alpha_2}
\Phi_{\alpha_1}\Big\rangle_{\!\!\SL}
=
\Big\langle
{\cal I}(\Phi_{\alpha_n})\dots
{\cal I}(\Phi_{\alpha_2})
{\cal I}(\Phi_{\alpha_1})\Big\rangle_{\!\!\LL}.
 $$
}

Let us remark that our motivation goes beyond a self contained CFT proof of the statement above. It characterizes only the simplest of the relations
suggested by  the AGT correspondence  \cite{Alfimov:2013cqa} and by the analogy with the rational cases \cite{Wyllard:2011mn}.
One may expect for instance an exact relation between the para-Liouville theories \cite{Bershtein:2010wz} and projected tensor product of several copies of the Liouville theory.
It seems that better understanding of all equivalences of this type would provide a new insight into underlying structures of  at least certain  class of nonrational CFT models.

The organization of the paper is as follows. In Section 2.1 the basic concepts of the free field realization of the NS sector are introduced mainly for the notational convenience.
In Section 2.2 a special attention is paid to the description of the reflection map and the reflected modes in the NS Verma modules which play an essential role in the constructions of \cite{Belavin:2011sw}. The properties of the states generated by solely  fermionic modes are summarized in Propositions 1 and 2 of Subsection 2.2. They are proven in Appendix A.

Section 3 is devoted to the  properties of basic chiral structures related to the chiral symmetry algebra ${\sf A}_{\NS}$ of the SL theory.
In Subsection 3.1 we review  decomposition (\ref{decva}) of the ${\sf A}_{\NS}$ Verma module into projected tensor product of Virasoro Verma modules constructed in \cite{Belavin:2011sw}.
In Subsection 3.2 we analyze 3-point ${\sf A}_{\NS}$ conformal blocks and derive
the central for the whole paper formula for the  blow up factor.
This is the main and technically the most involved result of the present paper.
It is based on Propositions 1 and 2 of Subsection 2.2 and on the formula for generalized Selberg integral derived in Appendix B.
In Subsection 3.3 we investigate the ${\sf A}_{\NS}$ algebra 4-point conformal blocks.
As a side result we obtain expressions for the NS super-Virasoro blocks in terms of the Virasoro ones and the blow up factors.

In Section 4 we complete our proof of the SL-LL equivalence for arbitrary correlation functions on the sphere in the GSO projected NS sector.
To this end we analyze in Subsection 4.1 relations between chiral structures of all the there theories involved.
It turns out that one can prove the SL-LL equivalence for the left and for the right chiral parts separately.
It is also remarkable that although the left and the right structure constants are different
the relations they satisfy involve exactly the same coefficients.
This is a consequence of two types of identities for Barnes gamma functions which are derived in Appendix C.
In Section 4.2 we analyze how the relations between chiral structures result in an equivalence of full CFT theories.
We construct an exact map between local fields and show it preserves all the 3-point functions and the factorization
of correlation functions on the sphere.

There are several possible continuations of the present paper.
The most obvious is a completion of the proof for the whole GSO projected models.
With the results of \cite{Schomerus:2012se} it seems that
an extension to the Ramond sector is straightforward. The same concerns
an extension to arbitrary closed surfaces which would require an analysis of 1-point toric functions.
More challenging is an extension to bordered surfaces.

The chiral structure constants played an essential role
in the calculations of the present paper.They also show up
in the relative normalization of the fusing matrix and the 6j-symbol of a continuous series of representations of ${\cal U}_q(sl(2,\mathbb{R}))$
\cite{Teschner:2001rv}. This relation and  the orthogonality properties of the 6j-symbols yield
a general expression for structure constants
satisfying the crossing bootstrap equation on the Liouville theory spectrum
(\cite{Teschner:2001rv}, formula (252)).
To what extend the chiral structure constants form fundamental building blocks of Liouville type models is by itself important problem
related to the questions of classification and new model building.
The double Liouville theory is the first example of nonrational models with the spectrum being
 diagonal in the continuous and non-diagonal in the discrete parameter.
It is interesting if there are other consistent models of this type.

Another possible topic is to investigate already mentioned relations between the para-Liouville theories
and projected tensor products of several copies of the Liouville theory which are well supported by
the AGT correspondence \cite{Bonelli:2011jx,Bonelli:2011kv,Wyllard:2011mn}.
For rational models another equivalence is known \cite{Crnkovic:1989ug}. It relates the tensor products of the ${\cal N}=2$ super minimal models and the Ising model
to projected tensor products of the ${\cal N}=1$ super minimal models and the parafermionic models.
Although  the AGT 4-dimensional counterpart is not expected in this case it would be very interesting
to check if there is a non-rational version of this correspondence.

\section{Free field representation of the NS sector}
\setcounter{equation}{0}
\subsection{Chiral vertex operators}

Let ${\cal F}_b$ be the bosonic Fock space generated by modes of the Heisenberg algebra
\begin{equation}
\label{Heisenberg:algebra:bosonic}
[{c}_m,{c}_n] = m\delta_{m+n,0},
\hskip 1cm
m,n \in {\mathbb Z}\setminus \{0\},
\end{equation}
out of the vacuum $\ket{0_b}$ satisfying ${c}_m\ket{0_b}=0\;{\rm for}\;m>0$. We denote by
${\cal F}_\NS$  the fermionic NS Fock space generated by modes of the algebra
\begin{equation}
\label{Heisenberg:algebra:fermionic}
\{\psi_k,\psi_l\} =\delta_{k+l,0},
\hskip 1cm
k,l\in\mathbb{Z}+
{\textstyle \frac12},
\end{equation}
out of the vacuum $\ket{0_f}$ obeying
\(
\psi_r\ket{0_f}= 0\;{\rm for}\;r>0.
\)
The super-scalar Hilbert space ${\cal H}_\NS$ of the NS sector can be introduced as a tensor product
$$
{\cal H}_\NS = {\cal H}_0 \otimes {\cal F}_b \otimes {\cal F}_{\NS},
$$
where
${\cal H}_0 \cong L^2({\mathbb R})$ is a representation space of the zero mode operators ${\p,{\sf q}},$ satisfying
\(
[{\p,{\sf q}}] = -i.
\)
A scalar product  in ${\cal H}_\NS$ is defined by imposing modes' conjugation properties
\begin{equation}
\label{hermitian:conjugation}
{\p}^{\dag} = {\p},
\hskip 1cm
{\sf q}^{\dag} = {\sf q},
\hskip 1cm
{c}_n^\dag = {c}_{-n},
\hskip 1cm
\psi_k^\dag = \psi_{-k},
\end{equation}
together with the normalization condition
\[
\langle\,{p_1}\!\ket{p_2} = \delta(p_1-p_2),
\]
where
$
\ket{p} = |p\drangle\otimes\ket{0_b}\otimes\ket{0_f}$ and $
{\p}|p\drangle = p|p\drangle.
$
The space  ${\cal H}_\NS$ can be seen as a direct integral of Hilbert spaces:
\begin{equation}
\label{Hdec}
{\cal H}_\NS= \int\limits_{\mathbb R}\! dp\, {\cal H}_p \;,\;\;\;\;\;{\cal H}_p = |p\drangle\otimes {\cal F}_b \otimes {\cal F}_{\NS},
\end{equation}
with the scalar product in ${\cal H}_p$ determined by conjugation properties (\ref{hermitian:conjugation}) and the normalization
$\langle\,p\ket{p}=1$.
Vectors $c_{-M} \psi_{-K} \ket{p}$ where
$$
\begin{array}{lllllllllll}
c_{-M}
&=&
c_{-m_j}\ldots c_{-m_1},&&
m_j \geqslant \ldots \geqslant  m_1,&&m_r\in \mathbb{N},
\\
\psi_{-K} &=&\psi _{-k_i}\ldots \psi_{-k_1},&&
k_i > \ldots  > k_1 ,&&
 k_s\in \mathbb{N}-\textstyle \frac12,
\end{array}
$$
form  an orthogonal basis in ${\cal H}_p,$
$$
\bra{p} \psi_{-K'}^\dagger c_{-M'}^\dagger c_{-M} \psi_{-K} \ket{p}= N_{M K} \delta_{M,M'}\delta_{K,K'}.
$$

The space  ${\cal H}_\NS$ carries two representations ($\pm$)
of the NS algebra with the same central charge $c = \frac32 + 3Q^2$:
\begin{eqnarray}
\label{NS:alebra:rep:1}
\nonumber
L_0(\pmp)
& = &
\sum\limits_{m \geqslant 1}{c}_{-m}{c}_m +
\sum\limits_{k\geqslant \frac12}k\,\psi_{-k}\psi_{k} +
\frac18Q^2 + \frac12 {\sf p}^2,
\\
L_n({\pmp})
& = &
\frac12\sum\limits_{m\neq 0,n}\!{c}_{n-m}{c}_m +
\frac12\sum\limits_{k\in {\mathbb Z} +\frac12}\hskip -4pt k\,\psi_{n-k}\psi_{k}+
\left(\frac{inQ}{2}\pm{\sf p}\right){c}_n,
\hskip 5mm  n \neq 0,
\\
\nonumber
G_k({\pmp}) & = & \sum\limits_{m\neq 0}{c}_{m}\psi_{k-m}+ (iQk \pm{\sf p})\psi_k,
\end{eqnarray}
and the same highest weight states
\begin{eqnarray*}
L_0({\pmp})\ket{p} &=&\Delta_p\ket{p},
\\
L_0({\pmp})\ket{-p} &=&\Delta_p\ket{-p},\;\;\;\;\Delta_p \;=\; \frac18 Q^2 + \frac12p^2,
\\
L_n({\pmp})\ket{p} &=& G_{k}({\pmp})\ket{p} = 0,
\\
L_n({\pmp})\ket{-p} &=& G_{k}({\pmp})\ket{-p} = 0, \;\; n,k > 0.
\end{eqnarray*}
Let us define the superscalar components
\begin{eqnarray*}
\varphi_<(z)
& = &
-i\sum\limits_{n=1}^\infty \frac{{c}_{-n}}{n}\,z^{n},
\hskip 1cm
\varphi_>(z)
\; = \;
i\sum\limits_{n = 1}^{\infty} \frac{{c}_n}{n}\, z^{-n},
\hskip 1cm
\psi(z)
\; = \;
\sum\limits_{k\in {\mathbb Z} + \frac12}\psi_k\, z^{-k-\frac12}
\end{eqnarray*}
and the ordered exponential
\begin{eqnarray}
\label{ordered:exponential}
{\sf E}^\alpha(z)
& = &
z^{-\Delta_\alpha}\,
{\rm e}^{\frac12\alpha{\sf q}}\,
{\rm e}^{\alpha\varphi_{<}(z)}\,
z^{-i\alpha{\sf p}}\,
{\rm e}^{\alpha\varphi_{>}(z)}\,
{\rm e}^{\frac12\alpha{\sf q}},
\end{eqnarray}
where $\Delta_\alpha = \frac12\alpha(Q-\alpha).$
Explicit calculations give:
\begin{eqnarray}
\nonumber
\left[{\sf p},{\sf E}^\alpha(z)\right]
& = &
-i\alpha{\sf E}^\alpha(z),
\\[4pt]
\nonumber
\left[L_n({\sf p}),{\sf E}^\alpha(z)\right]
& = &
z^n\big(z\partial_z +(n+1)\Delta_\alpha\big){\sf E}^\alpha(z),
\\[4pt]
\label{simple:commutators}
\left[G_k({\sf p}),{\sf E}^\alpha(z)\right]
& = &
-i\alpha\,z^{k+\frac12}\, \psi(z){\sf E}^\alpha(z),
\\[4pt]
\nonumber
\left[L_n({\sf p}),\psi(z){\sf E}^\alpha(z)\right]
& = &
z^n\big(z\partial_z +(n+1)\big(\Delta_\alpha+{\textstyle \frac12}\big)\big)\psi(z){\sf E}^\alpha(z),
\\[4pt]
\nonumber
\left\{G_k({\sf p}),\psi(z){\sf E}^\alpha(z)\right\}
& = &
\frac{i}{\alpha} z^{k-\frac12}
\big(z\partial_z +(2k+1)\Delta_\alpha\big){\sf E}^\alpha(z).
\end{eqnarray}
The ordered exponential is thus an NS super-primary field which can be seen as a family of maps
\[
{\sf E}^\alpha(z):\;\ {\cal H}_p\ \mapsto\ {\cal H}_{p-i\alpha}.
\]
In general
\[
{\sf E}^{\alpha_n}(z_n)\ldots{\sf E}^{\alpha_1}(z_1) :\;\
{\cal H}_p\ \mapsto\ {\cal H}_{p-i(\alpha_1+\ldots+\alpha_n)}.
\]
Iff the {\em neutrality condition}
\begin{equation}
\label{neutrality:condition}
q = p -i(\alpha_1+\ldots\alpha_n)
\end{equation}
holds the matrix elements of ordered exponentials
\[
\langle q| {\sf E}^{\alpha_n}(z_n)\ldots{\sf E}^{\alpha_1}(z_1)|p\rangle
\]
do not vanish and
coincide with the chiral correlators  of NS primary fields
\[
\langle\Delta_q| {\sf V}_{\Delta_{\alpha_n}}(z_n)\ldots{\sf V}_{\Delta_{\alpha_1}}(z_1)|\Delta_p\rangle.
\]

For $\alpha = b$ or $\alpha = \frac{1}{b}:$
\begin{eqnarray*}
\left[L_n({\sf p}),\psi(z){\sf E}^\alpha(z)\right]
& = &
z^n\big(z\partial_z +(n+1)\big)\psi(z){\sf E}^\alpha(z)
\; = \;
\partial_z\big(z^{n+1}\psi(z){\sf E}^\alpha(z)\big),
\\[4pt]
\left\{G_k({\sf p}),\psi(z){\sf E}^\alpha(z)\right\}
& = &
\frac{i}{\alpha} z^{k-\frac12}
\Big(z\partial_z +\big(k+{\textstyle\frac12}\big)\Big){\sf E}^\alpha(z)
\; = \;
\frac{i}{\alpha}
\partial_z\Big(z^{k+\frac12}{\sf E}^\alpha(z)\Big).
\end{eqnarray*}
Hence, for closed contours the screening charge operators
\[
{\sf Q}_b = \oint\!dz\ \psi(z){\sf E}^b(z)
\hskip 1cm {\rm and} \hskip 1cm
{\sf Q}_{\frac1b} = \oint\!dz\ \psi(z){\sf E}^{\frac1b}(z),
\]
satisfy the relations
\[
\big[L_m({\sf p}),{\sf Q}_b \big] \; = \; \big\{G_k({\sf p}),{\sf Q}_b\big\} \; = \; \big[L_m({\sf p}),{\sf Q}_{\frac1b} \big] \; = \; \big\{G_k({\sf p}),{\sf Q}_{\frac1b}\big\} \; = \; 0.
\]
In the free field model one can represent matrix element of an arbitrary NS chiral
primary field $V_{\Delta_\alpha}(z)$  in eight possible ways:
\begin{eqnarray}
\nonumber
\langle\Delta_q|V_{\Delta_\alpha}(z)|\Delta_p\rangle
& = &
\langle \pm q|{\sf E}^\alpha(z) {\sf Q}_b^r{\sf Q}_{\frac{1}{b}}^s|\pm p\rangle
\hskip .95cm
{\rm for}
\hskip .5cm
\pm q = \pm p -i\alpha -i\left(rb + sb^{-1}\right),
\\[-8pt]
\label{basic:rep:1}
\\
\nonumber
\langle\Delta_q|V_{\Delta_\alpha}(z)|\Delta_p\rangle
& = &
\langle \pm q|{\sf E}^{Q-\alpha}(z) {\sf Q}_b^r{\sf Q}_{\frac{1}{b}}^s|\pm p\rangle
\hskip .5cm
{\rm for}
\hskip .5cm
\pm q = \pm p +i\alpha - i\left((r+1)b + (s+1)b^{-1}\right),
\end{eqnarray}
with uncorrelated signs in front of $p$ and $q.$

Under similar restrictions on the ``momenta'' $p$ and $q,$ there also  exist free field representations of matrix elements of the operator
\[
*V_{\Delta_\alpha}(1)=
{\cal G}_{-\frac12}V_{\Delta_\alpha}(1)
=[{ G}_{-\frac12},V_{\Delta_\alpha}(1)] .
\]
Using (\ref{simple:commutators}) we get
\[
*V_{\Delta_\alpha}(1) = -i\alpha \psi(1) {\sf E}^\alpha(1)
\]
hence
\begin{eqnarray}
\label{basic:rep:2}
\langle\Delta_q|\!*\!V_{\Delta_\alpha}(z)|\Delta_p\rangle
& = &
-i\alpha\langle \pm q|\psi(1){\sf E}^\alpha(z) {\sf Q}_b^r{\sf Q}_{\frac{1}{b}}^s|\pm p\rangle
\end{eqnarray}
for
$\pm q = \pm p -i\alpha -i\left(rb + sb^{-1}\right)$ and
\begin{eqnarray}
\label{basic:rep:3}
\langle\Delta_q|\!*\!V_{\Delta_\alpha}(z)|\Delta_p\rangle
& = &
-i(Q-\alpha)\langle \pm q|\psi(1){\sf E}^{Q-\alpha}(z) {\sf Q}_b^r{\sf Q}_{\frac{1}{b}}^s|\pm p\rangle
\end{eqnarray}
for $\pm q = \pm p +i\alpha - i\left((r+1)b + (s+1)b^{-1}\right)\!.$ Since the fermion field $\psi(z)$ and the screening charges are odd with respect to the
fermion parity operator $(-1)^F$ defined by
\[
\left\{(-1)^F,\psi_k\right\} = \left[(-1)^F,c_m\right] = 0,
\]
the total number of screening charges appearing in (\ref{basic:rep:1}) must be even, while the total number of screening charges appearing in (\ref{basic:rep:2})
and in (\ref{basic:rep:3}) must be odd.

\subsection{Reflection map in the NS sector}

For each level $t\in \frac12\mathbb N$ we introduce the transition matrices  $S^t(\pm p)$:
\begin{eqnarray}
\label{transition_matrix}
\nonumber
 L_{-M}(\pmp)G_{-K}(\pmp) \ket{p}
&=&
\sum\limits_{|N|+|L|=t}  S^{\,t}_{NL,MK}(\pm p)\, c_{-N} \psi_{-L} \ket{p},
\\[-6pt]
\\[-6pt]
\nonumber
\nonumber
S^{\,t}_{NL,MK}(\pm p) &=& \frac{\bra{p} \psi_{-L}^\dagger  c_{-N}^\dagger L_{-M}(\p) G_{-K}(\p)  \ket{\pm p}}{N_{NL}}.
\end{eqnarray}
The matrix $S^{\,t}$ (with a different normalization)  was studied in \cite{Kato:1987qda} where the formula for its determinant was found. It takes the form
\begin{equation}
\label{kac_for_s}
\det S^t(\pm p) = {\rm const} \begin{array}[t]{c}
{\displaystyle\prod} \\[-6pt]
{\scriptscriptstyle
1 \leqslant\, rs \,\leqslant\, 2t}
\\[-8pt]
{\scriptscriptstyle
r + s\,\in\, 2{\mathbb N}
}
\end{array}\!\!\!\!
\left(  p\mp p_{rs}    \right)^{P_\NS(n-\frac{rs}{2})},\;\;\;\;\;p_{rs}=
i\!\left(r b +{s\over b}\right) ,\;\;\;\;Q=b+{1\over b},
\end{equation}
where multiplicities  $P_\NS(f)$ are defined by the generating function
$$
\sum\limits_{f\in\frac12\mathbb{N}}P_\NS(f)\,x^f= \prod\limits_{\scriptscriptstyle m\in \mathbb{N}}\frac{1}{1-x^m}
\prod\limits_{\scriptscriptstyle k\in \mathbb{N}-\frac12}(1+x^k)
 \; .
$$
For real $Q$ and $p$
\begin{equation}
\label{dagger}
L_n({ p})^\dagger = L_{-n}({ p}),
\hskip 1cm
G_{k}({ p})^{\dag} = G_{-k}({ p}),
\end{equation}
and $S^t(\pm p)$ are simply related to  the Gram matrix $B^t$ of the $t$-level Schapovalov form
\begin{eqnarray}
\nonumber
B^t(\Delta_p) &=& S^t(\pm p)^\dagger N^t S^t(\pm p),
\\[-6pt]
\\[-6pt]
\nonumber
B^{\,t}(\Delta_p)_{M'K',MK} &=& \bra{\Delta_p} G_{-K'}^\dagger L_{-M'}^\dagger L_{-M}G_{-K} \ket{\Delta_p}.
\label{gramm}
\end{eqnarray}

Let $\ket{\Delta_p}$ be the normalized highest weight vector in the NS Verma module ${\cal V}_{\Delta_p}$
of the highest weight ${\Delta_p}$. The maps
\begin{eqnarray}
\label{isom}
\nonumber
\imath_\pm (p) &:& \;{\cal V}_{\Delta_p} \ni L_{-M}G_{-K}\ket{\Delta_p} \to L_{-M}(\p)G_{-K}(\p) \ket{\pm p} \in {\cal H}_{\pm p},
\\[-6pt]
\\[-6pt]
\nonumber
\jmath_\pm (p) &:& \;{\cal V}_{\Delta_p} \ni L_{-M}G_{-K}\ket{\Delta_p} \to L_{-M}(-\p)G_{-K}(-\p) \ket{\pm p} \in {\cal H}_{\pm p}
\end{eqnarray}
are by construction morphisms of representations. Determinant formula (\ref{kac_for_s}) and hermitian
conjugation properties (\ref{dagger}) imply that for real $Q$ and $p$ these maps are unitary isomorphisms.
So is the composition
\begin{eqnarray*}
r(p) \;=\; \imath_-(p) \circ \imath_+(p)^{-1} &:&{\cal H}_{p} \to {\cal H}_{-p}
\end{eqnarray*}
 called the reflection map \cite{Zamolodchikov:1995aa,Teschner:2001rv}.\footnote{
In the present construction $r(p)\ket{p} = \ket{-p}$ which is in line with the symmetric normalization of the structure constants.
In the usual formulation of Liouville theory it is more convenient to normalize the reflection map
by the condition $r(p)\ket{p} = D_\NS \left( {Q\over 2}+ ip \right)\ket{-p}$ where $D_\NS$ denotes the NS reflection amplitude \cite{Rashkov:1996jx,Poghosian:1996dw}.}
Up to a multiplicative constant it is uniquely determined by the intertwining property
\begin{equation}
\label{intertwining}
L_m(-p) r(p) =r(p) L_m(p),\;\;\;G_k(-p) r(p) =r(p) G_k(p).
\end{equation}
One can study the relation between representations $ \{L({\scriptstyle +}{ p}),G({\scriptstyle +}{ p})\}$ and
$ \{L({\scriptstyle -}{ p}),G({\scriptstyle -}{p})\}$ introducing reflected modes in ${\cal H}_p$ \cite{Belavin:2011sw}
(see also \cite{Belavin:2011js} for a similar construction in the Virasoro case):
\begin{eqnarray}
\label{reflected modes}
c_m^{\rm \scriptscriptstyle R}(p)&=& r(p)^{-1} c_m \:r(p),
\;\;\;\;\;
\psi_k^{\rm \scriptscriptstyle R}(p)\;=\; r(p)^{-1} \psi_k\: r(p)
.
\end{eqnarray}
In principle they can be expressed as series of monomials in $c_m$ and $\psi_r$.
The general form of this transformation is  however not known. It can be calculated term by term from
intertwining property (\ref{intertwining}) which takes the form :
\begin{equation}
\label{equivalencerep}
\begin{array}{rcl}
&&\hspace{-150pt}
\frac12\sum\limits_{m\neq 0,n}\!{c}_{n-m}{c}_m +
\frac12\sum\limits_{k\in {\mathbb Z} +\frac12}\hskip -4pt k\,\psi_{n-k}\psi_{k}+
\left(\frac{inQ}{2}+{ p}\right){c}_n
\\
& = &
\frac12\sum\limits_{m\neq 0,n}\!{c}^\R_{n-m}{c}^\R_m +
\frac12\sum\limits_{k\in {\mathbb Z} +\frac12}\hskip -4pt k\,\psi^\R_{n-k}\psi^\R_{k}+
\left(\frac{inQ}{2}-{ p}\right){c}^\R_n
\\[15pt]
 {\displaystyle\sum\limits_{k\neq 0}}c_k\psi_{l-k} +\big(iQl+{ p}\big)\psi_l
&=&
{\displaystyle\sum\limits_{k\neq 0}}c^\R_k\psi^\R_{l-k} +\big(iQl-{ p}\big)\psi_l^\R
.
\end{array}
\end{equation}
The matrix of the reflection map $r^t(p)$ with respect to the basis $ c_{-M}\psi_{-K}\ket{p}$
$$
c^\R_{-N}(p)\psi^\R_{-L}(p)\ket{p}
=
\sum\limits_{N'L'}  r^t(p)_{N'L',NL} c_{-N'}\psi_{-L'} \ket{p}
$$
is simply related to the transition matrix
$$
r^t(p)= S^t(-p)^{-1}S^t(p).
$$
We shall also need expressions of the oscillation  bases $\{c_{-N}\psi_{-L}\ket{p}\}$, $\{c^\R_{-N}(p)\psi^\R_{-L}(p)\ket{p}\}$
in terms of $\{ L_{-M}(p)G_{-K}(p) \ket{p}\}$:
\begin{equation}
\label{osc_gen}
\begin{array}{rcl}
c_{-N}\psi_{-L}\ket{p}
&=&
\displaystyle
\sum\limits_{MK}  S^t(p)^{-1}_{MK,NL} L_{-M}( p)G_{-K}( p) \ket{p}
\\[12pt]
&=&
\displaystyle
\sum\limits_{MK,M'K'} N^t_{NL} \overline{S^t( p)}_{NL,M'K'}\,B^t(\Delta_p)^{-1}_{M'K',MK} L_{M}( p)G_{-K}( p) \ket{p},
\\[12pt]
c^\R_{-N}(p)\psi^\R_{-L}(p)\ket{p}
&=&
\displaystyle
\sum\limits_{MK}  S^t(-p)^{-1}_{MK,NL} L_{-M}(p)G_{-K}(p) \ket{p}
\\[12pt]
&=&
\displaystyle
\sum\limits_{MK,M'K'} N^t_{NL} \overline{S^t(-p)}_{NL,M'K'}\,B^t(\Delta_p)^{-1}_{M'K',MK} L_{M}(p)G_{-K}(p) \ket{p}.
\end{array}
\end{equation}

In the Fock space representation of an NS Verma module we used above  the modes
$c_m, \psi_k$ are elementary $p$-independent objects in contrast to their reflected counterparts
$c^\R_m(p), \psi^\R_k(p)$ which are complicated infinite series with coefficients being rational functions of momentum $p$.
A symmetric picture can be achieved by lifting  both sets of modes to the NS Verma module:
\begin{equation}
\label{tilde_modes}
\begin{array}{lllllllllll}
\tilde c_m (p) &=& \imath_+(p)^{-1} c_m \,\imath_+(p), && \tilde \psi_k (p) &=& \imath_+(p)^{-1} \psi_k\, \imath_+(p),
\\[2pt]
\tilde c^\R_m (p) &=& \imath_-(p)^{-1} c_m \,\imath_-(p), && \tilde \psi^\R_k (p) &=& \imath_-(p)^{-1} \psi_k\, \imath_-(p).
\end{array}
\end{equation}
In this formulation
\begin{equation}
\label{ref}
\tilde c^\R_m (p)=\tilde c_m (-p),\;\;\;\tilde \psi^\R_k (p)=\tilde \psi_k (-p),
\end{equation}
and the reflection map takes the form:
$$
\tilde r(p)= \imath_+(p)^{-1} \imath_-(p)\;:\;{\cal V}_{\Delta_p}\;\to\;{\cal V}_{\Delta_p}.
$$
Modes (\ref{tilde_modes}) can be used to
 construct two different orthogonal bases
in the NS Verma module ${\cal V}_{\Delta_p}$
\begin{equation}
\label{osc}
\begin{array}{rcl}
\tilde c_{-N}(p)\tilde \psi_{-L}(p)\ket{\Delta_p}
&=&
\displaystyle
\sum\limits_{MK}  S^t(p)^{-1}_{MK,NL} L_{-M}G_{-K} \ket{\Delta_p}
\\[12pt]
&=&
\displaystyle
\sum\limits_{MK,M'K'} N^n_{NL} \overline{S^t(p)}_{NL,M'K'}\,B^t(p)^{-1}_{M'K',MK} L_{-M}G_{-K} \ket{\Delta_p},
\\[12pt]
\tilde c^\R_{-N}(p)\tilde\psi^\R_{-L}(p)\ket{\Delta_p}
&=&
\displaystyle
\sum\limits_{MK}  S^t(-p)^{-1}_{MK,NL} L_{-M}G_{-K} \ket{\Delta_p}
\\[12pt]
&=&
\displaystyle
\sum\limits_{MK,M'K'} N^t_{NL} \overline{S^t(-p)}_{NL,M'K'}\,B^t(p)^{-1}_{M'K',MK} L_{-M}G_{-K} \ket{\Delta_p}.
\end{array}
\end{equation}
In a generic case (when the modules ${\cal V}_{\Delta_{rs}+{rs\over 2}}$ are irreducible) the coefficients in (\ref{osc}) have simple poles at
$p= p_{rs}$ and $p= -p_{rs}$ respectively.
In the following we shall need a more detailed information about the structure of pure fermionic states
$
\tilde \psi(p)_{-K}\ket{\Delta_p}.
$ This  can be conveniently summarized in the form of two propositions below. The proofs are  given in Appendix A.

\noindent {\bf Proposition 1} {\em The only possible singularities of
 the coefficients of the decomposition of the state $\tilde \psi(p)_{-k_m}\ldots \tilde\psi_{-k_1}(p)\ket{\Delta_p},\ k_m > \ldots > k_1,$
with respect to the base $L_{-N}G_{-L}\ket{\Delta_{p}}$
 are simple poles at
$$
p=p_{rs}, \;\;\;r+s\leqslant 2k_m+1,\;\;\; r,s \in {\mathbb N},\;\;\;r+s\in 2\mathbb{N}.
$$
}

Let us define polynomials
\begin{eqnarray}
\nonumber
\Omega(p,j) &=&
\left\{
\begin{array}{llllll}
 l^\NS(2ip+Q,2j)&&j>0\\
 l^\NS(-2ip+Q,-2j)&&j<0
 \end{array}\right.
\\
\label{defel}
l^\NS(x,n)&=&
\begin{array}[t]{c}
{\displaystyle\prod} \\[-5pt]
{\scriptscriptstyle
0 \leqslant r,s}
\\[-7pt]
{\scriptscriptstyle r+s <n}
\\[-7pt]
{\scriptscriptstyle
r+s\,\in\, 2{\mathbb N}
}
\end{array}
\left(x +rb + s{1\over b}\right).
\end{eqnarray}
Proposition 1 implies that all coefficients of the decompositions
\begin{equation}
\label{decompo}
\Omega(p,2k_m+1) \,\tilde \psi(p)_{-K}\ket{\Delta_p} =\sum a_{NL}(p) \, L_{-N}G_{-L}\ket{\Delta_p}
\end{equation}
are polynomials in $p$ variable. One can estimate  the degree of these polynomials.

\noindent {\bf Proposition 2} {\it Let $J=\{-{2j-1\over 2},\dots,-{1\over 2}\}$ and let $K\subset J$. Then
$$
\deg \left(\Omega(p,j) S^{t}(p)^{-1}_{NL,\emptyset K} \right)\leqslant \deg \Omega(p,j) - 2\#N-\#L
$$
where $ \;t=|K|$.
}


\section{Conformal blocks}
\setcounter{equation}{0}
\subsection{ Verma modules}

Let ${\cal V}_{\Delta_p}$ be the NS module with the central charge $c={3\over 2} +3Q^2, Q=b+b^{-1}$
and the highest weight
$
\Delta_p= {Q^2\over 8}+{p^2\over 2}.
$
We introduce another copy of the fermionic NS Fock space $\widetilde{ \cal F}_\NS$  generated by the modes
$$
\{f_k,f_{l}\} =\delta_{k+l,0},\;\;k,l\in\mathbb{Z}+\textstyle {1\over 2}
$$
out of the vacuum state $|\,{ 0_{\!\tilde f}}\,\rangle$, $f_k|\,{0_{\!\tilde f}}\,\rangle=0,\;k>0$.
The generators
\begin{equation}
\label{elef}
L^{\sf f}_n = \frac12\sum\limits_k k:\!f_{n-k}f_k\!:
\end{equation}
define on $\widetilde{ \cal F}_\NS$ a representation of the $c={1\over 2}$ Virasoro algebra which is a direct sum
of two irreducible highest weight representations
\begin{equation}
\label{dirsum}
\widetilde{\cal F}_\NS = {\sf V}_0\oplus {\sf V}_{1\over 2}
\end{equation}
with the highest weights $\Delta=0$ and $\Delta= {1\over 2}$, respectively.

We shall introduce an indefinite scalar product on
$\widetilde{ \cal F}_\NS$   by the relations
$$
f^\dagger_k = - f_{-k},\;\;\;\langle\,0_{\tilde f}\,|\,{0}_{\!\tilde f}\,\rangle= 1.
$$
With respect to this scalar product  
$
\left(L^{\sf f}_n\right)^\dagger = L^{\sf f}_{-n}
$
and direct sum (\ref{dirsum}) is orthogonal with the induced scalar product  positively definite on the first
summand and negatively definite on the second.

On the ${\sf A}_\NS$ algebra Verma module ${\cal A}_{\Delta_p}={\cal V}_{\Delta_p}\otimes \widetilde{ \cal F}_\NS$ one can construct two sets
of generators:
\begin{equation}
\label{virasoro gen}
\begin{array}{rcr}
L_n^{\li}
& = &\displaystyle
\frac{1}{1-b^2}L_n \;-\;\; \frac{1+2b^2}{1-b^2}L^{\sf f}_n
\; + \;
\;\frac{b}{1-b^2}\sum\limits_r f_{n-r}G_r,
\\[8pt]
L_n^{\gi}
& = &\displaystyle
\frac{1}{1-b^{-2}}L_n - \frac{1+2b^{-2}}{1-b^{-2}}L^{\sf f}_n + \frac{b^{-1}}{1-b^{-2}}\sum\limits_r f_{n-r}G_r.
\end{array}
\end{equation}
They form two mutually commuting Virasoro algebras \cite{Crnkovic:1989gy,Crnkovic:1989ug,Lashkevich:1992sb}
\begin{eqnarray*}
{[L_m^{\li}, L_n^{\li}]}&=& (m-n)L_{m+n}^{\li} + {c^{\li}\over 12} (m^3-m)\delta_{m+n,0},
\\
{[L_m^{\gi}, L_n^{\gi}]}&=& (m-n)L_{m+n}^{\gi} + {c^{\gi}\over 12} (m^3-m)\delta_{m+n,0},
\\
{[L_m^{\li}, L_n^{\gi}]}&=&0
\end{eqnarray*}
and satisfy the standard conjugation relations:
\begin{equation}
\label{conj vir}
\left(L^{\li}_n\right)^\dagger = L^{\li}_{-n},\;\;\;\;\left(L^{\gi}_n\right)^\dagger = L^{\gi}_{-n}.
\end{equation}
The corresponding central charges are given by (\ref{defblibgi}).

The problem of decomposing  the ${\sf Vir} \oplus {\sf Vir}$ representation
on ${\cal A}_{\Delta_p}$ into irreducible components has been analyzed in \cite{Belavin:2011sw}.
We shall briefly recall the main points of this derivation.
Using construction (\ref{osc}) in the NS Verma module ${\cal V}_{\Delta_p}$ one introduces a family of states:
\begin{equation}
\label{hwstates}
\begin{array}{rcll}
\ket{p,0}
&=&
 \ket{\Delta_p}\otimes |\,{0}_{\!\tilde f}\,\rangle \in {\cal A}_{\Delta_p},&
\\[6pt]
\ket{p,j}& = & \tilde\chi_{-\frac{2j-1}{2}}(p)\ldots \tilde\chi_{-\frac32}(p)\,\tilde\chi_{-\frac12}(p)\ket{p,0}&\hspace{0pt}{\rm for}\;j>0,
\\[6pt]
\ket{p,j} & = & \tilde\chi^\R_{-\frac{2|j|-1}{2}}(p)\ldots\tilde\chi^\R_{-\frac32}(p)\,\tilde\chi^\R_{-\frac12}(p)\ket{p,0}&\hspace{0pt}{\rm for}\;j<0,
\end{array}
\end{equation}
where
$
j\in \mathbb{Z}
$
and
\[
\tilde\chi_k(p) = f_k - i\tilde \psi_k(p),
\hskip 1cm
\tilde\chi^\R_k(p) = f_k - i\tilde \psi^\R_k(p),
\hskip 1cm
k \in {\mathbb Z} + \textstyle\frac12.
\]
The modes can be conveniently organized into local fields
\begin{equation}
\label{fields}
\begin{array}{llllllllll}
\tilde \psi(\xi ) |_{{\cal V}_{\Delta_p}\otimes \tilde {\cal F}_\NS}&=& \sum\limits_{k\in \mathbb{Z}+{1\over 2}} \tilde\psi_k(p)\xi^{-k-{1\over 2}},
&&
\tilde \psi^\R(\xi ) |_{{\cal V}_{\Delta_p}\otimes \tilde {\cal F}_\NS}&=& \sum\limits_{k\in \mathbb{Z}+{1\over 2}} \tilde\psi^\R_k(p)\xi^{-k-{1\over 2}},
\\
\tilde \chi(\xi ) |_{{\cal V}_{\Delta_p}\otimes \tilde {\cal F}_\NS}&=& \sum\limits_{k\in \mathbb{Z}+{1\over 2}} \tilde\chi_k(p)\xi^{-k-{1\over 2}},
&&
\tilde \chi^\R(\xi ) |_{{\cal V}_{\Delta_p}\otimes \tilde {\cal F}_\NS}&=& \sum\limits_{k\in \mathbb{Z}+{1\over 2}} \tilde\chi^\R_k(p)\xi^{-k-{1\over 2}}.
\end{array}
\end{equation}

Note that (\ref{ref}) implies
$
\tilde\chi^\R_k(p) =\tilde\chi_k(-p)
$ hence
$$
\ket{p,-j}=\ket{-p,j}.
$$
By construction
$$
\tilde\chi_k(p)^\dagger = -\tilde\chi_k(p),\;\;\;\tilde\chi^\R_k(p)^\dagger = -\tilde\chi^\R_k(p)
$$
and
$$
\{\tilde\chi_k(p),\tilde\chi_l(p)\} = \{\tilde\chi^\R_k(p),\tilde\chi^\R_l(p)\} = 0 \;\;\;{\rm for}\;{\rm all} \;\;k,l\in \mathbb{Z} + \textstyle {1\over 2}.
$$
It follows that for all $j\in \mathbb{Z}$ the states $\ket{p,j}$ are of zero norm
($
\langle\,p,j\,\ket{p,j}=0
$).

In the Fock space ${\cal F}_\NS\otimes\tilde{\cal F}_\NS$ the modes $\tilde\chi_k(p), \tilde\chi^\R_k(p)$ are  represented by
\[
\chi_k = f_k - i \psi_k,
\hskip 1cm
\chi^\R_k(p) = f_k - i \psi^\R_k(p),
\hskip 1cm
k \in {\mathbb Z} + \textstyle\frac12.
\]
With the help of both representations (\ref{equivalencerep}) one can compute the commutation relations
\begin{eqnarray*}
{[L_n^{\li}+L^{\gi}_n,\chi_r]}&=& -\left(\textstyle{1\over 2}n+r\right)\chi_{r+n},
\\
{[b L_n^{\li}+b^{-1}L^{\gi}_n,\chi_r]}&=& -\left((n+r)Q + ip \right)\chi_{r+n} +i \sum\limits_{m\neq 0} c_m \chi_{r+n-m},
\\
{[L_n^{\li}+L^{\gi}_n,\chi_r^\R]}&=& -\left(\textstyle{1\over 2}n+r\right)\chi_{r+n}^\R,
\\
{[b L_n^{\li}+b^{-1}L^{\gi}_n,\chi_r^\R]}&=& -\left((n+r)Q -i p \right)\chi_{r+n}^\R +i \sum\limits_{m\neq 0} c_m \chi_{r+n-m}^\R,
\end{eqnarray*}
where
\begin{eqnarray*}
L_n^{\li}+L_n^{\gi} &=& L_n + \frac12 \sum_{k=-\infty}^{\infty} k : f_{n-k} f_k:,
\\
b \, L_n^{\li}+ b^{-1} \, L_n^{\gi} &=& Q \sum_{k=-\infty}^{\infty} k : f_{n-k} f_k: - \sum_{k=-\infty}^{\infty}  f_{n-k} G_k.
\end{eqnarray*}
Using these formulae one shows that states (\ref{hwstates})
are highest weight states with respect to both Virasoro algebras
\begin{equation}
\label{hwconditions}
\begin{array}{llllllllll}
L^{\li}_0 \ket{p,j} &=& \Delta^{\li} (p, j) \ket{p, j},&&L^{\li}_n \ket{p, j} &=& 0 \;\;\; {\rm for}\;\;\; n> 0,
\\[6pt]
L^{\gi}_0 \ket{p,j} &=& \Delta^{\gi} (p, j) \ket{p, j},&&L^{\gi}_n \ket{p, j} &=& 0 \;\;\; {\rm for}\;\;\; n> 0,
\\[6pt]
\end{array}
\end{equation}
and
\begin{equation}
\label{defaliagi}
\begin{array}{rcl}
\Delta^{\li}(p, j) &=&\alpha^\li(Q^\li-\alpha^\li)
\;=\;
\frac14\left(Q^{\li}\right)^2 + \left(p^{\li} + \frac{i}{2}j\,b^{\li}\right)^2
\;=\;
 \frac{1}{1-b^2}\left(\frac{Q^2}{8} + \frac{\left(p + ij\,b\right)^2}{2}\right),
\\
\Delta^{\gi}(p, j)&=&
\alpha^\gi(\alpha^\gi-Q^\gi)
\;=\;
-\frac14\left(Q^{\gi}\right)^2 - \left(p^{\gi}+ \frac{i}{2}\,\frac{j}{b^{\gi}}\right)^2
\;=\;
\frac{1}{1-b^{-2}}\left(\frac{Q^2}{8} + \frac{\left(p + ij\, b^{-1}\right)^2}{2}\right),
\\[6pt]
\alpha^\li &=& \frac{\alpha}{\sqrt{2-2b^2}} \;=\;{Q^\li \over 2} +i p^\li,\;\;\;\;\;\;\;\;
p^{\li} \;=\; \frac{p}{\sqrt{2-2b^2}},
\\[6pt]
\alpha^\gi &=& \frac{b\,\alpha}{\sqrt{2-2b^2}} \;=\;{Q^\li \over 2} +i p^\gi,\;\;\;\;\;\;\;\;
p^{\gi} \;=\; \frac{b\,p}{\sqrt{2-2b^2}}.
\end{array}
\end{equation}
By explicit calculation of vectors
$\psi_{-{1\over 2}}(p)\ket{p},\psi^\R_{-{1\over 2}}(p)\ket{p}$
one can show that $\ket{p,j}$ and $\ket{p,-j}$ are linearly independent.
Since
\begin{equation}
\label{weights}
\Delta^{\li}(p,j)+ \Delta^{\gi}(p, j) =  {Q^2\over 8}+{p^2\over 2}+{j^2\over 2}= \Delta_p+{j^2\over 2}.
\end{equation}
one has two different Virasoro highest weight vectors on each ${j^2\over 2}$ level subspace of
${\cal A}_{\Delta_p}$ except $j=0$
where there is only one state $\ket{p,0}$.
We have shown that
$$
\bigoplus\limits_{j\in \mathbb{Z}} {\sf V}_{\Delta^{\li}(p, j)}\otimes {\sf V}_{\Delta^{\gi}(p, j)}
\; \subset \;
{\cal A}_{\Delta_p}.
$$
In order to show that the spaces above are equal one can calculate characters of the representations involved.
For real $p$ and $Q$ the modules are non-degenerate and
$$
\begin{array}{rllrlll}
\chi\left({\cal V}_{\Delta_p}\right)&=&\displaystyle q^{\Delta_p} \prod\limits_{n=1}^\infty {\,\;\;\;1+ q^{n-\frac12}\over 1-q^n},
\\[10pt]
\chi\left(\widetilde{\cal F}_{\NS}\right)&=& \displaystyle \prod\limits_{n=1}^\infty (1+ q^{n-\frac12}),
\\[10pt]
\chi\left({\sf V}_{\Delta^{\sigma}(p, k)}\right)&=&
\displaystyle q^{\Delta^\sigma(p,k)} \prod\limits_{n=1}^\infty {1\over 1-q^n},
&
\sigma= \li,\gi.
\end{array}
$$
Using relation (\ref{weights}) and the Jacobi triple identity  one gets
\begin{eqnarray*}
\sum\limits_{k=-\infty}^{+\infty} \chi\left({\sf V}_{\Delta^{\li}(p, k)}\right)\chi\left({\sf V}_{\Delta^{\gi}(p, k)}\right)
&=&
\chi\left({\cal V}_{\Delta_p}\right)\chi\left(\widetilde{\cal F}_{\NS}\right),
\end{eqnarray*}
hence
\begin{equation}
\nonumber
\bigoplus\limits_{j\in \mathbb{Z}} {\sf V}_{\Delta^{\li}(p, j)}\otimes {\sf V}_{\Delta^{\gi}(p, j)}
\; = \;
{\cal A}_{\Delta_p}
\end{equation}
which is just decomposition (\ref{decva}).
The equivalence above is a unitary isomorphism if we assume on the l.h.s. the scalar product such that
\begin{eqnarray}
\label{directsumprod}
\langle \nu^\li_{p,j}\otimes \nu^\gi_{p,j}\,|\,\nu^\li_{p,j'}\otimes \nu^\gi_{p,j'}\rangle &=& \langle\,p,j\,\ket{p,-j}\delta_{j+j',0}
\end{eqnarray}
where $\nu^\sigma_{p,j}$ denotes the highest weight state
in the Virasoro Verma module ${\sf V}_{\Delta^\sigma(p,j)}$ ($\sigma= \li,\gi$).
Let us stress that the skew form  of product (\ref{directsumprod}) is the only one consistent with the complex weights $\Delta^{\li}(p, k),
\Delta^{\gi}(p, k), k\neq 0$ and the hermiticity of $L^\li_0, L^\gi_0$. We shall discuss this point in some more details in Subsection 3.3.

\subsection{ Blow up factor}

The  three-point block $\rho^\A_\NS(\xi_3,\xi_2,\xi_1|z)$ with respect to the ${\sf A}_{\NS}$ symmetry algebra is defined as
a solution to the ${\sf A}_{\NS}$ Ward identities normalized by the conditions:
\begin{equation}
\label{A3blocks}
\rho^\A_\NS(\nu_{p_3},\nu_{p_2},\nu_{p_1}|z)
=
\rho^\A_\NS(\nu_{p_3},*\nu_{p_2},\nu_{p_1}|z)=1,
\end{equation}
 where $\nu_p= \ket{\Delta_p}\otimes |\,{0}_{\!\tilde f}\,\rangle $ and $* \nu_p= G_{-{1\over 2}}\nu_p$.

 Our goal in this subsection is to calculate
the block $\rho^\A_\NS(\xi_3,\xi_2,\xi_1|z)$ for $z=1$ and for arbitrary states
$$
\xi_{p,j}\equiv \ket{p,j}_n = \Omega(p,j)\ket{p,j},\;\;\;j\in\mathbb{Z},\;\;\;\Omega(p,0)=1.
$$
By definition
\begin{equation}
\label{blowupfactor}
\rho^\A_\NS(\xi_{p_3,j_3}, \xi_{p_2,j_2},\xi_{p_1,j_1}|1)
=\left\{
\begin{array}{llllllllll}
&
\displaystyle
\frac{{}_n\!\bra{p_3,j_3} V_{p_2,j_2}(1) \ket{p_1,j_1}_n}
{{}_n\!\bra{p_3,0} V_{p_2,0}(1) \ket{p_1,0}_n}
&
{\rm for}
&
j_1+j_2+j_3\in 2\mathbb{Z},
\\
[14pt]
&
\displaystyle
\frac{{}_n\!\bra{p_3,j_3} V_{p_2,j_2}(1) \ket{p_1,j_1}_n}
{{}_n\!\bra{p_3,0} *\!V_{p_2,0}(1) \ket{p_1,0}_n}
&
{\rm for}
&
j_1+j_2+j_3\in 2\mathbb{Z}+1,
\end{array}
\right.
\end{equation}
where  $V_{p,j}(z)$ are chiral vertex operators corresponding to the states $\ket{p,j}_n$ and
\[
*V_{p,0}(1)={\cal G}_{-{1\over 2}}V_{p,0}=[{ G}_{-{1\over 2}},V_{p,0}(1)].
\]
For $j=0$,
$
V_{p,0}(z)= V_{\Delta_p}\otimes 1
$
 where $V_{\Delta_p}(z)$ is  the super-Virasoro primary field corresponding to the highest weight state $\ket{\Delta_p}$.
An explicit expression for (\ref{blowupfactor}) was proposed in \cite{Belavin:2011sw} where it was called the blow up factor due to the role it plays on the four-dimensional side of the AGT correspondence.
We shall compute it employing the free field techniques. At the first step we show that (\ref{blowupfactor}) is a polynomial and calculate an upper bound on its degree.
This is based on the propositions of Subsection~2.2 proven in Appendix A. Then using possible free filed representations we find all zeros of this polynomial.
Finally in order to fix the overall constant we calculate (\ref{blowupfactor}) in a simple special case. The last step requires formulae for generalized Selberg integrals which are derived in Appendix B.

From decompositions (\ref{decompo}) it follows that (\ref{blowupfactor}) is a polynomial in the parameters $\alpha_i,\,i=1,2,3$ with all $\alpha_i$ dependence coming from
the Ward identities in the super-Liouville factor. To determine its order observe that for arbitrary multi-indices:
\begin{eqnarray*}
{\rm deg}_{\alpha_i}\left(
\frac{\bra{\Delta_{p_3}}G_{-K_3}^\dagger   L_{-M_3}^\dagger
{\cal L}_{-M_2} {\cal G}_{-K_2} V_{\Delta_{p_2}}(1)L_{-M_1}G_{-K_1} \ket{\Delta_{p_1}}}
{\bra{\Delta_{p_3}} V_{\Delta_{p_2}}(1)\ket{\Delta_{p_1}}}
\right)
& \leqslant &\sum\limits_{j=1}^3 2\# M_j + \#K_j
\end{eqnarray*}
for  $\sum\limits_{j=1}^3  \#K_j \in 2{\mathbb N}$ while
\begin{eqnarray*}
{\rm deg}_{\alpha_i}\left(
\frac{\bra{\Delta_{p_3}}G_{-K_3}^\dagger   L_{-M_3}^\dagger
{\cal L}_{-M_2} {\cal G}_{-K_2} V_{\Delta_{p_2}}(1)L_{-M_1}G_{-K_1} \ket{\Delta_{p_1}}}
{\bra{\Delta_{p_3}} *\!V_{\Delta_{p_2}}(1)\ket{\Delta_{p_1}}}
\right)
& \leqslant &\sum\limits_{j=1}^3 2\# M_j + \#K_j -1
\end{eqnarray*}
for  $\sum\limits_{j=1}^3  \#K_j \in 2{\mathbb N}-1.$
By the construction of $\ket{p,j}_n$  and Proposition 2 one gets:
\begin{eqnarray*}
{\rm deg}_{\alpha_i}\rho^\A_\NS(\xi_{p_3,j_3}, \xi_{p_2,j_2},\xi_{p_1,j_1}|1)
& \leqslant &j_1^2+j_2^2+j_3^2
\end{eqnarray*}
when $j_1+j_2+j_3$ is even and
\begin{eqnarray*}
{\rm deg}_{\alpha_i}\rho^\A_\NS(\xi_{p_3,j_3}, \xi_{p_2,j_2},\xi_{p_1,j_1}|1)
& \leqslant &j_1^2+j_2^2+j_3^2 -1
\end{eqnarray*}
when $j_1+j_2+j_3$ is odd.

There are several free field representations of the matrix element
\(
{}_n\!\bra{p_3,j_3} V_{p_2,j_2}(1) \ket{p_1,j_1}_n.
\)
Suppose that all $j_i$ are positive. The state
$\ket{p,j}_n$ can then be represented either as
\[
\Omega(p,j)\chi_{-\frac{2j-1}{2}} \ldots \chi_{-\frac12}\ket{p},
\]
or
\[
\Omega(p,j)\chi^\R_{-\frac{2j-1}{2}} (-p)\ldots \chi^\R_{-\frac12}(-p)\ket{-p}.
\]
There are also two different representations for the descendant field $V_{p,j}(1):$
\[
\frac{\Omega(p,j)}{(2\pi i)^j}
\oint\limits_{1}\!\frac{d\xi_j}{(\xi_j-1)^{j}}\ldots\oint\limits_{1}\! \frac{d\xi_1}{\xi_1-1}
\chi(\xi_j)\ldots\chi(\xi_1)
{\sf E}^{\alpha}(1){\sf Q}_b^r{\sf Q}_{\frac{1}{b}}^s,
\hskip 1cm
\alpha = \frac{Q}{2} + ip,
\]
and
\[
\hspace{-65pt}
\frac{\Omega(p,j)}{(2\pi i)^j}
\oint\limits_{1}\!\frac{d\xi_j}{(\xi_j-1)^{j}}\ldots\oint\limits_{1}\! \frac{d\xi_1}{\xi_1-1}
\chi^\R(\xi_j)\ldots\chi^\R(\xi_1)
{\sf E}^{Q-\alpha}(1){\sf Q}_b^r{\sf Q}_{\frac{1}{b}}^s,
\]
where
$$
\chi(\xi)|_{{\cal H}_{p}\otimes \widetilde{ \cal F}_\NS}= \sum\limits_{k\in \mathbb{Z}+{1\over 2}}\left(f_k-i\psi_k  \right) \xi^{-k-{1\over 2}},
\;\;\;
\chi^\R(\xi)|_{{\cal H}_{p}\otimes \widetilde{ \cal F}_\NS}= \sum\limits_{k\in \mathbb{Z}+{1\over 2}}\left(f_k-i\psi^\R_k(p)  \right) \xi^{-k-{1\over 2}}
$$
are the free field representations of the fields $\tilde \chi(\xi),\,\tilde\chi^R(\xi)$ (\ref{fields}).
In consequence there are eight distinct representations of the blow-up factor.

Let us first consider  the  one with no reflected fields. Suppose that
\(
p_3 = p_1 -i(\alpha_2 + rb + sb^{-1}).
\)
Then, for $j_1+j_2+j_3 \in 2{\mathbb N}\cup \{0\}:$
\begin{eqnarray}
\label{3pt:freefield:1a}
&&
\hskip -.5cm
\rho^\A_\NS(\xi_{p_3,j_3}, \xi_{p_2,j_2},\xi_{p_1,j_1}|1)
\; = \;
\frac{1}{(2\pi i)^{j_2}}
\frac{\Omega(-p_3,j_3)\Omega(p_2,j_2)\Omega(p_1,j_1)}{\langle p_3|{\sf E}^{\alpha_2}(1){\sf Q}_b^r{\sf Q}_{\frac{1}{b}}^s|p_1\rangle}
\\[2pt]
\nonumber
&\times &
\oint\limits_{1}\!\frac{d\xi_{j_2}}{(\xi_{j_2}-1)^{j_2}}\ldots\oint\limits_{1}\! \frac{d\xi_1}{\xi_1-1}
\langle p_3|
\chi^\dagger_{-\frac12}\ldots\chi^\dagger_{-\frac{2j_3-1}{2}}
\chi(\xi_{j_2})\ldots\chi(\xi_1)
{\sf E}^{\alpha_2}(1){\sf Q}_b^r{\sf Q}_{\frac{1}{b}}^s
\chi_{-\frac{2j_1-1}{2}}\ldots\chi_{-\frac12}
|p_1\rangle
\end{eqnarray}
while for $j_1+j_2+j_3 \in 2{\mathbb N}-1:$
\begin{eqnarray}
\label{3pt:freefield:1b}
&&
\hskip -.5cm
\rho^\A_\NS(\xi_{p_3,j_3}, \xi_{p_2,j_2},\xi_{p_1,j_1}|1)
\; = \;
\frac{i}{\alpha_2}\frac{1}{(2\pi i)^{j_2}}
\frac{\Omega(-p_3,j_3)\Omega(p_2,j_2)\Omega(p_1,j_1)}{\langle p_3|\psi(1){\sf E}^{\alpha_2}(1){\sf Q}_b^r{\sf Q}_{\frac{1}{b}}^s|p_1\rangle}
\\[2pt]
\nonumber
&\times &
\oint\limits_{1}\!\frac{d\xi_{j_2}}{(\xi_{j_2}-1)^{j_2}}\ldots\oint\limits_{1}\! \frac{d\xi_1}{\xi_1-1}
\langle p_3|
\chi^\dagger_{-\frac12}\ldots\chi^\dagger_{-\frac{2j_3-1}{2}}
\chi(\xi_{j_2})\ldots\chi(\xi_1)
{\sf E}^{\alpha_2}(1){\sf Q}_b^r{\sf Q}_{\frac{1}{b}}^s
\chi_{-\frac{2j_1-1}{2}}\ldots\chi_{-\frac12}
|p_1\rangle.
\end{eqnarray}

The correlator  in the integrand  of the r.h.s.\ of (\ref{3pt:freefield:1a}) and of (\ref{3pt:freefield:1b}) can be rewritten as an integral of product of
the bosonic
\[
\bra{0_b}\otimes \dlangle p_3|{\sf E}^{\alpha_2}(1){\sf E}^b(u_1)\ldots {\sf E}^b(u_r){\sf E}^{\frac1b}(u_{r+1})\ldots {\sf E}^{\frac1b}(u_{r+s})|p_1\drangle\otimes\ket{0_b},
\]
and the fermionic factor
\begin{equation}
\label{fermionic:correlator:1}
(-1)^{j_3}\bra{0}
\chi(\xi_1)\ldots\chi(\xi_N)\psi(u_1)\ldots\psi(u_{r+s})
\ket{0}
\end{equation}
with $N= j_1+j_2+j_3,\;\ket{0}=\ket{0_f}\otimes|0_{\tilde f}\rangle.$
From the commutation relations of modes one has
\begin{equation}
\label{fermionc:OPE:1}
\chi(w)\chi(z) \; \sim \; 0,
\hskip 1cm
\psi(w)\psi(z) \; \sim \; \frac{1}{w-z}
\hskip .7cm
{\rm and}
\hskip .7cm
\chi(w)\psi(z) \; \sim \; -\frac{i}{w-z}.
\end{equation}
This implies that   (\ref{fermionic:correlator:1})   vanishes for
$r+s < j_1+j_2+j_3$ and therefore the l.h.s.\ of (\ref{3pt:freefield:1a})
contains a factor
\begin{equation}
\label{factor:1}
\begin{array}[t]{c}
{\displaystyle\prod} \\[-5pt]
{\scriptscriptstyle
0 \leqslant r,s}
\\[-7pt]
{\scriptscriptstyle
r+s\,\in\, 2{\mathbb N}
}
\\[-7pt]
{\scriptscriptstyle r+s <j_1+j_2+j_3}
\end{array}
\hskip - 5mm
\left(p_3-p_1  +i(\alpha_2 + rb + sb^{-1})\right)
\hskip 7pt
\propto
\hskip 7pt
l^\NS
\left({\textstyle {1\over 2}}Q  +ip_1 +ip_2-ip_3,j_1+j_2+j_3\right),
\end{equation}
where $l^\NS(x,n)$ has been defined in (\ref{defel}) while the l.h.s.\ of (\ref{3pt:freefield:1a}) is proportional to
\begin{equation}
\label{factor:1b}
l^\R\left({\textstyle {1\over 2}}Q  +ip_1 +ip_2-ip_3,j_1+j_2+j_3\right)
\end{equation}
with
\begin{equation}
\label{defelR}
l^\R(x,n) =
\begin{array}[t]{c}
{\displaystyle\prod} \\[-5pt]
{\scriptscriptstyle
0 \leqslant r,s}
\\[-7pt]
{\scriptscriptstyle r+s <n}
\\[-7pt]
{\scriptscriptstyle
r+s\,\in\, 2{\mathbb N} -1
}
\end{array}
\left(x +rb + sb^{-1}\right).
\end{equation}
Let us now consider the free-field representations of the blow-up factor containing at least one reflected field.
Since the oscillators $\psi^\R_k(-p)$ are complicated functions of both the bosonic $c_n$ and the fermionic $\psi_k$
oscillators one cannot factorize the correlators appearing in the corresponding integrands into a bosonic and a fermionic parts.
For the same reason
the OPE of $
{\sf E}^{\alpha}(z)$
and $\chi^\R(w) $ does not vanish.
The free field representations $\psi^\R(w)$ and $\psi(z)$ of fields introduced in (\ref{fields}) commute with ${\sf p}.$
Their OPE cannot therefore contain any primary field of the form
${\sf E}^{\alpha}(z)$ with $\alpha\neq 0.$ The only admissible operators are thus even
members of the conformal family of the identity field. Since both $\psi^\R(w)$ and $\psi(z)$ are primary fields with conformal dimension $\frac12,$ one gets
$$
\chi^{\R}(w)\chi(z)|_{{\cal H}_{p}\otimes \widetilde{ \cal F}_\NS} \; \sim \; \frac{C(p)}{w-z},
$$
where $C(p)=-\langle\,p,-1\,\ket{\,p,1}\neq 0$.

Taking into account these properties of the reflected field $\psi^\R(w)$
one can still conclude that the free field  representation must vanish whenever the number of $\chi$ field insertions
exceeds the number of reflected fields and screening charges. It follows in particular that representation with three reflected fields
does not provide any direct information about possible zeros of the blow up factor.
The remaining six representations involve the following quotients
\begin{eqnarray}
\label{3pt:freefield:2}
{\langle p_3|\chi_{\frac12}\ldots\chi_{\frac{2j_3-1}{2}}
\chi(\xi_{j_2})\ldots\chi(\xi_1)
{\sf E}^{\alpha_2}(1){\sf Q}_b^r{\sf Q}_{\frac{1}{b}}^s
\chi^\R_{-\frac{2j_1-1}{2}}\ldots\chi^\R_{-\frac12}
|-p_1\rangle
\over
{\langle p_3|\underline{\hspace*{5pt}}{\sf E}^{\alpha_2}(1){\sf Q}_b^r{\sf Q}_{\frac{1}{b}}^s|-p_1\rangle}
}
\end{eqnarray}
for
\(
p_3 = -p_1 -i\left(\alpha_2 + rb + sb^{-1}\right),
\)
where $\underline{\hspace*{5pt}}{\sf E}^{\alpha_2}(1)$ denotes either ${\sf E}^{\alpha_2}(1)$ (when $j_1+j_2+j_3$ is even) or
$*{\sf E}^{\alpha_2}(1)$ (when $j_1+j_2+j_3$ is odd),
\begin{eqnarray}
\label{3pt:freefield:3}
{\langle p_3|
\chi_{\frac12}\ldots\chi_{\frac{2j_3-1}{2}}
\chi^\R(\xi_{j_2})\ldots\chi^\R(\xi_1)
{\sf E}^{Q-\alpha_2}(1){\sf Q}_b^r{\sf Q}_{\frac{1}{b}}^s
\chi_{-\frac{2j_1-1}{2}}\ldots\chi_{-\frac12}
|p_1\rangle
\over
{\langle p_3|\underline{\hspace*{5pt}}{\sf E}^{Q-\alpha_2}(1){\sf Q}_b^r{\sf Q}_{\frac{1}{b}}^s|p_1\rangle}}
\end{eqnarray}
for
\(
p_3 = p_1  -i\left(Q-\alpha_2+rb + sb^{-1}\right)
\)
,
\begin{eqnarray}
\label{3pt:freefield:4}
{\langle -p_3|
\chi^\R_{\frac12}\ldots\chi^\R_{\frac{2j_3-1}{2}}
\chi(\xi_{j_2})\ldots\chi(\xi_1)
{\sf E}^{\alpha_2}(1){\sf Q}_b^r{\sf Q}_{\frac{1}{b}}^s
\chi_{-\frac{2j_1-1}{2}}\ldots\chi_{-\frac12}
|p_1\rangle
\over
{\langle - p_3|\underline{\hspace*{5pt}}{\sf E}^{\alpha_2}(1){\sf Q}_b^r{\sf Q}_{\frac{1}{b}}^s|p_1\rangle}}
\end{eqnarray}
for
\(
-p_3 = p_1 -i\left(\alpha_2 +rb + sb^{-1}\right),
\)
\begin{eqnarray}
\label{3pt:freefield:5}
{\langle p_3|\chi_{\frac{2j_3-1}{2}}\ldots\chi_{\frac12}
\chi^\R(\xi_{j_2})\ldots\chi^\R(\xi_1)
{\sf E}^{Q-\alpha_2}(1){\sf Q}_b^r{\sf Q}_{\frac{1}{b}}^s
\chi^\R_{-\frac{2j_1-1}{2}}\ldots\chi^\R_{-\frac12}
|-p_1\rangle
\over
{\langle p_3|\underline{\hspace*{5pt}}{\sf E}^{Q-\alpha_2}(1){\sf Q}_b^r{\sf Q}_{\frac{1}{b}}^s|-p_1\rangle}
}
\end{eqnarray}
for
\(
p_3 = -p_1  -i\left(Q-\alpha_2+rb + sb^{-1}\right),
\)
\begin{eqnarray}
\label{3pt:freefield:6}
{\langle - p_3|
\chi^\R_{\frac12}\ldots\chi^\R_{\frac{2j_3-1}{2}}
\chi(\xi_{j_2})\ldots\chi(\xi_1)
{\sf E}^{\alpha_2}(1){\sf Q}_b^r{\sf Q}_{\frac{1}{b}}^s
\chi^\R_{-\frac{2j_1-1}{2}}\ldots\chi^\R_{-\frac12}
|-p_1\rangle
\over
{\langle - p_3|\underline{\hspace*{5pt}}{\sf E}^{\alpha_2}(1){\sf Q}_b^r{\sf Q}_{\frac{1}{b}}^s|-p_1\rangle}}
\end{eqnarray}
for
\(
-p_3 = -p_1 -i(\alpha_2 + rb + sb^{-1})
\)
and
\begin{eqnarray}
\label{3pt:freefield:7}
{\langle -p_3|
\chi^\R_{\frac12}\ldots\chi^\R_{\frac{2j_3-1}{2}}
\chi^\R(\xi_{j_2})\ldots\chi^\R(\xi_1)
{\sf E}^{Q-\alpha_2}(1){\sf Q}_b^r{\sf Q}_{\frac{1}{b}}^s
\chi_{-\frac{2j_1-1}{2}}\ldots\chi_{-\frac12}
|p_1\rangle
\over
{\langle - p_3|\underline{\hspace*{5pt}}{\sf E}^{Q-\alpha_2}(1){\sf Q}_b^r{\sf Q}_{\frac{1}{b}}^s|p_1\rangle}}
\end{eqnarray}
for
\(
-p_3 = p_1  -i\left(Q-\alpha_2+rb + sb^{-1}\right).
\)
They imply the  factors
\begin{eqnarray}
\label{factor:2}
&&
l^\sharp\hskip -2pt
\left({\textstyle {1\over 2}}Q  -ip_1 +ip_2-ip_3,j_2+j_3- j_1\right),
\\
&&
\label{factor:3}
l^\sharp\hskip -2pt
\left({\textstyle {1\over 2}}Q  +ip_1 -ip_2 -ip_3, j_1+j_3-j_2\right),
\\
&&
\label{factor:4}
l^\sharp\hskip -2pt
\left({\textstyle {1\over 2}}Q   +ip_1+ip_2+ip_3,j_1+j_2-j_3\right),
\\
\label{factor:5}
&&
l^\sharp\hskip -2pt
\left({\textstyle {1\over 2}}Q   -ip_1-ip_2-ip_3,j_3- j_1-j_2\right),
\\
&&
\label{factor:6}
l^\sharp\hskip -2pt
\left({\textstyle {1\over 2}}Q  -ip_1 +ip_2+ip_3,j_2-j_1-j_3\right),
\\
&&
\label{factor:7}
l^\sharp\hskip -2pt
\left({\textstyle {1\over 2}}Q   +ip_1-ip_2+ip_3,j_1-j_2-j_3\right),
\end{eqnarray}
respectively, where $\sharp=\scriptstyle{\rm NS}$ for even $j_1+j_2+j_3,$ while $\sharp = {\scriptstyle \rm R}$ for $j_1+j_2+j_3$ being odd.

For a given set of positive $j_1,j_2,j_3$ not all the representations contribute.
For instance
if the inequalities
\begin{equation}
\label{restrictions:on:excitation}
j_2+j_3 > j_1,
\hskip 1cm
j_3+j_1 > j_2,
\hskip 1cm
j_1+ j_2 > j_3,
\end{equation}
hold, (\ref{factor:1}) (resp.\ (\ref{factor:1b})) and  (\ref{factor:2}) -- (\ref{factor:4}) are the only products with non-empty ranges of integers $r,s$.
Since the number of pairs $(r,s)$ satisfying $ 0 \leqslant r+s < n \in 2{\mathbb N}$ with $r+s \in 2{\mathbb Z}$ is equal to $\frac14n^2$,
the total number of factors in (\ref{factor:1}) and  (\ref{factor:2}) -- (\ref{factor:4}) in the ``even'' case is equal to
\begin{eqnarray*}
\frac14(j_1+j_2+j_3)^2 + \frac14(j_2+j_3- j_1)^2 + \frac14 (j_1+j_2-j_3)^2 + \frac14 (j_1+j_3-j_2)^2
& = &
j_1^2 +j_2^2 + j_3^2.
\end{eqnarray*}
The product of terms (\ref{factor:1}) and  (\ref{factor:2}) -- (\ref{factor:4}) exhausts therefore the dependence of  blow-up factor (\ref{blowupfactor})
on all $\alpha_i.$ Hence, if inequalities (\ref{restrictions:on:excitation}) are satisfied,
\begin{eqnarray}
\label{blowupfactor:result:1}
&&
\hskip -1cm
\rho^\A_\NS(\xi_{p_3,j_3}, \xi_{p_2,j_2},\xi_{p_1,j_1}|1)
\; =\;C^{j_3}_{\;j_2j_1}(b)
\\[4pt]
\nonumber
&&\times\;
l^\NS\left({\textstyle {1\over 2}}Q  +ip_1+ip_2-ip_3,j_1+j_2+j_3\right)
l^\NS\left({\textstyle {1\over 2}}Q  -ip_1+ip_2-ip_3,j_2+j_3- j_1\right)
\\[4pt]
\nonumber
&&
\times\;
l^\NS\left({\textstyle {1\over 2}}Q +ip_1-ip_2 -ip_3,j_1+j_3-j_2\right)
l^\NS\left({\textstyle {1\over 2}}Q +ip_1+ip_2 +ip_3,j_1+j_2-j_3\right)
\end{eqnarray}
for $j_1+j_2+j_3 \in 2{\mathbb N}.$

Similarly, since the number of pairs $(r,s)$ satisfying $ 0 \leqslant r+s < n \in 2{\mathbb N}-1$ with $r,s \in 2{\mathbb Z}-1$ is equal to $\frac14(n^2-1),$ and
\begin{eqnarray*}
\frac14\left((j_1+j_2+j_3)^2-1\right)
+
\frac14\left((j_2+j_3-j_1)^2-1\right)
\\[4pt]
+
\frac14\left((j_1+j_3-j_2)^2-1\right)
+
\frac14\left((j_1+j_2-j_3)^2-1\right)
& = &
j_1^2+j_2^2+j_3^2 -1
\end{eqnarray*}
one gets that, if inequalities (\ref{restrictions:on:excitation}) are satisfied,
\begin{eqnarray}
\label{blowupfactor:result:1b}
&&
\hskip -1cm
\rho^\A_\NS(\xi_{p_3,j_3}, \xi_{p_2,j_2},\xi_{p_1,j_1}|1)
\; =\;C^{j_3}_{\;j_2j_1}(b)
\\[4pt]
\nonumber
&&\times\;
l^\R\left({\textstyle {1\over 2}}Q  +ip_1+ip_2-ip_3,j_1+j_2+j_3\right)
l^\R\left({\textstyle {1\over 2}}Q  -ip_1+ip_2-ip_3,j_2+j_3- j_1\right)
\\[4pt]
\nonumber
&&
\times\;
l^\R\left({\textstyle {1\over 2}}Q +ip_1-ip_2 -ip_3,j_1+j_3-j_2\right)
l^\R\left({\textstyle {1\over 2}}Q +ip_1+ip_2 +ip_3,j_1+j_2-j_3\right)
\end{eqnarray}
for $j_1+j_2+j_3 \in 2{\mathbb N}-1.$

Looking for non-empty ranges of products and counting the degree of the resulting polynomial one obtains formulae for the blow up factor for other ranges
of positive integers $j_1, j_2, j_3$:
\begin{eqnarray}
\label{blowupfactor:result:2}
&&
\hskip -1cm
\rho^\A_\NS(\xi_{p_3,j_3}, \xi_{p_2,j_2},\xi_{p_1,j_1}|1)
\; =\;C^{j_3}_{\;j_2j_1}(b)
\\[4pt]
\nonumber
&&\times\;
l^\sharp\hskip-3pt \left({\textstyle {1\over 2}}Q  +ip_1+ip_2-ip_3,j_1+j_2+j_3\right)
l^\sharp\hskip-3pt \left({\textstyle {1\over 2}}Q  -ip_1+ip_2-ip_3,j_2+j_3- j_1\right)
\\[4pt]
\nonumber
&&
\times\;
l^\sharp\hskip-3pt \left({\textstyle {1\over 2}}Q -ip_1+ip_2 +ip_3,j_2-j_1-j_3\right)
l^\sharp\hskip-3pt \left({\textstyle {1\over 2}}Q +ip_1+ip_2 +ip_3,j_1+j_2-j_3\right),
\end{eqnarray}
for $j_2 > j_1 + j_3$ (and, consequently $j_1 + j_2 > j_3$ and $j_2 + j_3 > j_1$),
\begin{eqnarray}
\label{blowupfactor:result:3}
&&
\hskip -1cm
\rho^\A_\NS(\xi_{p_3,j_3}, \xi_{p_2,j_2},\xi_{p_1,j_1}|1)
\; =\;C^{j_3}_{\;j_2j_1}(b)
\\[4pt]
\nonumber
&&\times\;
l^\sharp\hskip-3pt \left({\textstyle {1\over 2}}Q  +ip_1+ip_2-ip_3,j_1+j_2+j_3\right)
l^\sharp\hskip-3pt \left({\textstyle {1\over 2}}Q  +ip_1-ip_2+ip_3,j_1-j_2- j_3\right)
\\[4pt]
\nonumber
&&
\times\;
l^\sharp\hskip-3pt \left({\textstyle {1\over 2}}Q +ip_1-ip_2 -ip_3,j_1+j_3-j_2\right)
l^\sharp\hskip-3pt \left({\textstyle {1\over 2}}Q +ip_1+ip_2 +ip_3,j_1+j_2-j_3\right),
\end{eqnarray}
for
\(
j_1 > j_2 + j_3
\)
and
\begin{eqnarray}
\label{blowupfactor:result:4}
&&
\hskip -1cm
\rho^\A_\NS(\xi_{p_3,j_3}, \xi_{p_2,j_2},\xi_{p_1,j_1}|1)
\; =\;C^{j_3}_{\;j_2j_1}(b)
\\[4pt]
\nonumber
&&\times\;
l^\sharp\hskip-3pt \left({\textstyle {1\over 2}}Q  +ip_1+ip_2-ip_3,j_1+j_2+j_3\right)
l^\sharp\hskip-3pt \left({\textstyle {1\over 2}}Q  -ip_1+ip_2-ip_3,j_2+j_3- j_1\right)
\\[4pt]
\nonumber
&&
\times\;
l^\sharp\hskip-3pt \left({\textstyle {1\over 2}}Q +ip_1-ip_2 -ip_3,j_1+j_3-j_2\right)
l^\sharp\hskip-3pt \left({\textstyle {1\over 2}}Q -ip_1-ip_2 -ip_3,j_3-j_1-j_2\right),
\end{eqnarray}
for
\(
j_3 > j_1 + j_2.
\)

In order to determine the normalization constant $C^{j_3}_{\;j_2j_1}(b)$ we assume inequalities (\ref{restrictions:on:excitation}) and
use the representations (\ref{3pt:freefield:1a}) and (\ref{3pt:freefield:1b}) in the simplest non-vanishing
case , i.e.\ for $s=0,\ r = j_1+j_2+j_3 $ and
\[
ip_3 = ip_1 + \alpha_2 + Nb, \hskip 1cm N = j_1+j_2+j_3.
\]
In this case the second line of (\ref{3pt:freefield:1a})  and  (\ref{3pt:freefield:1b})  takes the form
\begin{eqnarray*}
I_{\rm\scriptscriptstyle num}\!\left[^{j_3}_{j_2j_1}\right]
& = &
(-1)^{j_3}\int\limits_0^1\!dt_1\int\limits_0^{t_1}\!dt_2 \ldots \int\limits_0^{t_{N-1}}\hskip -3pt dt_N
\bra{p_3}{\sf E}^{\alpha_2}(1){\sf E}^b(t_1)\ldots {\sf E}^b(t_N)\ket{p_1}
h^{j_3}_{\;j_2j_1}(t_1,\ldots,t_N)
\\
& = &
(-1)^{j_3}\int\limits_0^1\!dt_1\int\limits_0^{t_1}\!dt_2 \ldots \int\limits_0^{t_{N-1}}\hskip -3pt dt_N
\prod\limits_{k=1}^N t_k^{-b\alpha_1}(1-t_k)^{-b\alpha_2}
\hskip -10pt\prod\limits_{1\leqslant k < l\leqslant N}\hskip -10pt (t_k-t_l)^{-b^2}
h^{j_3}_{\;j_2j_1}(t_1,\ldots,t_N)
\end{eqnarray*}
where
\begin{eqnarray*}
h^{j_3}_{\;j_2j_1}(t_1,\ldots,t_N)
& = &
\frac{1}{(2\pi i)^{j_2}}
\oint\limits_{1}\!\frac{d\xi_{j_2}}{(\xi_{j_2}-1)^{j_2}}\ldots\oint\limits_{1}\! \frac{d\xi_1}{\xi_1-1} \ g^{j_3}_{\;j_2j_1}(t_1,\ldots,t_N),
\\[4pt]
g^{j_3}_{\;j_2j_1}(t_1,\ldots,t_N)
& = &
\langle 0|
\chi_{\frac12}\ldots\chi_{\frac{2j_3-1}{2}}
\chi(\xi_{j_2})\ldots\chi(\xi_1)
\psi(t_1)\ldots \psi(t_N)
\chi_{-\frac{2j_1-1}{2}}\ldots\chi_{-\frac12}
|0\rangle.
\end{eqnarray*}
The integral $h^{j_3}_{\;j_2j_1}(t_1,\ldots,t_N)$
 is  by definition  totally  antisymmetric in $t_k$ variables.
The integrand
$g^{j_3}_{\;j_2j_1}$ can be calculated by means of the Wick theorem with a help of OPE-s (\ref{fermionc:OPE:1}). Since all
the modes of the field $\chi$ anticommute with each other, the only contributions arise from the anticommutators between a mode of the field $\chi$ and a mode of the field $\psi.$
Indeed, had we taken into account a contribution from a ``pairing'' between the fields $\psi,$ we would have been left with a correlator containing only modes of the field $\chi,$ which vanishes. This shows that
\begin{equation}
\label{zeros}
h^{j_3}_{\;j_2j_1}(t_1,\ldots,t_N)\to 0 \;\;\;\;{\rm for}\;\;\;\;\;t_k \to t_l,\ 1 \leqslant k,l \leqslant N
\end{equation}
and the only possible singularities may arise at $t_l=1,0, \,l=1,\ldots,N$.
Since
\[
\frac{1}{2\pi i}\oint\limits_{1}\!\frac{d\xi_{j_2}}{(\xi_{j_2}-1)^{j_2}}\psi(t_l)\chi(\xi_{j_2})
\; \sim \;
-i\frac{1}{2\pi i}\oint\limits_{1}\!\frac{d\xi_{j_2}}{(\xi_{j_2}-1)^{j_2}}\frac{1}{t_l-\xi_{j_2}}
\; = \;
-\frac{i}{(t_l-1)^{j_2}}
\]
one has
\begin{equation}
\label{singularity1}
h^{j_3}_{\;j_2j_1}(t_1,\ldots,t_N) \sim (t_l-1)^{-j_2}\;\;\;\;{\rm  for}\;\;\;\; t_l \to 1,\ l =1,\ldots,N.
\end{equation}
The commutation relation
$
\left\{\psi(t_l),\chi_k\right\} = -i\,t_l^{k-\frac12}
$
implies
\begin{eqnarray}
&&h^{j_3}_{\;j_2j_1}(t_1,\ldots,t_N) \sim t_l^{-j_1} \;\;\;\;\;{\rm  for}\;\;\;\; t_l \to 0,\ l =1,\ldots,N,
\label{singularity0}
\\
&&h^{j_3}_{\;j_2j_1}(t_1,\ldots,t_N) \sim t_l^{j_3-1} \;\;\;\;{\rm  for}\;\;\;\; t_l \to \infty,\ l =1,\ldots,N.
\label{asymptotic}
\end{eqnarray}
The only totally antisymmetric function with zeros (\ref{zeros}), singularities (\ref{singularity1}), (\ref{singularity0}) and asymptotics (\ref{asymptotic}) reads
\[
h^{j_3}_{\;j_2j_1}(t_1,\ldots,t_N) = \sigma^{j_3}_{\;j_2j_1} \prod\limits_{k=1}^N t_k^{-j_1}(1-t_l)^{-j_2}\hskip -5pt \prod\limits_{1 \leqslant k < l \leqslant N}\hskip -5pt (t_k-t_l)
\]
where $ \sigma^{j_3}_{\;j_2j_1}$ does not depend on the variables $t_k.$

Taking into account sign factors we get from the definition of $h^{j_3}_{\;j_2j_1}(t_1,\ldots,t_N)$ that, up to sub-leading terms
\[
h^{j_3}_{\;j_2j_1}(t_1,\ldots,t_N)
\; = \;
\left\{
\begin{array}{rcl}
-it_N^{-j_1}h^{\;j_2}_{j_3\,j_1-1}(t_1,\ldots,t_{N-1}) &\; {\rm for}\; & t_N \to 0,
\\[10pt]
-i(-1)^{N+j_2-1}(t_N-1)^{-j_2}h^{j_2-1}_{j_3\,j_1}(t_1,\ldots,t_{N-1}) & {\rm for} & t_N \to 1,
\\[10pt]
-i(-1)^{N+j_2-1}t_N^{j_3-1}h^{j_2}_{j_3-1\,j_1}(t_1,\ldots,t_{N-1}) & {\rm for} & t_N \to \infty.
\end{array}
\right.
\]
Comparing this formula with the one above we get a recurrence relation uniquely determining $ \sigma^{j_3}_{\;j_2j_1}$. This yields
\begin{equation}
\label{h:odd:explicit:1}
h^{j_3}_{\;j_2j_1}(t_1,\ldots,t_N)
=
(-i)^N \prod\limits_{k=1}^N t_k^{-j_1}(1-t_l)^{-j_2}\hskip -5pt \prod\limits_{1 \leqslant k < l \leqslant N}\hskip -5pt (t_k-t_l).
\end{equation}
We conclude from (\ref{h:odd:explicit:1}) that
$I_{\rm\scriptscriptstyle num}\!\left[^{j_3}_{j_2j_1}\right]$ is expressed through a standard Selberg integral and thus can be calculated with the result
\begin{eqnarray}
\label{numerator}
&&
\hskip -1cm
I_{\rm\scriptscriptstyle num}\!\left[^{ j_3}_{j_2j_1}\right]\; =
\\[2pt]
\nonumber
& = &
\frac{(-1)^{j_3}}{N!}
\prod\limits_{k=0}^{N-1}
\frac{
\Gamma(1+(k+1)(g+1))\Gamma(1 -b\alpha_1-j_1 + k(g+1))\Gamma(1-b\alpha_2-j_2 + k(g+1))
}{
\Gamma(g+1)\Gamma\big( 2-b\alpha_1-j_1 -b\alpha_2-j_2+ (N-1+k)(g+1)\big)
},
\end{eqnarray}
where
$g = -\frac12b\,Q,
\,
N = j_1+j_2+j_3.
$

Suppose that $j_1+j_2+j_3$ is even. In the special case under consideration the matrix element $\bra{p_3}{\sf E}^{\alpha_2}(1){\sf Q}_b^N\ket{p_1}$
takes the form
\begin{eqnarray}
\nonumber
&&
\hskip -1.5cm
\bra{p_3}{\sf E}^{\alpha_2}(1){\sf Q}_b^{2m}\ket{p_1}
\; = \;
\int\limits_0^1\!dt_1\int\limits_0^{t_1}\!dt_2\ldots\hskip -5pt \int\limits_0^{t_{2m-1}}\hskip -5pt dt_{2m}\
\bra{p_3}{\sf E}^{\alpha_2}(1)\prod\limits_{k=1}^{2m}{\sf E}^b(t_k)\ket{p_3}\,
\bra{0_f}\prod\limits_{k=1}^{2m}\psi(t_k)\ket{0_f}
\\[-8pt]
\label{even:integral}
\\[-8pt]
\nonumber
& = &
\int\limits_0^1\!dt_1\int\limits_0^{t_1}\!dt_2\ldots\hskip -5pt \int\limits_0^{t_{2m-1}}\hskip -5pt dt_{2m}\
\prod\limits_{k=1}^{2m} t_k^{-b\alpha_1}(1-t_k)^{-b\alpha_2}
\hskip -5pt\prod\limits_{1\leqslant k < l \leqslant 2n}\hskip -5pt
(t_k-t_l)^{-b^2-1}\,
\hbox{\boldmath $P$\!}_{2m}(t_1,\ldots,t_{2m})
\end{eqnarray}
where
\begin{eqnarray*}
\hbox{\boldmath $P$\!}_{2m}(t_1,\ldots,t_{2m})
& = &
\bra{0_f}\prod\limits_{k=1}^{2m}\psi(t_k)\ket{0_f}
\hskip -5pt\prod\limits_{1\leqslant k < l \leqslant 2m}\hskip -5pt(t_k-t_l)
\; = \;
{\rm Pf}\left(\frac{1}{t_k-t_l}\right)
\hskip -2pt\prod\limits_{1\leqslant k <l \leqslant 2m}\hskip -5pt(t_k-t_l)
\end{eqnarray*}
is a totally symmetric polynomial. We calculate this generalization of the Selberg integral
 in Appendix B.
Using (\ref{modified:Selberg}) one obtains
\begin{eqnarray}
\label{second:long:integral}
&&
\hskip -.7cm
\bra{p_3}{\sf E}^{\alpha_2}(1){\sf Q}_b^{2m}\ket{p_1}
\; = \;
\\
\nonumber
&&
\hskip -5mm
\frac{1}{2^m}
\prod\limits_{q=0}^{m-1}\prod\limits_{j=1}^2
\frac{\Gamma(1+q + (2q+j)g)\Gamma(1+q-b\alpha_1+(2q+j-1)g)\Gamma(1+q-b\alpha_2+(2q+j-1)g)
}{
\Gamma(1+g)\Gamma\big(1+m+q-b(\alpha_1+\alpha_2) + (2(m+q) +j-2)g\big)}.
\end{eqnarray}
The properties of the  Euler gamma function allow to present the ratio of (\ref{numerator}) and (\ref{second:long:integral}) in the form
\begin{eqnarray}
\nonumber
&&
 (-1)^{j_3}
2^m \left(\frac{b}{2}\right)^{m^2}
l^\NS(-2mb,j_1+j_2+j_3)
\\
\nonumber
& \times &
(-1)^{j_1^2}\left(\frac{b}{2}\right)^{m(m-2j_1)}
\;
{l^\NS(-2p_1-2mb,j_2+j_3-j_1)\over l^\NS(Q+2ip_1,2j_1)}
\\[-6pt]
\label{the:left:hand:side}
\\[-6pt]
\nonumber
& \times &
(-1)^{j_2^2}\left(\frac{b}{2}\right)^{m(m-2j_2)}
{
l^\NS(-2ip_2-2mb,j_1+j_3-j_2)
\over
l^\NS(Q+ 2ip_2,2j_2)}
\\
\nonumber
& \times &
(-1)^{(m-j_3)^2} \left(\frac{b}{2}\right)^{n(n-2j_3)}
{
l^\NS(Q+2ip_1+2ip_2+2mb,j_1+j_2-j_3)
\over
l^\NS(-2ip_1-2ip_2-2mb,2j_3)
},
\end{eqnarray}
which completes the calculation of representation (\ref{3pt:freefield:1a}) of the l.h.s.\ of (\ref{blowupfactor:result:1}).
The r.h.s.\ of (\ref{blowupfactor:result:1}) takes the form
\begin{eqnarray}
\nonumber
C^{j_3}_{\;j_2j_1}(b)
&&
\hspace{-5pt}
l^\NS(-2mb,j_1+j_2+j_3)
\\
\nonumber
\times &&
\hspace{-5pt}
l^\NS(-2p_1-2mb,j_2+j_3-j_1)
\\[-8pt]
\label{the:right:hand:side}
\\[-8pt]
\nonumber
\times &&
\hspace{-5pt}
l^\NS(-2ip_2-2mb,j_1+j_3-j_2)
\\
\nonumber
\times &&
\hspace{-5pt}
l^\NS(Q+2ip_1+2ip_2+2mb,j_1+j_2-j_3).
\end{eqnarray}
Comparing (\ref{the:left:hand:side}) with (\ref{the:right:hand:side}) we get:
\begin{equation}
\label{blockconstant}
C^{j_3}_{\;j_2j_1}(b)
\; = \;
(-1)^{j_1+j_2-j_3\over 2}2^{j_1+j_2+j_3\over 2}  .
\end{equation}

To check this result in the case of odd $j_1+j_2+j_3 = 2m-1$  we observe that the matrix element
\begin{eqnarray}
\label{Selberg:odd:integral:1}
&&
\hskip -1cm
\bra{p_3}\psi(1){\sf E}^{Q}(1){\sf Q}_b^{2m-1}\ket{p_1} \; =
\\
\nonumber
& = &
\int\limits_0^1\!dt_{2m-1}\hskip -5pt \int\limits_0^{t_{2m-1}} \hskip -8pt dt_{2m-2}\ldots\int\limits_0^{t_2}\!dt_{1}
\prod\limits_{k=1}^{2m-1} t_k^{-b\alpha_1} (1-t_k)^{-bQ}
\hskip -15pt\prod\limits_{1\leqslant k<l \leqslant 2m-1}\hskip -15pt (t_l-t_k)^{2g}\,
\hbox{\boldmath $P$\!}^{\,(1)}_{2m-1}(t_{2m-1},\ldots,t_{1}),
\end{eqnarray}
where
\begin{eqnarray*}
\hbox{\boldmath $P$\!}^{\,(1)}_{2m-1}(t_{2m-1},\ldots,t_{1})
& = &
\hskip -10pt\prod\limits_{1\leqslant k<l \leqslant 2m-1}\hskip -10pt  (t_l-t_k)\,
\langle 0_f|\psi(1)\psi(t_{2m-1})\ldots \psi(t_{1})|0_f\rangle,
\end{eqnarray*}
can be calculated as a particular limit of integral (\ref{even:integral}). Let us rewrite (\ref{even:integral})
in the form
\begin{eqnarray*}
\nonumber
&&
\hskip -1.5cm
\bra{p_3}{\sf E}^{\alpha_2}(1){\sf Q}_b^{2m}\ket{p_1}
\; = \;
\\
\nonumber
& = &
\int\limits_0^1\!dt_{2m}\hskip -3pt \int\limits_0^{t_{2m}} \hskip -2pt dt_{2m-1}\ldots\int\limits_0^{t_2}\!dt_{1}
\prod\limits_{k=1}^{2m} t_k^{-b\alpha_1}(1-t_k)^{-b\alpha_2}
\hskip -5pt\prod\limits_{1\leqslant k < l \leqslant 2m}\hskip -5pt
(t_l-t_k)^{2g}\,
\hbox{\boldmath $P$\!}_{2m}(t_{2m},\ldots,t_{1}).
\end{eqnarray*}
For $t_{2m} \to 1:$
\begin{eqnarray*}
&&
\hskip -1cm
\prod\limits_{k=1}^{2m} t_k^{-b\alpha_1}(1-t_k)^{-b\alpha_2}
\hskip -5pt\prod\limits_{1\leqslant k < l \leqslant 2m}\hskip -5pt
(t_l-t_k)^{2g}\,
\hbox{\boldmath $P$\!}_{2m}(t_{2m},\ldots,t_{1})
\\
& = &
(1-t_{2m})^{-b\alpha_2}
\prod\limits_{k=1}^{2m-1} t_k^{-b\alpha_1}(1-t_k)^{-b\alpha_2+2g+1}
\hskip -15pt\prod\limits_{1\leqslant k < l \leqslant 2m-1}\hskip -15pt
(t_l-t_k)^{2g}\,
\hbox{\boldmath $P$\!}^{\,(1)}_{2m-1}(t_{2m-1},\ldots,t_{1}) + \ldots
\end{eqnarray*}
where the dots stand for terms sub-leading for $t_{2m}\to 1.$

Using the identity
\[
\lim\limits_{a\to -1_+}(1+a)\int\limits_{0}^1 (1-t)^ah(t)\,dt \; = \; h(1),
\]
valid for a function $h(t)$ left continuous at $t = 1,$ we thus get
\begin{eqnarray*}
\lim\limits_{\alpha_2 \to 1/b}(1-b\alpha_2)\bra{p_3}{\sf E}^{\alpha_2}(1){\sf Q}_b^{2m}\ket{p_1}
& = &
\bra{p_3}\psi(1){\sf E}^{Q}(1){\sf Q}_b^{2m-1}\ket{p_1}.
\end{eqnarray*}
Computing the limit one obtains
\begin{eqnarray}
\label{odd:integral:special:case}
&&
\hskip -3cm
\bra{p_3}\psi(1){\sf E}^{Q}(1){\sf Q}_b^{2m-1}\ket{p_1}
\; = \;
\frac{\Gamma(g)}{2^m}
\prod\limits_{q=1}^{m-1}\prod\limits_{j=1}^2\Gamma\big(q+(2q+j-1)g\big)
\\
\nonumber
& \times &
\prod\limits_{q=0}^{m-1}\prod\limits_{j=1}^2
\frac{
\Gamma\big(1+q+(2q+j)g\big)\Gamma\big(1+q-b\alpha_1 + (2q+j-1)g\big)
}{
\Gamma(1+g)\Gamma\big(m+q -b\alpha_1 +(2(m+q)+j-2)g\big)
}.
\end{eqnarray}
Comparing the ratio of (\ref{numerator}) and (\ref{odd:integral:special:case}) with formula
(\ref{blowupfactor:result:1b}), calculated at $ip_2 = \frac{Q}{2}$ and $ip_3 = ip_1 + Q + (2m-1)b,$ one can verify formula (\ref{blockconstant})
in the case when $j_1+j_2+j_3$ is odd.

Let us note that the cases when some or all $j$ indices are negative can be obtained either by direct analysis or by changing the sign of corresponding momenta.
The special cases when one or more $j$ indices go to zero can be derived using exactly the same method.
They can be seen as limiting cases of the situations considered above.
For instance in the case of $j_1=j_3, j_2=0$ one gets the formula for the scalar product
of normalized states
\begin{eqnarray}
\label{sprod}
_n\!\langle p,-j|p,j\rangle_n
&=&
{}_n\!\langle -p,j|p,j\rangle_n
=2^j
l^\NS(2ip,2j)l^\NS(Q+2ip,2j).
\end{eqnarray}

In Section 4 we shall need 3-point conformal blocks related to the 3-point functions
\begin{equation}
\label{3pointblocks}
\gamma^\A_\NS(\xi_{p_3,j_3}, \xi_{p_2,j_2},\xi_{p_1,j_1}|1)
=
\left\{
\begin{array}{llllllllll}
&
\displaystyle
\frac{\langle V_{p_3,j_3}(\infty) V_{p_2,j_2}(1) V_{p_1,j_1}(0)\rangle}
{\langle V_{p_3,0}(\infty) V_{p_2,0}(1) V_{p_1,0}(0)\rangle}
&
{\rm for}
&
j_1+j_2+j_3\in 2\mathbb{Z},
\\
[14pt]
&
\displaystyle
\frac{\langle V_{p_3,j_3}(\infty) V_{p_2,j_2}(1) V_{p_1,j_1}(0)\rangle}
{\langle V_{p_3,0}(\infty) *V_{p_2,0}(1) V_{p_1,0}(0)\rangle}
&
{\rm for}
&
j_1+j_2+j_3\in 2\mathbb{Z}+1.
\end{array}
\right.
\end{equation}
They can be expressed in terms of the 3-point block related to the matrix elements of the vertex operators (\ref{blowupfactor})
as follows
$$
\gamma^\A_\NS(\xi_{p_3,j_3}, \xi_{p_2,j_2},\xi_{p_1,j_1}|1)
=
(-1)^{j_3}\rho^\A_\NS(\xi_{-p_3,j_3}, \xi_{p_2,j_2},\xi_{p_1,j_1}|1).
$$
In the case when  all $j$  are positive and inequalities (\ref{restrictions:on:excitation})
are satisfied, using (\ref{blowupfactor:result:1}) and (\ref{blockconstant}) one gets
for instance
\begin{equation}
\label{3pb}
\gamma^\A_\NS(\xi_{p_3,j_3}, \xi_{p_2,j_2},\xi_{p_1,j_1}|1)
=
(-2)^{j_1+j_2+j_3\over 2}
\left\{
\begin{array}{llllllll}
{\cal B}^\NS_{ j_3  j_2 j_1}(\alpha_3,\alpha_2,\alpha_1)
&
{\rm for}
&
j_1+j_2+j_3\in 2\mathbb{Z},
\\[10pt]
{\cal B}^\R_{ j_3  j_2 j_1}(\alpha_3,\alpha_2,\alpha_1)
&
{\rm for}
&
j_1+j_2+j_3\in 2\mathbb{Z}+1,
\end{array}
\right.
\end{equation}
where
\begin{equation}
\label{defB}
\begin{array}{rrrrr}
{\cal B}^\sharp_{ j_3  j_2 j_1}(\alpha_3,\alpha_2,\alpha_1)
&=&
l^\sharp(\alpha_1+\alpha_2+\alpha_3-Q,j_1+j_2+j_3)
\\
&&\times\;
l^\sharp(\alpha_1+\alpha_2-\alpha_3,j_1+j_2-j_3)
\\
&&\times\;
l^\sharp(\alpha_1+\alpha_3-\alpha_2,j_1+j_3-j_2)
\\
&&\times\;
l^\sharp(\alpha_2+\alpha_3-\alpha_1,j_2+j_3-j_1)&\!\!,&\hspace{30pt}\sharp\,=\,\scriptstyle{\rm NS},\, \scriptstyle{\rm R}.
\end{array}
\end{equation}

\subsection{4-point conformal  blocks}

The states on the l.h.s.\ of (\ref{mrel}) are organized in terms of ${\sf A}_\NS$ Verma modules ${\cal A}_{\Delta_p}={\cal V}_{\Delta_p}\otimes \widetilde{ \cal F}_\NS$.
In order to compare both sides of (\ref{mrel}) it is convenient to choose a basis in ${\cal A}_{\Delta_p}$ consistent with direct sum decomposition (\ref{decva}):
\begin{equation}
\label{conbasis}
L_{-M}^{\li}L_{-N}^{\gi}       \ket{p,j}_n
\end{equation}
where $M,N$ are arbitrary ordered, integer multi-indices and $j\in \mathbb{Z}$.
Vertex operators related to the states $\ket{p,j}_n$ take the  form $(j>0)$:
\begin{eqnarray*}
V_{p,j}(z)
& = &
\Omega(p,j)
\oint\limits_{z}\!\frac{dw_j}{(w_j-z)^{j}}\ldots\oint\limits_{z}\! \frac{dw_1}{w_1-z}
\chi(w_j)\ldots\chi(w_1)
V_{p,0}(z),
\\
V_{p,-j}(z)
& = &
\Omega(p,-j)
\oint\limits_{z}\!\frac{dw_j}{(w_j-z)^{j}}\ldots\oint\limits_{z}\! \frac{dw_1}{w_1-z}
\chi^\R(w_j)\ldots\chi^\R(w_1)
V_{p,0}(z).
\end{eqnarray*}
For all $j\in \mathbb{Z}$ they are primary
with respect to the energy momentum tensors
\begin{eqnarray}
\label{TL:definition}
T^\sigma(z) & = & \sum\limits_{n\in\mathbb Z} {z^{-n-2}}{L^\sigma_n},
\hskip 1cm
\sigma = \li, \gi.
\end{eqnarray}
Indeed one can check by explicit calculations that the OPE
\begin{eqnarray}
\label{OPE:TVj}
T^\sigma(z)V_{p,j}(w)
& \sim &
\frac{1}{(z-w)^2}\,\Delta^\sigma(p,j) V_{p,j}(w) + \frac{1}{z-w}\,\big[L_{-1}^\sigma,V_{p,j}(w)\big],
\hskip 1cm
\sigma = \li, \gi.
\end{eqnarray}
holds. By standard contour arguments (\ref{OPE:TVj}) implies
\begin{eqnarray}
\label{comm:realtions:Vir}
\left[L^\sigma_n,V_{p,j}(z)\right]
& = &
(n+1)z^n\Delta^\sigma(p,j)V_{p,j}(1) + z^n\left[L^\sigma_{-1},V_{p,j}(1)\right].
\end{eqnarray}
For $L^\sigma_0$ eigenstates $L^\sigma_0\ket{\zeta_1} = \Delta_1^\sigma \ket{\zeta_1}, L^\sigma_0\ket{\zeta_3} = \Delta_3^\sigma \ket{\zeta_3}$,
one has in particular
\begin{eqnarray}
\label{com1}
\bra{\zeta_3}\left[L^\sigma_{-1},V_{p_2,j_2}(z)\right]\ket{\zeta_1}
& = &
\left(\Delta_3^\sigma-\Delta^\sigma(p_2,j_2)-\Delta_1^\sigma\right)\bra{\zeta_3}V_{p_2,j_2}(z)\ket{\zeta_1}.
\end{eqnarray}
It follows from (\ref{OPE:TVj}) and (\ref{com1}) that calculating the three-point conformal block in basis (\ref{conbasis}) one can use the conformal Ward identities
of the corresponding Liouville theories. This yields the relation
\begin{eqnarray}
\nonumber
&&\hspace{-120pt}\rho^\A_\NS(L_{-M_3}^{\li}L_{-N_3}^{\gi}  \xi_{p_3,j_3}, L_{-M_2}^{\li}L_{-N_2}^{\gi}  \xi_{p_2,j_2},L_{-M_1}^{\li}L_{-N_1}^{\gi}  \xi_{p_1,j_1}|z)
\\
\label{3blockrel}
&=&\rho_\li(L_{-M_3}\nu^\li_{p_3,j_3}, L_{-M_2}\nu^\li_{p_2,j_2},L_{-M_1}\nu^\li_{p_1,j_1}|z)
\\
\nonumber
&\times&
\rho_\gi(L_{-N_3}\nu^\gi_{p_3,j_3},\, L_{-N_2}\nu^\gi_{p_2,j_2},\,L_{-N_1}\nu^\gi_{p_1,j_1}|z)
\\
\nonumber
&\times&\rho^\A_\NS(\xi_{p_3,j_3}, \xi_{p_2,j_2},\xi_{p_1,j_1}|z)
\end{eqnarray}
where $\rho_\sigma$ is the three-point conformal block for the Virasoro algebra with the central charge $c^\sigma$ ($\sigma= \li,\gi$).

Relation (\ref{3blockrel}) gives rise to relations between higher conformal blocks. As an example we consider the spheric four-point blocks.
As in the case of  the NS algebra \cite{Hadasz:2006sb} one has four types of the ${\sf A}_\NS$ algebra blocks.
For each type there is one even,
\begin{eqnarray*}
\mathcal{T}^1_{\Delta}
\left[^{\underline{\hspace{3pt}}\,\Delta_3
\;\underline{\hspace{3pt}}\,\Delta_2}_{\hspace{3pt}\,\Delta_4
\;\hspace{3pt}\, \Delta_1} \right]\!
(z)
    &=&
z^{\Delta - \underline{\hspace{3pt}}\,\Delta_2 - \Delta_1} \left( 1 +
\sum_{m\in \mathbb{N}}\! z^m\,
    T^m_{ \Delta}
    \!\left[^{\underline{\hspace{3pt}}\,\Delta_3
\;\underline{\hspace{3pt}}\,\Delta_2}_{\hspace{3pt}\,\Delta_4
\;\hspace{3pt}\, \Delta_1} \right]
     \right),
\end{eqnarray*}
and one odd,
\begin{eqnarray*}
\mathcal{T}^{\frac{1}{2}}_{\Delta}
\!\left[^{\underline{\hspace{3pt}}\,\Delta_3
\;\underline{\hspace{3pt}}\,\Delta_2}_{\hspace{3pt}\,\Delta_4
\;\hspace{3pt}\, \Delta_1} \right]\!
(z)
    &=&
z^{\Delta - \underline{\hspace{3pt}}\,\Delta_2 - \Delta_1 }\hskip -5pt
\sum_{k\in \mathbb{N}- \frac{1}{2}} \hskip -5pt z^k\,
        T^k_{ \Delta}
        \!\left[^{\underline{\hspace{3pt}}\,\Delta_3
\;\underline{\hspace{3pt}}\,\Delta_2}_{\hspace{3pt}\,\Delta_4
\;\hspace{3pt}\, \Delta_1} \right],
\end{eqnarray*}
conformal block. Although in correlation functions the even and the odd blocks show up separately it is convenient for our present purposes  to combine them
into a single function
\begin{eqnarray*}
\mathcal{T}_{\Delta}
\!\left[^{\underline{\hspace{3pt}}\,\Delta_3
\;\underline{\hspace{3pt}}\,\Delta_2}_{\hspace{3pt}\,\Delta_4
\;\hspace{3pt}\, \Delta_1} \right]\!
(z)
    &=&
\mathcal{T}^{1}_{\Delta}
\!\left[^{\underline{\hspace{3pt}}\,\Delta_3
\;\underline{\hspace{3pt}}\,\Delta_2}_{\hspace{3pt}\,\Delta_4
\;\hspace{3pt}\, \Delta_1} \right]\!
(z)
+
\mathcal{T}^{\frac{1}{2}}_{\Delta}
\!\left[^{\underline{\hspace{3pt}}\,\Delta_3
\;\underline{\hspace{3pt}}\,\Delta_2}_{\hspace{3pt}\,\Delta_4
\;\hspace{3pt}\, \Delta_1} \right]\!
(z).
\end{eqnarray*}
In the formulae above $\underline{\hspace{4pt}}\,\Delta_i$  stand for the conformal weight $\Delta_i$ of the highest weight state $\nu_i$
or the conformal weight $*\Delta_i=\Delta_i +{1\over 2}$ of the state $*\nu_p= G_{-{1\over 2}}\nu_i$.
An explicit expression for coefficients depends on the basis used in the factorization of the corresponding four-point chiral correlator.
For the basis
$$
L_{-M}G_{-K} f_{-N} \nu_p
$$
all states with nonzero $f$ excitations drop out from the factorization formula and one gets
\begin{eqnarray}
\label{block:definition}
&&
T^{t}_{\Delta}
\!\left[^{\underline{\hspace{3pt}}\,\Delta_3
\;\underline{\hspace{3pt}}\,\Delta_2}_{\hspace{3pt}\,\Delta_4
\;\hspace{3pt}\, \Delta_1} \right] \; =
\\
\nonumber
&& =
\hspace*{-20pt}
\begin{array}[t]{c}
{\displaystyle\sum} \\[2pt]
{\scriptstyle
|K|+|M| = |L|+|N| = t
}
\end{array}
\hspace*{-20pt}
\rho_{\NS} (\nu_4, \underline{\hspace{4pt}}\,\nu_3 , L_{-M}G_{-K}\nu_{\Delta}|1 )
\ B^t({\Delta})^{-1}_{MK,NL}  \
\rho_{\NS} (L_{-N}G_{-L}\nu_{\Delta},  \underline{\hspace{4pt}}\,\nu_2 , \nu_1|1 ),
\end{eqnarray}
where {\small $B^t({\Delta})^{-1}$} is the matrix inverse to Gram matrix (\ref{gramm}) in the NS Verma module ${\cal V}_\Delta$
and $\rho_\NS$ denotes the three-point ${\cal N}=1$ superconformal block.\footnote{In order to simplify notation we use the same symbol $\nu_i$ for the state
 $\ket{\Delta_i}$ and $\ket{\Delta_i}\otimes\ket{0_{\tilde f}}$.} It follows that all four-point blocks of the algebra
${\sf A}_\NS$ in the tensor product exactly coincide with the ${\cal N}=1$ superconformal blocks in the super-Liouville factor:
\begin{eqnarray}
\label{iden}
\mathcal{T}_{\Delta}
\!\left[^{\underline{\hspace{3pt}}\,\Delta_3
\;\underline{\hspace{3pt}}\,\Delta_2}_{\hspace{3pt}\,\Delta_4
\;\hspace{3pt}\, \Delta_1} \right]\!
(z)
&=&
\mathcal{F}_{\Delta}
\!\left[^{\underline{\hspace{3pt}}\,\Delta_3
\;\underline{\hspace{3pt}}\,\Delta_2}_{\hspace{3pt}\,\Delta_4
\;\hspace{3pt}\, \Delta_1} \right]\!
(z)
\;=\;
\mathcal{F}^{1}_{\Delta}
\!\left[^{\underline{\hspace{3pt}}\,\Delta_3
\;\underline{\hspace{3pt}}\,\Delta_2}_{\hspace{3pt}\,\Delta_4
\;\hspace{3pt}\, \Delta_1} \right]\!
(z)
+
\mathcal{F}^{\frac{1}{2}}_{\Delta}
\!\left[^{\underline{\hspace{3pt}}\,\Delta_3
\;\underline{\hspace{3pt}}\,\Delta_2}_{\hspace{3pt}\,\Delta_4
\;\hspace{3pt}\, \Delta_1} \right]\!
(z).
\end{eqnarray}
where on the r.h.s.\ the notation of \cite{Hadasz:2006sb} was used.

On the other hand one can factorize on basis (\ref{conbasis}). With identification (\ref{iden}) this gives relations between
superconformal blocks in the NS sector and Virasoro conformal blocks.
In order to make this relation precise one has to extend the notion of the four-point Virasoro conformal block to complex intermediate weights.
If one insists on the conjugation rules
$$
L^\dagger_{n}=L_{-n}
$$
the only consistent scalar product on ${\cal V}_\Delta\oplus {\cal V}_{\bar\Delta}$ is given by the pairing
\begin{equation}
\label{consistentscalar}
\langle \bar\Delta\,|\,\Delta \rangle =\overline{\langle\Delta\,|\,\bar\Delta  \rangle }= c,
\end{equation}
where $c$ is a non-vanishing complex number which we set 1 in the following.
The Gram matrix takes the off-diagonal form with the complex conjugated blocks
\begin{eqnarray*}
B^m({\Delta})_{\bar M,N}&=& \langle \bar\Delta\,|\,L_{\bar M}L_{-N} \,|\,\Delta\rangle,
\\
B^m({\bar \Delta})_{ M,\bar N}&=& \langle \Delta\,|\,L_{ M}L_{-\bar N} \,|\,\bar\Delta\rangle
.
\end{eqnarray*}
The identity operator on ${\cal V}_\Delta\oplus {\cal V}_{\bar\Delta}$ can be expressed as
$$
{\rm id} = \sum\limits_{N,\bar M} L_{-N} |\,\Delta\rangle B^{-1}(\Delta)_{N,\bar M} \langle\bar \Delta\,|L_{\bar M}
+
\sum\limits_{N,\bar M} L_{-\bar N} |\,\bar \Delta\rangle B^{-1}(\bar\Delta)_{\bar N,M} \langle\Delta\,|L_{M}.
$$
For each  conjugate pair of complex weights $\Delta, \bar \Delta$ it is then natural to introduce  two
four-point conformal blocks:
\begin{eqnarray*}
\mathcal{F}_{\Delta}
\!\left[^{\,\Delta_3\;\,\Delta_2}
_{\,\Delta_4\;\, \Delta_1} \right]\!
(z)
    &=&
z^{\Delta - \Delta_2 - \Delta_1} \left( 1 +
\sum_{m\in \mathbb{N}} \! z^m\,
    F^m_{ \Delta}
   \!\left[^{\,\Delta_3\;\,\Delta_2}
_{\,\Delta_4\;\, \Delta_1} \right]
     \right),
\\
F^m_{ \Delta}
   \!\left[^{\,\Delta_3\;\,\Delta_2}
_{\,\Delta_4\;\, \Delta_1} \right]
&=&
\begin{array}[t]{c}
{\displaystyle\sum} \\[2pt]
{\scriptstyle
|N| = |\bar M| = m
}
\end{array}
\hspace*{-10pt}
\rho (\nu_4, \nu_3 , L_{-N}\nu_{ \Delta}|1 )
\ B^t({\Delta})^{-1}_{N,\bar M}  \
\rho (L_{-\bar M}\nu_{\bar\Delta},  \nu_2 , \nu_1|1 )
\end{eqnarray*}
and its $\bar \Delta$ counterpart.
With this definition one has
\begin{eqnarray*}
\mathcal{F}_{\Delta_p}
\!\left[^{\Delta_3 \;\Delta_2}_{\Delta_4 \;\Delta_1} \right]\!
(z)
&=&
\sum\limits_{j\in \mathbb{Z}}
A_{\Delta(p,j)}\!\left[^{\Delta_3 \;\Delta_2}_{\Delta_4 \;\Delta_1} \right]\,
\mathcal{F}^{\li}_{\Delta^{\li}(p, j)}
\!\left[^{\Delta^\li_3 \;\Delta^\li_2}_{\Delta^\li_4 \;\Delta^\li_1} \right]\!
(z)\
\mathcal{F}^{\gi}_{\Delta^{\gi}(p, j)}
\!\left[^{\Delta^{\gi}_3 \;\Delta^{\gi}_2}_{\Delta^{\gi}_4 \;\Delta^{\gi}_1} \right]\!
(z)\ z^{j^2\over 2} ,
\\
A_{\Delta(p,j)}\!\left[^{\Delta_3 \;\Delta_2}_{\Delta_4 \;\Delta_1} \right]
&=&
{
\rho^\A_\NS(\xi_{p_4,0}, \xi_{p_3,0},\xi_{p,j}|1)
\rho^\A_\NS(\xi_{p,-j}, \xi_{p_2,0},\xi_{p_1,0}|1)
\over _n\langle\, p,-j\ket{p,j}_n}
,
\end{eqnarray*}
where we use the simplified notation $\Delta^\li_i=\Delta^\li(p_i,0) $, $\Delta^\gi_i=\Delta^\gi(p_i,0) $.
Formulae for the other types of blocks follow from the relation
\begin{equation}
\label{ghalf}
\textstyle G_{-{1\over 2}}\ket{\Delta_p}\otimes |\,{0}_{\!\tilde f}\,\rangle = {1\over 2}\ket{p,1}_n -\left(ip+{Q\over 2}\right) \ket{\Delta_p}\otimes f_{-{1\over 2}}|\,{0}_{\!\tilde f}\,\rangle.
\end{equation}
For blocks with a single star the $f$ excitations drop out and one has
\begin{eqnarray*}
\mathcal{F}_{\Delta_p}
\!\left[^{\Delta_3 *\Delta_2}_{\Delta_4 \;\;\Delta_1} \right]\!
(z)
&=&
\sum\limits_{j\in \mathbb{Z}}
A_{\Delta(p,j)}\!\left[^{\Delta_3 *\Delta_2}_{\Delta_4 \;\;\Delta_1} \right]\,
\mathcal{F}^{\li}_{\Delta^{\li}(p, j)}
\!\left[^{\Delta^\li_3 *\Delta^\li_2}_{\Delta^\li_4 \;\;\Delta^\li_1} \right]\!
(z)\
\mathcal{F}^{\gi}_{\Delta^{\gi}(p, j)}
\!\left[^{\Delta^{\gi}_3 *\Delta^{\gi}_2}_{\Delta^{\gi}_4 \;\;\Delta^{\gi}_1} \right]\!
(z)\ z^{j^2\over 2} ,
\\
A_{\Delta(p,j)}\!\left[^{\Delta_3 *\Delta_2}_{\Delta_4 \;\;\Delta_1} \right]
&=&
{1\over 2}
{
\rho^\A_\NS(\xi_{p_4,0}, \xi_{p_3,0},\xi_{p,j}|1)
\rho^\A_\NS(\xi_{p,-j}, \xi_{p_2,1},\xi_{p_1,0}|1)
\over _n\langle\, p,-j\ket{p,j}_n}
,
\end{eqnarray*}
where $*\Delta^\li_i=\Delta^\li(p_i,1) $, $*\Delta^\gi_i=\Delta^\gi(p_i,1) $.

The relations for double star blocks is slightly more complicated.
First of all relation (\ref{ghalf}) implies
\begin{eqnarray*}
\mathcal{F}_{\Delta_p}
\!\left[^{*\Delta_3 *\Delta_2}_{\;\;\Delta_4\;\; \Delta_1} \right]\!
(z)
&=&
{1\over 4}\mathcal{T}_{\Delta_p}
\!\left[^{\Delta(p_3,1)\,\Delta(p_2,1)}_{\;\;\;\;\;\Delta_4
\;\;\;\;\;\Delta_1} \right]\!
(z)
+
{2\Delta_p \over 1-z} \mathcal{F}_{\Delta_p}
\!\left[^{\Delta_3
\;\Delta_2}_{\Delta_4
\;\Delta_1} \right]\!
(z).
\end{eqnarray*}
Then using (\ref{3blockrel}) one gets
\begin{eqnarray*}
\mathcal{F}_{\Delta_p}
\!\left[^{*\Delta_3 *\Delta_2}_{\;\;\Delta_4 \;\;\Delta_1} \right]\!
(z)
&=&
\sum\limits_{j\in \mathbb{Z}}
A_{\Delta(p,j)}\!\left[^{*\Delta_3 *\Delta_2}_{\;\;\Delta_4 \;\;\Delta_1} \right]\,
\mathcal{F}^{\li}_{\Delta^{\li}(p, j)}
\!\left[^{*\Delta^\li_3 *\Delta^\li_2}_{\;\;\Delta^\li_4 \;\;\Delta^\li_1} \right]\!
(z)\
\mathcal{F}^{\gi}_{\Delta^{\gi}(p, j)}
\!\left[^{*\Delta^{\gi}_3 *\Delta^{\gi}_2}_{\;\;\Delta^{\gi}_4 \;\;\Delta^{\gi}_1} \right]\!
(z)\ z^{j^2\over 2}
\\
&&\hspace{-70pt}+\;
{2\Delta_p \over 1-z}
\sum\limits_{j\in \mathbb{Z}}
A_{\Delta(p,j)}\!\left[^{\Delta_3 \;\Delta_2}_{\Delta_4 \;\Delta_1} \right]\,
\mathcal{F}^{\li}_{\Delta^{\li}(p, j)}
\!\left[^{\Delta^\li_3 \;\Delta^\li_2}_{\Delta^\li_4 \;\Delta^\li_1} \right]\!
(z)\
\mathcal{F}^{\gi}_{\Delta^{\gi}(p, j)}
\!\left[^{\Delta^{\gi}_3 \;\Delta^{\gi}_2}_{\Delta^{\gi}_4 \;\Delta^{\gi}_1} \right]\!
(z)\ z^{j^2\over 2} ,
\\
A_{\Delta(p,j)}\!\left[^{*\Delta_3 *\Delta_2}_{\;\;\Delta_4 \;\;\Delta_1} \right]
&=&
{1\over 4}
{
\rho^\A_\NS(\xi_{p_4,0}, \xi_{p_3,1},\xi_{p,j}|1)
\rho^\A_\NS(\xi_{p,-j}, \xi_{p_2,1},\xi_{p_1,0}|1)
\over _n\langle\, p,-j\ket{p,j}_n}.
\end{eqnarray*}

\section{Equivalence }

\subsection{Chiral structure constants}

It was shown in Subsection 3.1 that the map
\begin{equation}
\label{map}
L_{-M}^{\li}L_{-N}^{\gi}   \ket{p,j}_n\;\;\;\longrightarrow\;\;\;
L_{-M}^{\li}|\,\nu^\li_{p,j}\rangle
\otimes
L_{-N}^{\gi}   |\,\nu^\gi_{p,j}\rangle
\end{equation}
is an
isomorphism  of
${\sf Vir} \oplus {\sf Vir}$ representations.
It has its counterpart for the corresponding  chiral operators
$$
{\cal L}_{-M}^{\li}{\cal L}_{-N}^{\gi}    V_{p,j}\;\;\;\longrightarrow\;\;\;
{\cal L}_{-M}^{\li}V^\li_{p,j} \otimes
{\cal L}_{-N}^{\gi}V^\gi_{p,j}.
$$
It provides an equivalence of the theories if all three point functions of corresponding operators are equal.
We have shown in the previous subsection that for $L_n^\li, L_n^\gi$ generators one can use the same Virasoro Ward identities on both sides of the correspondence.
It is then enough to check the correlators of the operators $V_{p,j}$. One obviously has the same situation in the right sector.

The following identities are responsible for the relations between the chiral structure constants\footnote{We give a simple derivation of these identities along with other useful formulae in Appendix C}:
\begin{eqnarray}
\label{ridentity1}
{\Gamma_\bli (\ali )\over \Gamma_\bgi (\agi + \bgi)}
&=& D(b)
b^{-{b^2\over 1-b^2} {\alpha(Q-\alpha)\over 4}}
\textstyle\left({1-b^2\over 2}\right)^{-{\alpha(Q-\alpha)\over 8}+{1\over 4}}
\Gamma^\NS_b\left(\alpha\right),
\\\label{ridentity2}
{\Gamma_\bli (\ali +{1\over 2}\bli)\over \Gamma_\bgi (\agi+ {1\over 2}(\bgi)^{-1}+\bgi)}
&=& D(b)
b^{-{b^2\over 1-b^2} {\alpha(Q-\alpha)\over 4}  +{b\alpha\over 2-2b^2} -{b^2\over 4-4b^2}       }
\textstyle
\left({1-b^2\over 2}\right)^{-{\alpha(Q-\alpha)\over 8}+{1\over 8}}
\Gamma^\R_b\left(\alpha\right),
\\
\label{ridentity3}
{\Gamma_\bli (\ali -{1\over 2}\bli)\over \Gamma_\bgi (\agi- {1\over 2}(\bgi)^{-1}+\bgi)}
&=& D(b)
b^{-{b^2\over 1-b^2} {\alpha(Q-\alpha)\over 4}  -{b\alpha\over 2-2b^2} +{2+b^2\over 4-4b^2}}
\textstyle\left({1-b^2\over 2}\right)^{-{\alpha(Q-\alpha)\over 8}+{1\over 8} }
\Gamma^\R_b\left(\alpha\right),
\\
\nonumber
D(b)&=& {\sqrt{2\pi }\,\Gamma_\bli (Q^\li)
\over \Gamma_\bgi (\bgi+(\bgi)^{-1})
\textstyle\Gamma_b\left(Q\right)\Gamma_b\left({Q\over 2}\right)}.
\end{eqnarray}
The first one is the chiral versions of the identity for upsilon functions
proposed in \cite{Belavin:2011sw}.
Using the identities above
one gets
\begin{eqnarray}
\label{LC}
{\sf C}^\li_b(\alpha^\li_3,\alpha^\li_2,\alpha^\li_1){\sf C}^\gi_\bgi(\alpha^\gi_3,\alpha^\gi_2,\alpha^\gi_1)
& = &
F(b)
{\sf C}^\NS_b(\alpha_3,\alpha_2,\alpha_1),
\\
\label{RC}
\bar {\sf C}^\li_b(\alpha^\li_3,\alpha^\li_2,\alpha^\li_1)\bar {\sf C}^\gi_\bgi(\alpha^\gi_3,\alpha^\gi_2,\alpha^\gi_1)
& = &
F(b)
\bar {\sf C}^\NS_b(\alpha_3,\alpha_2,\alpha_1),
\\
\label{LD}
\textstyle
{\sf C}^\li_\bli(\alpha^\li_3,\alpha^\li_2+{\bli\over 2},\alpha^\li_1)
{\sf C}^\gi_\bgi(\alpha^\gi_3,\alpha^\gi_2+{1\over 2\bgi},\alpha^\gi_1)
&=&
{2\sqrt{2}\over \sqrt{2(2\alpha_2)(Q-2\alpha_2)}}
 F(b)
{\sf D}^\NS_b(\alpha_3,\alpha_2,\alpha_1),
\\
\label{RD}
\textstyle
\bar {\sf C}^\li_\bli(\alpha^\li_3,\alpha^\li_2+{\bli\over 2},\alpha^\li_1)
\bar {\sf C}^\gi_\bgi(\alpha^\gi_3,\alpha^\gi_2+{1\over 2\bgi},\alpha^\gi_1)
&=&
{2\sqrt{2}\over \sqrt{2(2\alpha_2)(Q-2\alpha_2)}}
 F(b)
\bar {\sf D}^\NS_b(\alpha_3,\alpha_2,\alpha_1),
\end{eqnarray}
where
$$
F(b)=D(b)\,b^{-{b^2\over 1-b^2} {Q^2\over 4}}
\textstyle\left({1-b^2\over 2}\right)^{-{Q^2\over 8}+{1\over 4}}.
$$

We shall now analyze the three point chiral functions of operators corresponding to arbitrary $\ket{p,j}_n$ states.
For clarity of presentation  we restrict ourselves to  the case when  all $j$ and $\bar j$ are positive and inequalities (\ref{restrictions:on:excitation})
are satisfied. Let us first consider the even case
$
j_1+j_2+j_3
\;\in\; 2\mathbb{N}.
$
Using formulae (\ref{evenj}) and (\ref{oddj}) of Appendix C and properties of Euler gamma functions one gets
\begin{eqnarray*}
&&
{
{\sf C}^\li_\bli(\alpha^\li_3+{j_3\bli\over 2},\alpha^\li_2+{j_2\bli\over 2},\alpha^\li_1+{j_1\bli\over 2})
{\sf C}^\gi_\bgi(\alpha^\gi_3+{j_3\over 2\bgi},\alpha^\gi_2+{j_2\over 2\bgi},\alpha^\gi_1+{j_1\over 2\bgi})
\over
{\sf C}^\li_\bli(\alpha^\li_3,\alpha^\li_2,\alpha^\li_1)
{\sf C}^\gi_\bgi(\alpha^\gi_3,\alpha^\gi_2,\alpha^\gi_1)
}
\\[10pt]
&=&
{
\bar{\sf C}^\li_\bli(\alpha^\li_3+{j_3\bli\over 2},\alpha^\li_2+{j_2\bli\over 2},\alpha^\li_1+{j_1\bli\over 2})
\bar{\sf C}^\gi_\bgi(\alpha^\gi_3+{j_3\over 2\bgi},\alpha^\gi_2+{j_2\over 2\bgi},\alpha^\gi_1+{j_1\over 2\bgi})
\over
\bar{\sf C}^\li_\bli(\alpha^\li_3,\alpha^\li_2,\alpha^\li_1)
\bar{\sf C}^\gi_\bgi(\alpha^\gi_3,\alpha^\gi_2,\alpha^\gi_1)
}
\\[10pt]
&=&
\left(\prod\limits_{k=1}^3 {(-1)^{ j_k^2\over 2}\over \sqrt{l(2\alpha_k,2 j_k)l(2\alpha_k-Q,2 j_k)}}\right)\,
{\cal B}^\NS_{ j_3  j_2 j_1}(\alpha_3,\alpha_2,\alpha_1),
\end{eqnarray*}
with
$
{\cal B}^\NS_{ j_3  j_2 j_1}(\alpha_3,\alpha_2,\alpha_1)
$ defined in (\ref{defB}).
Together with (\ref{LC}) and (\ref{RC}) it yields
\begin{eqnarray*}
&&
{
{\sf C}^\li_\bli(\alpha^\li_3+{j_3\bli\over 2},\alpha^\li_2+{j_2\bli\over 2},\alpha^\li_1+{j_1\bli\over 2})
{\sf C}^\gi_\bgi(\alpha^\gi_3+{j_3\over 2\bgi},\alpha^\gi_2+{j_2\over 2\bgi},\alpha^\gi_1+{j_1\over 2\bgi})
\over
{\sf C}^\NS_b(\alpha_3,\alpha_2,\alpha_1)
}
\\
&=&
{
\bar{\sf C}^\li_\bli(\alpha^\li_3+{j_3\bli\over 2},\alpha^\li_2+{j_2\bli\over 2},\alpha^\li_1+{j_1\bli\over 2})
\bar{\sf C}^\gi_\bgi(\alpha^\gi_3+{j_3\over 2\bgi},\alpha^\gi_2+{j_2\over 2\bgi},\alpha^\gi_1+{j_1\over 2\bgi})
\over
\bar{\sf C}^\NS_b(\alpha_3,\alpha_2,\alpha_1)
}
\\[10pt]
&=&
\left(\prod\limits_{k=1}^3 {(-1)^{ j_k^2\over 2}\over \sqrt{l(2\alpha_k,2 j_k)l(2\alpha_k-Q,2 j_k)}}\right)\,
{\cal B}^\NS_{ j_3  j_2 j_1}(\alpha_3,\alpha_2,\alpha_1)F(b).
\end{eqnarray*}
The comparison  with formula (\ref{3pb}) for the conformal block suggests the rescaling
\begin{eqnarray}
\label{scaling}
\zeta_{p,j}\equiv \ket{p,j}_{nn}&=& {1\over \sqrt{2^j l(2ip + Q,2 j)l(2ip,2 j)}} \ket{p,j}_{n}.
\end{eqnarray}
This gives the normalization
$
_{nn}\langle\,p,-j\,\ket{p,j}_{nn}=1
$
which is in line with the natural normalization we have chosen in Subsection 3.3 for Virasoro Verma modules over complex weights
(\ref{consistentscalar}). If we assume this scalar product in direct sum (\ref{decva}) the maps
\begin{eqnarray*}
I\;:\;
L_{-M}^{\li}L_{-N}^{\gi}    \ket{\zeta_{p,j}}&\longrightarrow&
L_{-M}^{\li}|\,\nu^\li_{p,j}\rangle
\otimes
L_{-N}^{\gi}   |\,\nu^\gi_{p,j}\rangle ,
\\
\bar I\;:\;
L_{-\bar M}^{\li}L_{-\bar N}^{\gi}   \ket{\bar\zeta_{p,j}}&\longrightarrow&
L_{-\bar M}^{\li}|\,\bar \nu^\li_{p,j}\rangle
\otimes
L_{-\bar N}^{\gi}   |\,\bar \nu^\gi_{p,j}\rangle ,
\end{eqnarray*}
become unitary isomorphisms.

For the new states one has
\begin{eqnarray}
\nonumber
&&
{
{\sf C}^\li_\bli(\alpha^\li_3+{j_3\bli\over 2},\alpha^\li_2+{j_2\bli\over 2},\alpha^\li_1+{j_1\bli\over 2})
{\sf C}^\gi_\bgi(\alpha^\gi_3+{j_3\over 2\bgi},\alpha^\gi_2+{j_2\over 2\bgi},\alpha^\gi_1+{j_1\over 2\bgi})
\over
F(b){\sf C}^\NS_b(\alpha_3,\alpha_2,\alpha_1)
}
\\
\label{chiraleven}
&=&
{
\bar{\sf C}^\li_\bli(\alpha^\li_3+{j_3\bli\over 2},\alpha^\li_2+{j_2\bli\over 2},\alpha^\li_1+{j_1\bli\over 2})
\bar{\sf C}^\gi_\bgi(\alpha^\gi_3+{j_3\over 2\bgi},\alpha^\gi_2+{j_2\over 2\bgi},\alpha^\gi_1+{j_1\over 2\bgi})
\over
F(b)\bar{\sf C}^\NS_b(\alpha_3,\alpha_2,\alpha_1)
}
\\[10pt]
\nonumber
&=&
\gamma^\A_\NS(\zeta_{p_3,j_3}, \zeta_{p_2,j_2},\zeta_{p_1,j_1}|1).
\end{eqnarray}

In the odd case
$
j_1+j_2+j_3
\;\in\; 2\mathbb{N}+1
$
 one obtains
\begin{eqnarray*}
&&
{
{\sf C}^\li_\bli(\alpha^\li_3+{j_3\bli\over 2},\alpha^\li_2+{j_2\bli\over 2},\alpha^\li_1+{j_1\bli\over 2})
{\sf C}^\gi_\bgi(\alpha^\gi_3+{j_3\over 2\bgi},\alpha^\gi_2+{j_2\over 2\bgi},\alpha^\gi_1+{j_1\over 2\bgi})
\over
{\sf C}^\li_\bli(\alpha^\li_3,\alpha^\li_2+{\bli\over 2},\alpha^\li_1)
{\sf C}^\gi_\bgi(\alpha^\gi_3,\alpha^\gi_2+{1\over 2\bgi},\alpha^\gi_1)
}
\\[10pt]
&=&
{
\bar{\sf C}^\li_\bli(\alpha^\li_3+{j_3\bli\over 2},\alpha^\li_2+{j_2\bli\over 2},\alpha^\li_1+{j_1\bli\over 2})
\bar{\sf C}^\gi_\bgi(\alpha^\gi_3+{j_3\over 2\bgi},\alpha^\gi_2+{j_2\over 2\bgi},\alpha^\gi_1+{j_1\over 2\bgi})
\over
\bar{\sf C}^\li_\bli(\alpha^\li_3,\alpha^\li_2+{\bli\over 2},\alpha^\li_1)
\bar{\sf C}^\gi_\bgi(\alpha^\gi_3,\alpha^\gi_2+{1\over 2\bgi},\alpha^\gi_1)
}
\\[10pt]
&=&
\left(\prod\limits_{k=1}^3 {(-1)^{ j_k^2\over 2}\over \sqrt{l(2\alpha_k,2 j_k)l(2\alpha_k-Q,2 j_k)}}\right)
\sqrt{2\alpha_2(Q-2\alpha_2)}\,
{\cal B}^\R_{ j_3  j_2 j_1}(\alpha_3,\alpha_2,\alpha_1) F(b)
\end{eqnarray*}
where
$
{\cal B}^\R_{ j_3  j_2 j_1}(\alpha_3,\alpha_2,\alpha_1)
$ is given by (\ref{defB}).
The counterpart of (\ref{chiraleven}) reads
\begin{eqnarray}
\nonumber
&&
{
{\sf C}^\li_\bli(\alpha^\li_3+{j_3\bli\over 2},\alpha^\li_2+{j_2\bli\over 2},\alpha^\li_1+{j_1\bli\over 2})
{\sf C}^\gi_\bgi(\alpha^\gi_3+{j_3\over 2\bgi},\alpha^\gi_2+{j_2\over 2\bgi},\alpha^\gi_1+{j_1\over 2\bgi})
\over
2\sqrt{2} F(b){\sf D}^\NS_b(\alpha_3,\alpha_2,\alpha_1)
}
\\
\label{chiralodd}
&=&
{
\bar{\sf C}^\li_\bli(\alpha^\li_3+{j_3\bli\over 2},\alpha^\li_2+{j_2\bli\over 2},\alpha^\li_1+{j_1\bli\over 2})
\bar{\sf C}^\gi_\bgi(\alpha^\gi_3+{j_3\over 2\bgi},\alpha^\gi_2+{j_2\over 2\bgi},\alpha^\gi_1+{j_1\over 2\bgi})
\over
2\sqrt{2} F(b)\bar{\sf D}^\NS_b(\alpha_3,\alpha_2,\alpha_1)
}
\\[10pt]
\nonumber
&=&
\gamma^\A_\NS(\zeta_{p_3,j_3}, \zeta_{p_2,j_2},\zeta_{p_1,j_1}|1).
\end{eqnarray}
It follows from (\ref{chiraleven}) and (\ref{chiralodd}) that after an appropriate overall rescaling  of the NS chiral structure constants
the maps
$
I,
\bar  I$
provide  equivalence of chiral correlators in the left and in the right chiral sector, respectively.

\subsection{Correlation functions}

We shall now compare  the full correlation functions on both sides of the correspondence.
If one chooses the standard scalar product on the left-right tensor products
$$
\langle\,\xi\otimes\bar \xi\,\ket{\chi\otimes \bar \chi}
=
\langle\,\xi\,\ket{\chi}\langle\,\bar\xi\,\ket{\bar\chi}
$$
on both sides of the correspondence, then the map
 ${\cal I}: {\cal H}^{\rm \scriptscriptstyle SL} \to {\cal H}^{\rm \scriptscriptstyle LL}$,
$$
{\cal I} \;:\;
\zeta_{p,j}\otimes \bar \zeta_{p,\bar j} \;\longrightarrow  \;
\left\{
\begin{array}{rlll}
\nu^\li_{p,j,\bar j} \otimes \bar \nu^\gi_{p,j,\bar j}
 &{\rm for}& j,\bar j \in 2\mathbb{Z},
\\[10pt]
i\,\nu^\li_{p,j,\bar j} \otimes  \bar \nu^\gi_{p,j,\bar j}
&{\rm for}& j,\bar j \in 2\mathbb{Z}+1,
\end{array}
\right.
$$
where
$$
\nu^\li_{p,j,\bar j} \equiv \nu^\li_{p,j}\otimes \bar \nu^\li_{p,\bar j},
\;\;\;\;\;\;\
\nu^\gi_{p,j,\bar j} \equiv \nu^\gi_{p,j}\otimes \bar \nu^\gi_{p,\bar j},
$$
is a unitary isomorphism.
Its counterpart for the local fields
\begin{equation}
\label{equivalence}
{\cal I} \;:\;
\Phi_{p,j,\bar j}  \;\longrightarrow  \;
\left\{
\begin{array}{rlll}
\Phi^\li_{p,j,\bar j}  \otimes\Phi^\gi_{p,j,\bar j}
&{\rm for}& j,\bar j \in 2\mathbb{Z},
\\[10pt]
i\,\Phi^\li_{p,j,\bar j}  \otimes\Phi^\gi_{p,j,\bar j}
 &{\rm for}& j,\bar j \in 2\mathbb{Z}+1,
\end{array}
\right.
\end{equation}
is expected to provide the SL-LL equivalence if all correlation functions of corresponding operators are equal.
The choice of phase in (\ref{equivalence}) may seem strange at this stage
but it is in fact indispensable
as we shall see in the following.

If we restrict ourselves to the correlation functions on the sphere it is enough to show that all 3-point functions coincide and
that the factorization procedures are the same. For the first part let us recall that
 for $L_n^\li, L_n^\gi$ generators one can use the same Virasoro Ward identities on both sides of the correspondence.
It is then sufficient  to check the 3-point functions of the operators $\Phi_{p,j,\bar j}$.

Let us  first verify  the equivalence for the superprimary fields $\Phi_{p,0,0}$.
From definitions (\ref{DOZZ:threepoint}), (\ref{GMM:threepoint}), (\ref{constant1}) and  chiral relations (\ref{LC}), (\ref{RC})
one has\footnote{For the sake of clarity we  drop locations of fields which are assumed to be the standard ones $(0,0),(1,1),(\infty,\infty)$.}
\begin{eqnarray*}
{
\Big\langle
\Phi^\li_{p_3,0,0}
\Phi^\li_{p_2,0,0}
\Phi^\li_{p_1,0,0}
\Big\rangle_{\!\!\li}
\Big\langle
\Phi^\gi_{p_3,0,0}
\Phi^\gi_{p_2,0,0}
\Phi^\gi_{p_1,0,0}
\Big\rangle_{\!\!\gi}
\over
\Big\langle
\Phi_{p_3,0,0}
\Phi_{p_2,0,0}
\Phi_{p_1,0,0}
\Big\rangle_{\!\!\SL}
}
&=&F(b)^2 {M^\li_\bli M^\gi_\bgi \over M^\NS_b}.
\end{eqnarray*}
where
$$
F(b)^2= {\Upsilon_\bgi (\bgi)
{\textstyle\Upsilon_b\left(b\right)\Upsilon_b\left({Q\over 2}\right)}
\over
\Upsilon_\bli (\bli)
}
b^{-{b^2\over 1-b^2} {Q^2\over 2}}
\left({1-b^2\over 2}\right)^{-{Q^2\over 4}+{1\over 2}}.
$$
This yields the relative normalization condition (\ref{normcond})
which we assume is satisfied.

Let us now conisder 3-point function with one excitation. From  definition (\ref{constant2}), normalization (\ref{scaling})
and  relation (\ref{ghalf}) one has on the SL side
$$
\Big\langle
\Phi_{p_3,0,0}
\Phi_{p_2,1,1}
\Phi_{p_1,0,0}
\Big\rangle_{\!\!\SL}
=
i{8\over {2(2\alpha_2)(Q-2\alpha_2)}} M^\NS_b
{\sf D}^\NS_b(\alpha_3,\alpha_2,\alpha_1)
\bar {\sf D}^\NS_b(\alpha_3,\alpha_2,\alpha_1).
$$
The calculations on the LL side can be readily  done using chiral formulae (\ref{LD}), (\ref{RD}). If condition (\ref{normcond})
is satisfied one gets
\begin{eqnarray*}
{i
\Big\langle
\Phi^\li_{p_3,0,0}
\Phi^\li_{p_2,1,1}
\Phi^\li_{p_1,0,0}
\Big\rangle_{\!\!\li}
\Big\langle
\Phi^\gi_{p_3,0,0}
\Phi^\gi_{p_2,1,1}
\Phi^\gi_{p_1,0,0}
\Big\rangle_{\!\!\gi}
\over
\Big\langle
\Phi_{p_3,0,0}
\Phi_{p_2,1,1}
\Phi_{p_1,0,0}
\Big\rangle_{\!\!\SL}
}
&=&1.
\end{eqnarray*}
Note that this partially explains our choice of phase in (\ref{equivalence}). Using essentially the same calculations one checks the formula above
for the three other cases $(j_2,\bar j_2)=(1,-1),(-1,1),(-1,-1)$.

We shall  now consider the three point function of operators $\Phi_{p,j,\bar j}$ corresponding to arbitrary $\zeta_{p,j}\otimes \zeta_{p,\bar j}$ states.
For simplicity we restrict ourselves to  the case when  all $j$ and $\bar j$ are positive and inequalities (\ref{restrictions:on:excitation})
are satisfied. Let us first assume that the sum of all left $j$ indices is even\footnote{With the GSO projection assumed the right indices satisfy the same condition.}
$$
j_1+j_2+j_3
\;\in\; 2\mathbb{N}.
$$
There are two subcases: all left $j$ indices  are even or two of them are odd and one is even.
In the first subcase one has on the SL side\footnote{The same left and right chiral Ward identities are assumed so the left and the right 3-point blocks are the same function.}
\begin{eqnarray*}
\Big\langle
\Phi_{p_3,j_3,\bar j_3}
\Phi_{p_2,j_2,\bar j_2}
\Phi_{p_1,j_1,\bar j_1}
\Big\rangle_{\!\!\SL}
\\
&&
\hspace{-120pt}
=\;
C^\NS_b(\alpha_3,\alpha_2,\alpha_1)
\gamma^\A_\NS(
\zeta_{p_3,j_3}
\zeta_{p_2,j_2}
\zeta_{p_1,j_1}
|1)
\gamma^\A_\NS(
\bar\zeta_{p_3,\bar j_3}
\bar\zeta_{p_2,\bar j_2}
\bar\zeta_{p_1,\bar j_1}
|1).
\end{eqnarray*}
Then by chiral formulae (\ref{chiraleven}) one obtains
\begin{eqnarray*}
{
\Big\langle
\Phi^\li_{p_3,j_3,\bar j_3}
\Phi^\li_{p_2,j_2,\bar j_2}
\Phi^\li_{p_1,j_1,\bar j_1}
\Big\rangle_{\!\!\li}
\Big\langle
\Phi^\gi_{p_3,j_3,\bar j_3}
\Phi^\gi_{p_2,j_2,\bar j_2}
\Phi^\gi_{p_1,j_1,\bar j_1}
\Big\rangle_{\!\!\gi}
\over
\Big\langle
\Phi_{p_3,j_3,\bar j_3}
\Phi_{p_2,j_2,\bar j_2}
\Phi_{p_1,j_1,\bar j_1}
\Big\rangle_{\!\!\SL}
}
&=&1.
\end{eqnarray*}
In the case of two odd and one even $j$'s one has instead
\begin{eqnarray*}
\Big\langle
\Phi_{p_3,j_3,\bar j_3}
\Phi_{p_2,j_2,\bar j_2}
\Phi_{p_1,j_1,\bar j_1}
\Big\rangle_{\!\!\SL}
\\
&&
\hspace{-120pt}
=\;
- C^\NS_b(\alpha_3,\alpha_2,\alpha_1)
\gamma^\A_\NS(
\zeta_{p_3,j_3}
\zeta_{p_2,j_2}
\zeta_{p_1,j_1}
|1)
\gamma^\A_\NS(
\bar\zeta_{p_3,\bar j_3}
\bar\zeta_{p_2,\bar j_2}
\bar\zeta_{p_1,\bar j_1}
|1)
\end{eqnarray*}
with the minus sign coming from the anti-commutation of the odd left and the odd right excitations.
On the other hand due to our choice of phase in (\ref{equivalence}) one gets on the LL side the factor $i^2$
so the 3-point functions on both sides coincide.

Let us now turn to the odd case
$$
j_1+j_2+j_3
\;\in\; 2\mathbb{N}+1.
$$
As before it splits into two subcases:   one left index is odd and two are even or all of them are odd.
In the first case the splitting of the 3-point function into $A$-algebra conformal blocks does not develop any extra sign:
\begin{eqnarray*}
\Big\langle
\Phi_{p_3,j_3,\bar j_3}
\Phi_{p_2,j_2,\bar j_2}
\Phi_{p_1,j_1,\bar j_1}
\Big\rangle_{\!\!\SL}
\\
&&
\hspace{-120pt}
=\;
\tilde C^\NS_b(\alpha_3,\alpha_2,\alpha_1)
\gamma^\A_\NS(
\zeta_{p_3,j_3}
\zeta_{p_2,j_2}
\zeta_{p_1,j_1}
|1)
\gamma^\A_\NS(
\bar\zeta_{p_3,\bar j_3}
\bar\zeta_{p_2,\bar j_2}
\bar\zeta_{p_1,\bar j_1}
|1).
\end{eqnarray*}
Then by chiral formulae (\ref{chiralodd}) one obtains
\begin{eqnarray*}
{i\,
\Big\langle
\Phi^\li_{p_3,j_3,\bar j_3}
\Phi^\li_{p_2,j_2,\bar j_2}
\Phi^\li_{p_1,j_1,\bar j_1}
\Big\rangle_{\!\!\li}
\Big\langle
\Phi^\gi_{p_3,j_3,\bar j_3}
\Phi^\gi_{p_2,j_2,\bar j_2}
\Phi^\gi_{p_1,j_1,\bar j_1}
\Big\rangle_{\!\!\gi}
\over
\Big\langle
\Phi_{p_3,j_3,\bar j_3}
\Phi_{p_2,j_2,\bar j_2}
\Phi_{p_1,j_1,\bar j_1}
\Big\rangle_{\!\!\SL}
}
&=&1.
\end{eqnarray*}
In the subcase of all odd indices one has the extra minus sign on the SL side and the
extra $i^2$ factor on the LL side, so again the 3-point functions coincide.
All cases involving other inequalities and non-positive indices can be analyzed in the same way. The final result
can be compactly written as follows
$$
\Big\langle
\Phi_{p_3,j_3,\bar j_3}
\Phi_{p_2,j_2,\bar j_2}
\Phi_{p_1,j_1,\bar j_1}
\Big\rangle_{\!\!\rm \scriptscriptstyle SL} =
\Big\langle
{\cal I}(\Phi_{p_3,j_3,\bar j_3})
{\cal I}(\Phi_{p_2,j_2,\bar j_2})
{\cal I}(\Phi_{p_1,j_1,\bar j_1})
\Big\rangle_{\!\!\rm \scriptscriptstyle LL}.
$$
Let us stress the role the phase $i$ introduced in (\ref{equivalence}) plays in the formula above ---
it reproduces the factor $i$ in the NS structure constant $\tilde C^\NS_b$ (\ref{constant2})
and the extra minus signs coming  from the anti-commutation of odd objects on the SL side.

The last step of our proof of the SL-LL equivalence is to compare the factorization of correlation functions.
To this end we choose in ${\cal H}^{\rm \scriptscriptstyle SL}$ the basis
\begin{equation}
\label{goodbasis}
L^\li_{-M}L^\gi_{-N}
\,\ket{\zeta_{p,j}}\otimes \bar L^\li_{-\bar M}\bar L^\gi_{-\bar N}\,\ket{\bar\zeta_{p,\bar j}}.
\end{equation}
The Virasoro generators in the formula above are all even operators and the map $\cal I$ is an isomorphisms of
${\sf Vir} \oplus {\sf Vir}\oplus \bar{\sf Vir} \oplus \bar{\sf Vir}$ representations one thus can safely drop
them from the subsequent formulae.
Let us consider for instance the factorization of the 4-point function on basis (\ref{goodbasis}).
In the simplified notation  it takes the form
\begin{eqnarray*}
\Big\langle
\Phi_4
\Phi_3
\Phi_2
\Phi_1
\Big\rangle_{\!\!\SL}
&=&
\int_p\sum\limits_{j,\bar j \in 2\mathbb{Z}}
\Big\langle
\Phi_4
\Phi_3
\Phi_{p,j,\bar j}
\Big\rangle_{\!\!\SL}
\Big\langle
\Phi_{-p,j,\bar j}\,
\Phi_2
\Phi_1
\Big\rangle_{\!\!\SL}
\\
&-&
\int_p\sum\limits_{j,\bar j \in 2\mathbb{Z}+1}
\Big\langle
\Phi_4
\Phi_3
\Phi_{p,j,\bar j}
\Big\rangle_{\!\!\SL}
\Big\langle
\Phi_{-p,j,\bar j}\,
\Phi_2
\Phi_1
\Big\rangle_{\!\!\SL}.
\end{eqnarray*}
On the LL side one has
\begin{eqnarray*}
&&
\hspace{-50pt}
\Big\langle
{\cal I}(\Phi_4)
{\cal I}(\Phi_3)
{\cal I}(\Phi_2)
{\cal I}(\Phi_1)
\Big\rangle_{\!\!\LL}
\\
&=&
\int_p\sum\limits_{j,\bar j \in 2\mathbb{Z}}
\Big\langle
{\cal I}(\Phi_4)
{\cal I}(\Phi_3)
\Phi^\li_{p,j,\bar j}  \otimes\Phi^\gi_{p,j,\bar j}
\Big\rangle_{\!\!\LL}
\Big\langle
\Phi^\li_{-p,j,\bar j}  \otimes\Phi^\gi_{-p,j,\bar j}\,
{\cal I}(\Phi_2)
{\cal I}(\Phi_1)
\Big\rangle_{\!\!\LL}
\\
&-&
\int_p\sum\limits_{j,\bar j \in 2\mathbb{Z}+1}
\Big\langle
{\cal I}(\Phi_4)
{\cal I}(\Phi_3)\,
i\,\Phi^\li_{p,j,\bar j}  \otimes\Phi^\gi_{p,j,\bar j}
\Big\rangle_{\!\!\LL}
\Big\langle
i \,\Phi^\li_{-p,j,\bar j}  \otimes\Phi^\gi_{-p,j,\bar j}\,
{\cal I}(\Phi_2)
{\cal I}(\Phi_1)
\Big\rangle_{\!\!\LL}
\\
&=&
\int_p\sum\limits_{j,\bar j \in 2\mathbb{Z}}
\Big\langle
{\cal I}(\Phi_4)
{\cal I}(\Phi_3)
{\cal I}(\Phi_{p,j,\bar j})
\Big\rangle_{\!\!\LL}
\Big\langle
{\cal I}(\Phi_{-p,j,\bar j}) \,
{\cal I}(\Phi_2)
{\cal I}(\Phi_1)
\Big\rangle_{\!\!\LL}
\\
&-&
\int_p\sum\limits_{j,\bar j \in 2\mathbb{Z}+1}
\Big\langle
{\cal I}(\Phi_4)
{\cal I}(\Phi_3)\,
{\cal I}(\Phi_{p,j,\bar j})
\Big\rangle_{\!\!\LL}
\Big\langle
{\cal I}(\Phi_{-p,j,\bar j}) \,
{\cal I}(\Phi_2)
{\cal I}(\Phi_1)
\Big\rangle_{\!\!\LL}.
\end{eqnarray*}
One thus gets the exact equivalence of the 4-point functions. It is also clear that the simple mechanism above works for
any factorization of any $n$-point function on the sphere as well.

\section*{Acknowledgments}

The work  was financed by the NCN grant DEC2011/01/B/ST1/01302.
We would like to thank Paulina Suchanek for numerous discussions and for shearing with us her understanding of the Ramond sector.
We would also like to thank Micha{\l} Pawe{\l}kiewicz for discussions on chiral structure constants in Liouville theories
and for suggesting to us a simple proof of identity (\ref{ridentity1}).

\appendix
\label{appendixA}
\section{Fermionic state properties }
In this appendix we shall prove the propositions of Sect. 2.2.


\renewcommand{\theequation}{A.\arabic{equation}}

{\bf Proof of Prop.1:}

In the coefficients of the decomposition
\begin{eqnarray*}
\tilde \psi_{-L}(p)\ket{\Delta_p}
&=&
\displaystyle
\sum\limits_{MK}  S^n(p)^{-1}_{MK,\emptyset L} L_{-M}G_{-K} \ket{\Delta_p}
\\[12pt]
&=&
\displaystyle
\sum\limits_{MK,M'K'} N^n_{NL} \overline{S^n(p)}_{\emptyset ,M'K'}\,B^n(p)^{-1}_{M'K',MK} L_{-M}G_{-K} \ket{\Delta_p}
\end{eqnarray*}
the poles arise due to  the existence
of a singular vector
$
\ket{\chi_{rs} }\in {\cal V}_{\Delta_{rs}}
$
on level ${rs}$.
Let $\chi_{rs}^{NL}$ denote the coefficients of $\ket{\chi_{rs}}$ in the basis $L_{-N}G_{-L}\ket{\Delta_{rs}},$
\begin{eqnarray}
\label{decomposition}
\ket{\chi_{rs}} &=& D_{rs} \ket{\Delta_{rs}}\ ,
\\
\nonumber
D_{rs}&=& \sum_{N,L}\chi_{rs}^{NL}L_{-N}G_{-L}\ .
\end{eqnarray}
We normalize $\ket{\chi_{rs}}$ such that the coefficient at $(L_{-1})^{rs}\ket{\Delta_{rs}}$
is equal 1. For ${rs}<|K|$ consider  vectors of the form
$$
L_{-N}G_{-L}\,D_{rs} \ket{\Delta_p}\,,
\hskip 5mm
|N|+|L|= |K| - \textstyle {rs}\,,
$$
so that
$
\ket{\chi_{rs}} = \lim_{p\to p_{rs}} D_{rs} \ket{\Delta_p}.
$
The set of these vectors can be always extended to a full basis in ${\cal V}_\Delta^{|K|}$.
Working in such a basis and using the properties of the Gram matrix $B^{|K|}_{c, \Delta}$
and its inverse \cite{Hadasz:2006sb} one gets
\begin{eqnarray}
\label{res:1}
&&\hspace{-35pt}
 \lim\limits_{p\to p_{rs}} (p-p_{rs})\,\tilde \psi(p)_{-K}\ket{\Delta_p}  \\
\nonumber
&=&
{A_{rs}\over 2p_{rs}}\sum
 \left(\lim\limits_{p\to p_{rs}}
   \bra{\Delta_p} D_{rs}^\dagger G_{-L}^\dagger L_{-N}^\dagger  \tilde \psi(p)_{-K} \ket{\Delta_p}
   \right)
   \ \left[B^{|K|-\frac{rs}{2}}_{c, \Delta_{rs}+\frac{rs}{2}}\right]^{NL,N'L'} \!
   L_{-N'}G_{-L'}\ket{\chi_{rs}} ,
\end{eqnarray}
where
\begin{eqnarray*}
\label{A:rs:1}
A_{rs}
&=&
\lim_{\Delta\to\Delta_{rs}}
\left(\frac{\bra{\Delta_{rs}} D_{rs}^\dagger D_{rs} \ket{\Delta_{rs}}}{\Delta - \Delta_{rs}}
\right)^{-1}.
\\
\nonumber
&=&
2^{rs-2}
\prod_{m=1-r}^r
\prod_{n=1-s}^s
\left(m b + {n}{b}^{-1}\right)^{-1}
\end{eqnarray*}
and $m+n \in 2{\mathbb Z}, \; (m,n) \neq (0,0),(r,s).$ Using unitary isomorphism (\ref{isom})
one easily checks that the limit in (\ref{res:1}) exists and is finite
\begin{eqnarray*}
\lim\limits_{p\to p_{rs}}
   \bra{\Delta_p} D_{rs}^\dagger G_{-L}^\dagger L_{-N}^\dagger  \tilde \psi(p)_{-K} \ket{\Delta_p}
&=&\lim\limits_{p\to p_{rs}}
   \bra{p} D(p)_{rs}^\dagger G(p)_{-L}^\dagger L(p)_{-N}^\dagger  \psi_{-K} \ket{p}
   \\
&=&
   \bra{p_{rs}} D(p_{rs})_{rs}^\dagger G(p_{rs})_{-L}^\dagger L(p_{rs})_{-N}^\dagger  \psi_{-K} \ket{p_{rs}},
\end{eqnarray*}
where $D(p)_{rs}= \sum_{N,L}\chi_{rs}^{NL}L(p)_{-N}G(p)_{-L}.$ Hence (\ref{res:1}) takes the form
\begin{eqnarray}
\label{res:2}
&&\hspace{-20pt}
 \lim\limits_{p\to p_{rs}} (p-p_{rs})\,\tilde \psi(p)_{-K}\ket{\Delta_p}  \\
\nonumber
&=&
{A_{rs}\over 2p_{rs}}\sum
 \bra{\chi(p_{rs})}  G(p_{rs})_{-L}^\dagger L(p_{rs})_{-N}^\dagger  \psi_{-K} \ket{p_{rs}}
   \ \left[B^{|K|-\frac{rs}{2}}_{c, \Delta_{rs}+\frac{rs}{2}}\right]^{NL,N'L'} \!
   L_{-N'}G_{-L'}\ket{\chi_{rs}} ,
\end{eqnarray}
where $\bra{\chi(p_{rs})}  = \bra{p_{rs}}D^\dagger_{rs}(p_{rs})$.

We shall show that due to special properties of pure fermionic states some of limits (\ref{res:2}) vanish.
Let us first observe that up to an overall normalization
the singular vectors $\bra{\chi(p_{rs})}$
can be constructed by means of the screening charges \cite{Kato:1987qda}:
\begin{equation}
\label{hws:formal:a}
\bra{p_{rs}-irb}{\sf Q}_b^r
\ = \
\oint\limits_{\infty}dz_r\oint\limits_{{\cal C}_{r-1}}dz_{r-1}\ldots\oint\limits_{{\cal C}_1}dz_1\
\bra{p_0}\psi(z_1){\sf E}^b(z_1)\ldots\psi(z_r){\sf E}^b(z_r)
\end{equation}
for $r \leqslant s$ or
\begin{equation}
\label{hws:formal:b}
\bra{p_{rs}-isb^{-1}}{\sf Q}_{\frac{1}{b}}^s
\ = \
\oint\limits_{\infty}dz_s\oint\limits_{{\cal C}_{s-1}}dz_{s-1}\ldots\oint\limits_{{\cal C}_1}dz_1\
\bra{p_0}\psi(z_1){\sf E}^{\frac{1}{b}}(z_1)\ldots\psi(z_s){\sf E}^{\frac{1}{b}}(z_s)
\end{equation}
for $s \leqslant r,$
where all the integration contours are closed\footnote{This  condition restricts  momenta to the discrete subset $\{p_{rs}\}.$}. It follows
 that the singular vector $\bra{\chi(p_{rs})}$ contains at most
min$\{r,s\}$ fermionic excitations (oscillators).

Suppose that $r\leqslant s$ and  consider matrix elements of the form
\[
\langle \chi(p_{rs})|G_{L}L_{N}\psi_{-\frac{2j-1}{2}}\ldots\psi_{-\frac32}\psi_{-\frac12}|p_{rs}\rangle
\]
where, for  notational convenience the argument $p_{rs}$ of generators
$L_{n}$ and $G_{l}$ is suppressed. Since for all $n > 0:$
\[
L_n\psi_{-\frac{2j-1}{2}}\ldots\psi_{-\frac32}\psi_{-\frac12}|p_{rs}\rangle = 0,
\]
we can restrict ourselves  to matrix elements of the form
\begin{equation}
\label{matrix:element:of:interest}
\langle \chi(p_{rs})|G_{l_1}\ldots G_{l_m}\psi_{-\frac{2j-1}{2}}\ldots\psi_{-\frac32}\psi_{-\frac12}|p_{rs}\rangle
\end{equation}
where $l_1 < l_2 < \ldots < l_m$ and
\begin{equation}
\label{level:matching}
\frac{rs}{2} + \sum\limits_{i=1}^m l_i = \frac12 j^2.
\end{equation}
Since
\[
\{G_l,\psi_k\}
\; = \;
(1-\delta_{l+k})c_{l+k} + \big(l- iQk\big)\delta_{l+k}
\]
the state
\[
G_{l_1}\ldots G_{l_m}\psi_{-\frac{2j-1}{2}}\ldots\psi_{-\frac32}\psi_{-\frac12}|p_{rs}\rangle
\]
contains at least $j-m$ fermionic excitations. On the other hand the
state $\langle \chi(p_{rs})|$ contains at most $r$ fermionic excitations. The inequality
\begin{equation}
\label{second:condition}
r \geqslant k-m
\end{equation}
is  thus a necessary condition for (\ref{matrix:element:of:interest}) not to vanish.
Since all $l_i$ are different
\begin{eqnarray*}
\sum\limits_{i=1}^m l_i
& \geqslant &
\frac12\sum\limits_{i=1}^m (2i-1)
\; = \;
\frac12m^2
\; \geqslant \;
\frac12(j-r)^2
\end{eqnarray*}
where the second inequality follows from (\ref{second:condition}). Using (\ref{level:matching}) we thus get
\begin{eqnarray*}
\frac12\left(j^2-rs\right)
& = &
\sum\limits_{i=1}^m l_i
\; \geqslant \;
\frac12(j-r)^2
\end{eqnarray*}
or, equivalently
\begin{eqnarray}
\label{the:final:condition}
r+s & \leqslant & 2j.
\end{eqnarray}
It follows that all the states at the level $\frac12j^2 > \frac{rs}{2},$ generated from $\bra{\chi(p_{rs})}$ by the
operators $G_l$ and $L_n$ are orthogonal to the state $\psi_{-\frac{2j-1}{2}}\ldots\psi_{-\frac32}\psi_{-\frac12}|p_{rs}\rangle$
{\em unless} condition (\ref{the:final:condition}) is satisfied.

Consider now a matrix element of the form
\begin{equation}
\label{matrix:element:of:interest:2}
\langle \chi(p_{rs})|G_{L}L_{N}\psi_{-J'}|p_{rs}\rangle
\end{equation}
where $K'$ is some subset of $J = \{\frac{2j-1}{2},\ldots,\frac32,\frac12\}.$ Since
\[
[L_m,\psi_k]
\; = \;
-\left(\frac12m + k\right)\psi_{m+k},
\]
the state $L_{N}\psi_{-K'}|p_{rs}\rangle$ (if non-zero) is a combination of states of the form $\psi_{-K}|p_{rs}\rangle,$ where again $K$ is a subset of $J.$
We can thus restrict ourselves to matrix elements of the form
\begin{equation}
\label{matrix:element:of:interest:3}
\langle \chi(p_{rs})|G_{L}\psi_{-K}|p_{rs}\rangle.
\end{equation}
Suppose that for $r+s > 2k$ there exists a multi-index  $L$ such that (\ref{matrix:element:of:interest:3}) is non-zero.
As a polynomial in $b$ and $b^{-1}$ the matrix element (\ref{matrix:element:of:interest:3}) has the large $b$ leading term
\[
\langle \chi(p_{rs})|G_{L}\psi_{-K}|p_{rs}\rangle
\stackrel{b \to \infty}{=}
{\cal N}b^{n_0} + {\cal O}\left(b^{n_0-1}\right), \hskip 1cm {\cal N}\neq 0.
\]
We shall now calculate the large $b$ leading term of the collerator
\begin{eqnarray}
\label{temp:corr}
\langle \chi(p_{rs})|G_{L}G_{J\setminus K}\psi_{-J}|p_{rs}\rangle
\end{eqnarray}
where $J\setminus K$ is the multi-index composed of those indices which complete $K$ to $J.$
To this end it is sufficient to keep only
 the ``linear'' parts of the generators
\[
G_k(p_{rs}) \sim  ib\left(k+\frac{r}{2}\right)\psi_k.
\]
This yields
\begin{eqnarray*}
\langle \chi(p_{rs})|G_{L}G_{J\setminus K}\psi_{-K}|p_{rs}\rangle
& \stackrel{b\to \infty}{=} &
\pm i^{\# J - \#K} \prod\limits_{l\in J\setminus K}\left(l+\frac{r}{2}\right)\,{\cal N}\, b^{n_0 + \# L - \#J}\left(1+ {\cal O}\left(b^{-1}\right)\right)
\end{eqnarray*}
and therefore the correlator on the l.h.s.\ does not vanish in contradiction with our previous considerations.
This completes the proof in the general case.

{\bf Proof of Prop.2}

The aim is to calculate the upper bound for the degree of the coefficients of the decomposition
$$
\Omega(p,j) \,\tilde \psi(p)_{-K}\ket{\Delta_p} =\sum \Omega(p,j) S^{n}(p)^{-1}_{NL,\emptyset K} \, L_{-N}G_{-L}\ket{\Delta_p}
$$
where $K\subset J=\{ {2j-1\over 2},\dots, {1\over 2}\}$.
Our strategy is to consider this equation in the free field representation
$$
\Omega(p,j) \, \psi_{-K}\ket{p} =\sum \Omega(p,j) S^{n}(p)^{-1}_{NL,\emptyset K} \, L(p)_{-N}G(p)_{-L}\ket{p}
$$
and express all the objects involved it terms of new variables with well defined scaling properties.
Let $\epsilon_1,\epsilon_2$ be real positive parameters such that
$
b=\sqrt{\epsilon_1\over \epsilon_2}.
$
We introduce the rescaled oscillators and the momentum
\[
d_n=\sqrt{\epsilon_1\epsilon_2} c_n,\;\;\;{\sf q}=\sqrt{\epsilon_1\epsilon_2} {\sf p},
\]
and the generators
\begin{eqnarray}
\label{NS:alebra:rep}
\nonumber
l_0
& = &
\sum\limits_{m \geqslant 1}{d}_{-m}{d}_m +
\epsilon_1\epsilon_2\sum\limits_{k\geqslant \frac12}k\,\psi_{-k}\psi_{k} +
\frac18(\epsilon_1+\epsilon_2)^2 + \frac12 {\sf q}^2\;=\;\epsilon_1\epsilon_2L_0({p}),
\\
l_n
& = &
\frac12\sum\limits_{m\neq 0,n}\!{d}_{n-m}{d}_m +
{\epsilon_1\epsilon_2\over 2}\sum\limits_{k\in {\mathbb Z} +\frac12}\hskip -4pt k\,\psi_{n-k}\psi_{k}+
\frac{1}{2}\left(in\epsilon_1+in\epsilon_2+ 2{\sf q}\right){d}_n
\\
&=& \epsilon_1\epsilon_2L_n({p}),
\\
\nonumber
g_k & = & \sum\limits_{m\neq 0}{d}_{m}\psi_{k-m}+ (ik\epsilon_1 +ik\epsilon_2 +{\sf q})\phi_k\;=\; \sqrt{\epsilon_1\epsilon_2}G_k({p}).
\end{eqnarray}
They satisfy the modified commutation relations
\begin{eqnarray}
\nonumber
[{d}_m,{d}_n] &=& m\epsilon_1\epsilon_2\delta_{m+n},
\\
\label{mNS}
\nonumber
\left[l_m,l_n\right] & = & (m-n)\epsilon_1\epsilon_2 l_{m+n} +\frac{\epsilon_1\epsilon_2(\epsilon_1\epsilon_2+ (\epsilon_1+\epsilon_2)^2)}{12}m\left(m^2-1\right)\delta_{m+n},
\\
\left[l_m,g_k\right] & = &\frac{m-2k}{2}\epsilon_1\epsilon_2 g_{m+k},
\\
\nonumber
\left\{g_k,g_l\right\} & = & 2 l_{k+l} + \frac{(\epsilon_1\epsilon_2+ (\epsilon_1+\epsilon_2)^2)}{3}\left(k^2 -\frac14\right)\delta_{k+l}.
\end{eqnarray}
Since
\begin{eqnarray*}
l_{-N}g_{-L}\ket{p} &=& \sum
   {\bra{p} \psi_{-K}^\dagger  d_{-M}^\dagger l_{-N} g_{-L}  \ket{p}\over m^n_{NL}}        d_{-M}\psi_{-K}\ket{p},
\\
m^n_{M K} &=& \bra{p} \psi_{-K}^\dagger d_{-M}^\dagger d_{-M} \psi_{-K} \ket{p}= (\epsilon_1\epsilon_2)^{\# M}N_{MK},
\end{eqnarray*}
one has
\begin{equation}
\label{eles}
d_{-M}\psi_{-K}\ket{p} =\sum s^{n}(q)^{-1}_{NL,M K} \, l_{-N}g_{-L}\ket{p}
\end{equation}
where
\begin{eqnarray}
\label{s}
s(q)^{\,n}_{MK,NL}&=&{\bra{p} \psi_{-K}^\dagger  d_{-M}^\dagger l_{-N}g_{-L} \ket{p}\over (\epsilon_1\epsilon_2)^{\# M}N_{MK}}.
\end{eqnarray}
We shall show that matrix elements $ s(q)^{\,n}_{MK,NL}$ are polynomials in all variables $\epsilon_1,\epsilon_2,q$.
Using the commutation relations
\begin{eqnarray*}
[l_n,d_m]
& = &
-m(1-\delta_{m+n})\epsilon_1\epsilon_2 d_{m+n} -{1\over 2} m\epsilon_1\epsilon_2 \big(2q-im\epsilon_1-im\epsilon_2 \big)\delta_{n+m},
\\
{[l_n,\psi_s]}
& = &
-\left(\frac12n + s\right)\epsilon_1\epsilon_2\psi_{n+s},
\\
{[g_k,d_m]}
& = &
-m\epsilon_1\epsilon_2\psi_{k+m},
\\
\{g_k,\psi_l\}
& = &
\sqrt{\epsilon_1\epsilon_2}(1-\delta_{k+l})d_{k+l} + \sqrt{\epsilon_1\epsilon_2}\big(q- il\epsilon_1-il\epsilon_2\big)\delta_{r+s},
\end{eqnarray*}
one can calculate the numerator
of (\ref{s})
$$
\bra{p} \psi_{-K}^\dagger  d_{-M}^\dagger l_{-N}g_{-L} \ket{p}
$$
moving  the bosonic oscillators $d_{-M}^\dagger=d_{m_1}\dots d_{m_j} $ one by one to the right.
Moving $d_{m_j}$ gives a sum of terms, each term  containing the commutator of $d_{m_j}$ with $l_n$ or $ g_l$
and hence  the factor $\epsilon_1\epsilon_2$. Moving subsequent oscillators $d_{m_{j'}},j'<j$ to the right yields new
terms involving commutators with generators, yielding again the factor $\epsilon_1\epsilon_2$, and oscillators resulting from the previous steps.
The commutation with the oscillators is nonzero only if $d_{m_{j-1}}$ meets its conjugated counterpart
and gives  the factor $\epsilon_1\epsilon_2$. Each step contributes therefore the factor $\epsilon_1\epsilon_2$.
If after moving all bosonic oscillators to the right  the result is nonzero it has the overall factor
$\epsilon_1^j\epsilon_2^j$ which cancels against the denominator of (\ref{s}).

Consider now the equation
\begin{equation}
\label{small}
\omega(q,j)\psi_{-K}\ket{p} =\sum \omega(q,j)s^{n}(q)^{-1}_{NL,\emptyset K} \, l_{-N}g_{-L}\ket{p}
\end{equation}
where
\begin{eqnarray*}
\omega(q,j) &=&
\begin{array}[t]{c}
{\displaystyle\prod} \\[-5pt]
{\scriptscriptstyle
1 \leqslant n,m}
\\[-7pt]
{\scriptscriptstyle n+m \leqslant 2\,j}
\\[-7pt]
{\scriptscriptstyle
n+m\,\in\, 2{\mathbb N}
}
\end{array}
\left(2qi  +(n-1)\epsilon_1 + (m-1)\epsilon_2\right).
\end{eqnarray*}
Under the scaling
$$
d_n \to \lambda d_n,\;\;\;\psi_l\to \psi_l,\;\;\;\epsilon_1\to \lambda \epsilon_1,\;\;\;\epsilon_2\to \lambda \epsilon_2,
$$
one has  the homogeneous transformation laws
$$
l_n \to \lambda^2 l_n,\;\;\;g_l\to \lambda g_l,\;\;\; \omega(q,j) \to \lambda^{\,j^2} \omega(q,j).
$$
As the r.h.s.\ of (\ref{small}) is a combination of linearly independent vectors and the l.h.s.\ scales as $\lambda^{j^2}$
the scaling rule for each coefficient must be
$$
\omega(q,j)s^{n}(q)^{-1}_{NL,\emptyset K} \to \lambda^{j^2-2\# N -\# L}\omega(q,j)s^{n}(q)^{-1}_{NL,\emptyset K}.
$$
If $\omega(q,j)s^{n}(q)^{-1}_{NL,\emptyset K}$ is a polynomial in variables $q,\epsilon_1,\epsilon_2$ then the scaling
implies
$$
\deg_p \left(\Omega(p,j) S^{n}(p)^{-1}_{NL,\emptyset K} \right)=
\deg_q \left(\omega(q,j) s^{n}(q)^{-1}_{NL,\emptyset K} \right)
\leq j^2 - 2\#N-\#L.
$$
We shall show that this is indeed the case. To this end let us first calculate the determinant of $s^n$.
Since,
\begin{eqnarray*}
S^{\,n}_{NL,MK}( p) &=&
(\epsilon_1\epsilon_2)^{{1\over 2}\#N - \# M-{1\over 2}\#K}
 s^{n}(q)_{NL,M K}
\end{eqnarray*}
it follows from relation (\ref{kac_for_s})  that the whole $q$ dependence of $\det s^n$ is given by the factor
$$
 \begin{array}[t]{c}
{\displaystyle\prod} \\[-6pt]
{\scriptscriptstyle
1 \leqslant\, rs \,\leqslant\, 2n}
\\[-8pt]
{\scriptscriptstyle
r + s\,\in\, 2{\mathbb N}
}
\end{array}\!\!\!\!
\left(  2q- i r\epsilon_1 - is\epsilon_2   \right)^{P_\NS(n-{rs\over 2})}.
$$
In order to find the full $\epsilon_1,\epsilon_2$ dependence we calculate the determinant of the matrix
$$
b^{\,n}_{M'K',MK} = \bra{\Delta_p} g_{-K'}^\dagger l_{-M'}^\dagger l_{-M}g_{-K} \ket{\Delta_p}.
$$
Up to a numerical factor it is given by
\begin{equation}
\label{det1}
\det b^n \propto \det m^n
 \begin{array}[t]{c}
{\displaystyle\prod} \\[-6pt]
{\scriptscriptstyle
1 \leqslant\, rs \,\leqslant\, 2n}
\\[-8pt]
{\scriptscriptstyle
r + s\,\in\, 2{\mathbb N}
}
\end{array}\!\!\!\!
\left(\left(  2q- i r\epsilon_1 - is\epsilon_2   \right)\left(  2q+ i r\epsilon_1 + is\epsilon_2   \right)\right)^{P_\NS(n-{rs\over 2})}.
\end{equation}
where $m^n$ is the diagonal matrix
$
m^n_{M'K',M K} = m^n_{MK}
\delta_{M'K',M K}.
$
This easily follows from the Kac formula for matrix $B^n$
$$
\det B^n \propto
 \begin{array}[t]{c}
{\displaystyle\prod} \\[-6pt]
{\scriptscriptstyle
1 \leqslant\, rs \,\leqslant\, 2n}
\\[-8pt]
{\scriptscriptstyle
r + s\,\in\, 2{\mathbb N}
}
\end{array}\!\!\!\!
\left(\left(  2p- i rb - is{1\over b}  \right)\left(  2p+ i rb + is{1\over b}   \right)\right)^{P_\NS(n-{rs\over 2})}
$$
and  the relations
\begin{eqnarray*}
B^{\,n}(\Delta_p)_{M'K',MK} &=&(\epsilon_1\epsilon_2)^{-\#M'-{1\over 2}\#K'-\#M -{1\over 2}\#K }b^{\,n}_{M'K',MK},
\\
\deg_{\Delta_p} B^n &=&  \sum (\# M+\# K).
\end{eqnarray*}
Since
$$
 b^n= (s^n)^\dagger m^n s^n
$$
formula (\ref{det1}) implies that up to a numerical factor
$$
\det (s^n) \propto
\begin{array}[t]{c}
{\displaystyle\prod} \\[-6pt]
{\scriptscriptstyle
1 \leqslant\, rs \,\leqslant\, 2n}
\\[-8pt]
{\scriptscriptstyle
r + s\,\in\, 2{\mathbb N}
}
\end{array}\!\!\!\!
\left(  2q- i r\epsilon_1 - is\epsilon_2   \right)^{P_\NS(n-{rs\over 2})}.
$$
The matrix elements $s^{n}(q)^{-1}_{NL,\emptyset K} $ are given by
\begin{equation}
\label{quotient}
s^{n}(q)^{-1}_{NL,\emptyset K} ={C[s^{n}_{\emptyset K, NL}] \over \det s^n}
\end{equation}
where $C[s^{n}_{\emptyset K, NL}]$ denotes  the cofactor of the matrix element $s^n_{\emptyset K, NL}$.
 $C[s^{n}_{\emptyset K, NL}]$ is a sum o products of matrix elements $s^n_{M' K', N'L'}$ and therefore a polynomial in variables $\epsilon_1, \epsilon_2, q$.
 It follows from Prop.1 that the only singularities of the coefficients $S^{n}(p)^{-1}_{NL,\emptyset K} $ are simple poles at zeros of $\Omega(p,j)$.
The same is true for $s^{n}(q)^{-1}_{NL,\emptyset K} $ and $\omega(q,j)$. But this means cancelation of many factors between
the nominator and the denominator in (\ref{quotient}). The only possible linear factors left in the denominator are those entering $\omega(q,j)$,
hence
$$
\omega(q,j) s^{n}(q)^{-1}_{NL,\emptyset K}
$$
is a polynomial in all variables.

\section{Generalized Selberg integral }
\setcounter{equation}{0}
\renewcommand{\theequation}{B.\arabic{equation}}
Our  task is to calculate the integral
\begin{eqnarray}
\label{the:integral}
I_N(A,B;g)
& = &
\int\limits_{0}^{1}\!dt_1\ldots \int\limits_{0}^{1}\!dt_N\
\prod\limits_{k=1}^N\ t_k^A(1-t_k)^B
\hskip -5pt \prod\limits_{1 \leqslant k < l \leqslant N}\hskip -5pt |t_k-t_l|^{2g}\
\hbox{\boldmath $P$\!}_N(t_1,\ldots,t_{N})
\end{eqnarray}
for even $N =2m$  and where
\begin{eqnarray}
\label{P:polynom}
\hbox{\boldmath $P$\!}_N(t_1,\ldots,t_{N})
& = &
\langle\psi(t_1)\ldots\psi(t_N)\rangle
\hskip -5pt\prod\limits_{1\leqslant k < l\leqslant N}\hskip -5pt (t_k-t_l)
\; = \;
{\rm pf}\left({\hbox{\boldmath $A$}}\right)\hskip -5pt\prod\limits_{1\leqslant k < l\leqslant N}\hskip -5pt (t_k-t_l),
\end{eqnarray}
with
\[
A^k{}_l = \left\{
\begin{array}{cl}
0 & {\rm for}\ k = l,
\\[4pt]
\frac{1}{t_k-t_l}& {\rm for}\ k \neq l,
\end{array}
\right.
\]
is a symmetric polynomial. Our derivation parallels the original Selberg method
(see \cite{Forrester} for
a pedagogical discussion of various methods of deriving the Selberg formula).
Integral (\ref{the:integral}) has been already calculated long time ago (although in a different way)
in \cite{AlvarezGaume:1991bj}. It is also a special case of a more general formula discussed in \cite{Bershtein:2010wz}.
We present a simpler  derivation  mainly for the completeness of the present paper.

Since for $t_1\to t_2:$
\begin{eqnarray*}
\langle\psi(t_1)\psi(t_2)\psi(t_3)\ldots\psi(t_N)\rangle
& = &
\frac{1}{t_1-t_2}\langle\psi(t_3)\ldots\psi(t_N)\rangle + {\rm finite},
\end{eqnarray*}
the polynomials $\hbox{\boldmath $P$\!}_N$ satisfy the ``clustering property'':
\begin{eqnarray}
\label{clustering:property}
\hbox{\boldmath $P$\!}_N(x,x,t_1,\ldots,t_{N-2})
& = &
\prod\limits_{k=1}^{N-2}(x-t_k)^2 \hbox{\boldmath $P$\!}_{N-2}(t_1,\ldots,t_{N-2})
\end{eqnarray}
which is an important ingredient of the presented calculation.

Let $g$ be a natural number. Then,
\begin{equation}
\label{D:definition}
\hbox{\boldmath $D$\!}_N(t_1,\ldots,t_{N})
=
\hskip -5pt \prod\limits_{1 \leqslant k < l \leqslant N}\hskip -5pt (t_k-t_l)^{2g}\
\hbox{\boldmath $P$\!}_N(t_1,\ldots,t_{N})
\end{equation}
is a symmetric, uniform polynomial. From (\ref{D:definition}) and (\ref{P:polynom}) one infers its behavior under scaling and inverting of all arguments
\begin{eqnarray}
\label{D:prop:1}
\hbox{\boldmath $D$\!}_N(\Lambda t_1,\ldots,\Lambda t_N) &=& \Lambda^{N(N-1)g + \frac12N(N-2)}\, \hbox{\boldmath $D$\!}_N(t_1,\ldots,t_{N})
\\
\label{D:prop:2}
\hbox{\boldmath $D$\!}_N(t_1^{-1},\ldots,t_{N}^{-1}) &=& \prod\limits_{i=1}^{N}t_i^{1- (N-1)(2g+1)}\,\hbox{\boldmath $D$\!}_N(t_1,\ldots,t_{N}).
\end{eqnarray}
The polynomial $\hbox{\boldmath $D$\!}_N$ can be presented as
\begin{eqnarray}
\label{expansion}
\hbox{\boldmath $D$\!}_N(t_1,\ldots,t_{N})
& = &
\sum\limits_{\nu_j} c_{\nu_1,\ldots,\nu_N}\,t_1^{\nu_1}\ldots t_N^{\nu_N}.
\end{eqnarray}
Since the expansion coefficients $c_{\nu_1,\ldots,\nu_N}$ are totally symmetric with respect to permutation of their
indices, we can rewrite the sum as
\begin{eqnarray*}
\hbox{\boldmath $D$\!}_N(t_1,\ldots,t_{N})
& = &
\sum\limits_{\nu_j} c_{\nu_1,\ldots,\nu_N}t_{(1}^{\nu_1}\ldots t_{N)}^{\nu_N}
\end{eqnarray*}
where the indices $\nu_k$ are ordered
\begin{equation}
\label{ordering}
\nu_1\leqslant\nu_2\leqslant \ldots \leqslant \nu_N
\end{equation}
and the bracket denotes symmetrization.

The crucial point of the Selberg method is to find a lower and an upper bound on the possible
values of $\nu_k.$ Scaling property (\ref{D:prop:1}) gives
\begin{equation}
\label{wzor:sumacyjny}
\sum\limits_{k=1}^{N}\nu_k \; = \; N(N-1)g +  \frac12N(N-2)
\end{equation}
and together with  (\ref{ordering}) yields
\begin{equation}
\label{maximal}
\nu_N \geqslant (N-1)g+\frac12(N-2) .
\end{equation}
One also has
\[
\sum\limits_{l=1}^{N-k}\nu_l +\nu_k^{\rm\scriptscriptstyle max}\geqslant  N(N-1)g + \frac12N(N-2) ,
\hskip 1cm
k = 0,1,\ldots, N-1
\]
where $\nu_0^{\rm\scriptscriptstyle max}=0$ and for $k>0$, $\nu_k^{\rm\scriptscriptstyle max}$ denotes the maximal joint degree of the polynomial
$\hbox{\boldmath $D$\!}_N$ in the variables
$t_{N-k+1},\ldots t_{N}$:
\[
\hbox{\boldmath $D$\!}_N(t_1,\ldots,t_{N-k},\Lambda t_{N-k+1},\ldots, \Lambda t_{N})
\; = \;
\Lambda^{\nu_k^{\rm\scriptscriptstyle max}}\hbox{\boldmath $Q$\!}_N(t_1,\ldots,t_{N}) + {\cal O}\big(\Lambda^{\nu_k^{\rm\scriptscriptstyle max}-1}\big).
\]
Let
\begin{equation}
\label{parameters:appendix}
N = 2m,
\hskip 1cm
k = N-2p-j,\;\;
p = 0,\ldots,m-1,\;\ j = 1,2
\end{equation}
then
\begin{equation}
\label{nu:max:1}
\begin{array}{rcllllll}
\nu_{k}^{\rm\scriptscriptstyle max}
& = &
(2m-2p-1)(m+p)(2g+1) - m+p
&\hskip 10pt {\rm for} & j=1,
\\[6pt]
\nu_{k}^{\rm\scriptscriptstyle max}
& = &
2(m-p-1)(2m+2p+1)g +2(m-p-1)(m+p)
&\hskip 10pt {\rm for}& j=2.
\end{array}
\end{equation}
This gives
\begin{eqnarray*}
\sum\limits_{l=1}^{2p+1}\nu_l
& \geqslant &
2m(2m-1)g + 2m(m-1) - (2m-2p-1)(m+p)(2g+1) + m-p
\\[4pt]
& = &
(2p+1)\left((2p+1-1)g + p - \frac{p}{2p+1}\right)
\; \geqslant \;
(2p+1)\Big((2p+1-1)g + p \Big),
\\[4pt]
\sum\limits_{l=1}^{2p+2}\nu_l
& \geqslant &
2m(2m-1)g + 2m(m-1) - 2(m-p-1)(2m+2p+1)g -2(m-p-1)(m+p)
\\[4pt]
& = &
(2p+2)\Big((2p+2-1)g + p \Big),
\end{eqnarray*}
or, equivalently
\begin{eqnarray}
\sum\limits_{l=1}^{2p+j}\nu_l & \geqslant & (2p+j)\Big((2p+j-1)g + p \Big).
\end{eqnarray}
Taking into account the ordering of the indices $\nu_l$ we thus get
\begin{eqnarray}
\label{bound:1}
\nu_{2p+j} & \geqslant & (2p+j-1)g + p
\end{eqnarray}
which for $p=m-1,\ j=2$ agrees with (\ref{maximal}).

To obtain the upper bound on the index $\nu_{2p+j}$ we use property (\ref{D:prop:2}). It gives
\begin{eqnarray*}
\hbox{\boldmath $D$\!}_N (t_1,\ldots,t_{N})
& = &
\prod\limits_{j=1}^{N}t_j^{(N-1)(2g+1)-1}\,\hbox{\boldmath $D$\!}_N(t_1^{-1},\ldots,t_{N}^{-1})
\\[4pt]
& = &
\prod\limits_{j=1}^{N}t_j^{(N-1)(2g+1)-1}\,
\sum\limits_{\nu_1\leqslant\ldots\leqslant \nu_N}
c_{\nu_1,\ldots,\nu_N}
t_{(1}^{-\nu_1}\ldots t_{N)}^{-\nu_N}
\\[4pt]
& = &
\sum\limits_{\nu_1\leqslant\ldots\leqslant \nu_N}
c_{\nu_1,\ldots,\nu_N}
t_{(1}^{\nu'_N}\ldots t_{N)}^{\nu'_1}
\;\; =
\sum\limits_{\nu'_1\leqslant\ldots\leqslant \nu'_N}
c_{\nu'_1,\ldots,\nu'_N}
t_{(1}^{\nu'_1}\ldots t_{N)}^{\nu'_N},
\end{eqnarray*}
where
\[
\nu_l' = (N-1)(2g+1)-1 - \nu_{N+1-l}.
\]
Inequality (\ref{bound:1}) thus yields
\begin{eqnarray}
\label{bound:2:b}
\nu_{2p+j}
& \leqslant &
(m+p-1) + \big(2(m+p) +j-2\big)g.
\end{eqnarray}

Upon inserting (\ref{expansion}) into (\ref{the:integral}), the well-known formula for the Euler Beta gives
\begin{eqnarray*}
I_N(A,B;g)
& = &
\sum\limits_{\{\nu_k\}}
c_{\nu_1,\ldots\,\nu_N}
\prod\limits_{k=1}^N
\int\limits_{0}^{1} t_k^{\nu_i+A}(1-t_k)^B\,dt_k
=
\sum\limits_{\{\nu_k\}}
c_{\nu_1,\ldots\,\nu_N}
\prod\limits_{k=1}^N
\frac{\Gamma(1+A+\nu_k)\Gamma(1+B)}{\Gamma(2+A+B+\nu_k)}
\\
& = &
\hskip -5pt\sum\limits_{\nu_1\leqslant\ldots\leqslant\nu_N}\hskip -5pt
c_{\nu_{(1},\ldots\,\nu_{N)}}
\prod\limits_{k=1}^N
\frac{\Gamma(1+A+\nu_k)\Gamma(1+B)}{\Gamma(2+A+B+\nu_k)}.
\end{eqnarray*}
Using (\ref{bound:1}) we get
\begin{eqnarray*}
\Gamma(1+A+\nu_{2p+j})
& = &
\frac{\Gamma(1+A+\nu_{2p+j})}{\Gamma(1+A+(2p+j-1)g + p)}\, \Gamma(1+A+(2p+j-1)g + p)
\end{eqnarray*}
where
\begin{eqnarray*}
\frac{\Gamma(1+A+\nu_{2p+j})}{\Gamma(1+A+(2p+j-1)g + p)}
& = &
\big(A+\nu_{2p+j}\big)\big(A+\nu_{2p+j}-1\big)\ldots\big(1+A+(2p+j-1)g + p\big)
\end{eqnarray*}
is a polynomial in $A$ of degree $\nu_{2p+j}-(2p+j-1)g - p.$ This gives
\begin{eqnarray}
\label{polonom:expr:1}
\prod\limits_{p=0}^{m-1}\prod\limits_{j=1}^2\Gamma(1+A+\nu_{2p+j})
& = &
w_{\nu_1,\ldots,\nu_N}(A)\prod\limits_{p=0}^{m-1}\prod\limits_{j=1}^2\Gamma(1+A+(2p+j-1)g + p)
\end{eqnarray}
where
\begin{eqnarray*}
w_{\nu_1,\ldots,\nu_N}(A)
& = &
\prod\limits_{p=0}^{m-1}\prod\limits_{j=1}^2\
\frac{\Gamma(1+A+\nu_{2p+j})}{\Gamma(1+A+(2p+j-1)g + p)}
\end{eqnarray*}
is a polynomial in $A$ of degree (see (\ref{wzor:sumacyjny})):
\begin{eqnarray*}
\sum\limits_{p=0}^{m-1}\sum\limits_{j=1}^2\big(\nu_{2p+j}-(2p+j-1)g - p\big)
& = &
m(2m-1)g +m(m-1).
\end{eqnarray*}
Similarly, applying (\ref{bound:2:b}) we get
\begin{eqnarray}
\label{polonom:expr:2}
\prod\limits_{p=0}^{m-1}\prod\limits_{j=1}^2\frac{1}{\Gamma(2+A+B+\nu_{2p+j})}
& = &
u_{\nu_1,\ldots\nu_N}(A+B)
\\
\nonumber
&&\hspace{-60pt} \times \;
\prod\limits_{p=0}^{m-1}\prod\limits_{j=1}^2\
\frac{1}{\Gamma\big(2+A+B+(m+p-1) +(2(m+p) +j-2)g\big)}
\end{eqnarray}
where
\begin{eqnarray*}
u_{\nu_1,\ldots\nu_N}(A+B)
& = &
\prod\limits_{p=0}^{m-1}\prod\limits_{j=1}^2\
\frac{\Gamma\big(2+A+B+(m+p-1) +(2(m+p) +j-2)g\big)}{\Gamma(2+A+B+\nu_{2p+j})}
\end{eqnarray*}
is a polynomial in $A,B,$ of degree
\begin{eqnarray*}
\sum\limits_{p=0}^{m-1}\sum\limits_{j=1}^2
\big((m+p-1) +(2(m+p) +j-2)g - \nu_{2p+j}\big)
& = &
m(2m-1)g +m(m-1).
\end{eqnarray*}
Finally, we can write
\begin{eqnarray}
\label{polonom:expr:3}
\prod\limits_{p=0}^{m-1}\prod\limits_{j=1}^2\ \Gamma(1+B)
& = &
\frac{1}{Q_m(B)}\prod\limits_{p=0}^{m-1}\prod\limits_{j=1}^2\ \Gamma(1+B+(2p+j-1)g + p)
\end{eqnarray}
where
\begin{eqnarray*}
Q_m(B)
& = &
\prod\limits_{p=0}^{m-1}\prod\limits_{j=1}^2\,\frac{\Gamma(1+B+(2p+j-1)g + p)}{\Gamma(1+B)}
\end{eqnarray*}
is a polynomial in $B$ of degree
\begin{eqnarray*}
\sum\limits_{p=0}^{m-1}\sum\limits_{j=1}^2\big((2p+j-1)g + p\big)
& = &
m(2m-1)g +m(m-1).
\end{eqnarray*}
Using (\ref{polonom:expr:1}) -- (\ref{polonom:expr:3}) we thus get
\begin{eqnarray*}
&&
\hskip -2.5cm
\prod\limits_{k=1}^N
\frac{\Gamma(1+A+\nu_k)\Gamma(1+B)}{\Gamma(2+A+B+\nu_k)}
\; = \;
\frac{
w_{\nu_1,\ldots,\nu_N}(A)u_{\nu_1,\ldots,\nu_N}(A+B)
}{
Q_m(B)}
\\
& \times &
\prod\limits_{p=0}^{m-1}\prod\limits_{j=1}^2
\frac{\Gamma(1+A+(2p+j-1)g + p)\Gamma(1+B+(2p+j-1)g + p)
}{
\Gamma\big(2+A+B+(m+p-1) +(2(m+p) +j-2)g\big)}
\end{eqnarray*}
and consequently
\begin{equation}
\label{intermediate:1}
I_N(A,B;g)
=
\frac{P_m(A,B)}{Q_m(B)}\,
\prod\limits_{p=0}^{m-1}\prod\limits_{j=1}^2
\frac{\Gamma(1+A+(2p+j-1)g + p)\Gamma(1+B+(2p+j-1)g + p)
}{
\Gamma\big(2+A+B+(m+p-1) +(2(m+p) +j-2)g\big)}
\end{equation}
where
\begin{eqnarray*}
P_m(A,B)
& = &
\sum\limits_{\nu_1\leqslant\ldots\leqslant\nu_N}\,
c_{\nu_{(1},\ldots,\nu_{N)}}\,w_{\nu_1,\ldots,\nu_N}(A)u_{\nu_1,\ldots,\nu_N}(A+B).
\end{eqnarray*}
Since the coefficients $c_{\nu_1,\ldots,\nu_N}$ do not depend on $A$ and $B,$ $P_m(A,B)$ is a polynomial in $A$ of degree
not greater than $2m(2m-1)g +2m(m-1)$ and a polynomial in $B$ of degree not greater than $m(2m-1)g +m(m-1).$

In view of an obvious symmetry of integral (\ref{the:integral}) with respect to the exchange of $A$ and $B,$
\[
I_N(A,B;g) = I_N(B,A;g),
\]
formula (\ref{intermediate:1}) gives
\begin{eqnarray}
\label{ratios}
\frac{P_m(A,B)}{Q_m(B)}
& = &
\frac{P_m(B,A)}{Q_m(A)}
\end{eqnarray}
The r.h.s.\ of (\ref{ratios}) is a polynomial in $B,$ the same must be therefore true for the l.h.s. Since the degree of
$P_m(A,B)$ as a polynomial in $B$ is bounded by the degree of $Q_m(B),$ this polynomial is of a zero degree and thus $B-$independent.
In the same way one shows that ratio (\ref{ratios}) is also $A-$independent. We can thus write
\[
\frac{P_m(A,B)}{Q_m(B)}
=
C_m(g)
\]
and consequently
\begin{equation}
\label{intermediate:2}
I_N(A,B;g)
=
C_m(g)\,
\prod\limits_{p=0}^{m-1}\prod\limits_{j=1}^2
\frac{\Gamma(1+A+(2p+j-1)g + p)\Gamma(1+B+(2p+j-1)g + p)
}{
\Gamma\big(2+A+B+(m+p-1) +(2(m+p) +j-2)g\big)}.
\end{equation}
Our next task it to derive (and solve) a recurrence relation for $C_m(g).$ To this end let us write
\begin{equation}
\label{the:integral:2}
I_N(A,B;g)
=
N!\,
\int\limits_{0}^{1}\!dt_1\int\limits_{t_1}^1\!dt_2\ldots \int\limits_{t_{N-1}}^{1}\!dt_N\ \hbox{\boldmath $P$}_N(t_1,\ldots,t_N)
\prod\limits_{k=1}^N\ t_k^A(1-t_k)^B\hskip -5pt \prod\limits_{1 \leqslant k < l \leqslant N}\hskip -5pt (t_l-t_k)^{2g}
\end{equation}
and change the integration variables
in (\ref{the:integral:2}):
\[
t_1 = \tau\xi_1, \;\; t_2 = \tau\xi_2,\;\; \xi_1 + \xi_2 = 1,\;\;\; t_i\to t_{i-2},\;\; i = 3,4,\ldots N.
\]
Since
\begin{eqnarray*}
\int\limits_0^1\!dt_1\int\limits_{t_1}^1\!dt_2\ f(t_1,t_2)
& = &
\int\limits_0^1\!d\tau\,\tau\! \int\limits_0^{\frac12}\!d\xi_1\ f(\tau\xi_1,\tau\xi_2)
+
\int\limits_1^2\!d\tau\,\tau\!\! \int\limits_{1-\frac{1}{\tau}}^{\frac12}\!\!\!d\xi_1\  f(\tau\xi_1,\tau\xi_2),
\hskip 10mm
\xi_2 = 1-\xi_1,
\end{eqnarray*}
we get
\begin{eqnarray*}
I_N(A,B;g)
& = &
N!\,\int\limits_0^1\!d\tau\,\tau^{2A+2g+1}
\int\limits_0^\frac12\!d\xi_1  \prod\limits_{j=1}^2\,\xi_j^A (1-\tau\xi_j)^B|\xi_1-\xi_2|^{2g}\,
J_{N-2}(A,B;g|\tau,\xi_j)
\\
& + &
N!\,\int\limits_1^2\!d\tau\,\tau^{2A+2g+1}\!\!
\int\limits_{1-\frac{1}{\tau}}^{\frac12}\!\!\!d\xi_1\ \prod\limits_{j=1}^2\,\xi_j^A (1-\tau\xi_j)^B|\xi_1-\xi_2|^{2g}\,
J_{N-2}(A,B;g|\tau,\xi_j)
\end{eqnarray*}
where
\begin{eqnarray*}
J_{N-2}(A,B;g|\tau,\xi_j)
& = &
\int\limits_{\tau\xi_2}^{1}\!\!\!dt_1\int\limits_{t_1}^{1}\!\!dt_2\ldots\!\int\limits_{t_{N-3}}^{1}\!\!\!\!dt_{N-2}\,
\prod\limits_{k=1}^{N-2}t_k^A(1-t_k)^B\prod\limits_{j=1}^2|t_k-\tau\xi_j|^{2g}
\prod\limits_{1=k<l}^{N-2}|t_k-t_l|^{2g}
\\
&&
\hskip 4cm \times \;
\hbox{\boldmath $P$\!}_N (\tau\xi_1,\tau\xi_2,t_1,\ldots,t_{N-2}).
\end{eqnarray*}
Applying the identity
\begin{equation}
\lim\limits_{a\to (-1)^+} (1+a)\int_0^\Lambda\!dx\ x^a f(x) \; = \; f(0)
\end{equation}
we get
\begin{eqnarray}
\label{another:intermediate:formula}
\lim\limits_{A\to -g-1}(A+g+1)I_N(A,B;g)
& = &
N!\,X_2(g)\,
J_{N-2}(-g-1,B;g|0,\xi_j)
\end{eqnarray}
where
\begin{eqnarray*}
X_2(g)
& = &
\frac12
\int\limits_0^{\frac12}\!d\xi_1 \prod\limits_{j=1}^2\,\xi_j^{-g-1}\ |\xi_1-\xi_2|^{2g}
\; = \;
\frac14
\int\limits_0^{1}\!d\xi_1 \prod\limits_{j=1}^2\,\xi_j^{-g-1}\ |\xi_1-\xi_2|^{2g},
\hskip 1cm
\xi_2 = 1-\xi_1,
\end{eqnarray*}
and we applied the clustering property of the polynomial $\hbox{\boldmath $P$\!}_N.$
Further
\begin{eqnarray*}
J_{N-2}(-g-1,B;g|0,\xi_j)
& = &
\int\limits_0^{1}\!\!dt_1\ldots\!\!\!\int\limits_{t_{N-3}}^{1}\hskip -7ptdt_{N-2}
\prod\limits_{k=1}^{N-2}t_k^{1+3g}(1-t_k)^B
\hskip -5pt \prod\limits_{1=k<l}^{N-2}\hskip -2pt |t_k-t_l|^{2g}
\hbox{\boldmath $P$\!}_{N-2} (t_1,\ldots,t_{N-2})
\\
& = &
\frac{1}{(N-2)!}I_{N-2}(1+3g,B;g).
\end{eqnarray*}

To determine the value of $X_2(g)$ note that, since
\[
\hbox{\boldmath $P$\!}_2(t_1,t_2) = 1,
\]
the integral $I_2(A,B;g)$ is just a ``standard'' Selberg integral with a well-known value
\begin{eqnarray*}
I_2(A,B;g)
& = &
S_2(A,B;g)
\; = \;
\prod\limits_{j=0}^{1}
\frac{
\Gamma(A + 1 + jg)\Gamma(B+ 1 + jg)\Gamma(1+(j+1)g)
}{
\Gamma\big(A+B + 2 + (1+j)g\big)\Gamma(1+g)
},
\end{eqnarray*}
and consequently
\begin{eqnarray*}
X_2(g)
& = &
\frac{1}{2!}\lim\limits_{A\to -g-1}(A+g+1)S_2(A,B;g)
\; = \;
\frac{\Gamma(-g)}{2}\prod\limits_{j=1}^2 \frac{\Gamma(1+jg)}{\Gamma(1+g)}
\; = \;
\frac12\frac{\Gamma(-g)\Gamma(1+2g)}{\Gamma(1+g)},
\end{eqnarray*}
so that (\ref{another:intermediate:formula}) takes the form
\begin{eqnarray}
\label{C:recurrence:1}
\lim\limits_{A\to -g-1}(A+g+1)I_N(A,B;g)
& = &
\frac{N(N-1)}{2}\frac{\Gamma(-g)\Gamma(1+2g)}{\Gamma(1+g)}\,
I_{N-2}(1+3g,B;g).
\end{eqnarray}
Using in this equality the r.h.s.\ of formula (\ref{intermediate:2}) one gets after a short calculation
\begin{eqnarray*}
C_m(g)
& = &
\frac{2m(2m-1)}{2}\frac{\Gamma((2m-1)g+m)\Gamma(2mg + m)}{\Gamma^2(1+g)}\,
C_{m-1}(g)
\end{eqnarray*}
which implies
\begin{eqnarray*}
C_m(g)
& = &
\frac{(2m)!}{2^m}
\prod\limits_{p=1}^{m-1}\prod\limits_{j=1}^2 \frac{\Gamma(1+p + (2p+j)g)}{\Gamma(1+g)}\,
C_{1}(g).
\end{eqnarray*}
The fact that for $m=1$  integral (\ref{the:integral}) is just a Selberg integral yields
\[
C_1(g) = \frac{\Gamma(1+g)\Gamma(1+2g)}{\Gamma^2(1+g)}
\]
and
\begin{eqnarray}
\label{normalisation}
C_m(g)
& = &
\frac{(2m)!}{2^m}
\prod\limits_{p=0}^{m-1}\prod\limits_{j=1}^2 \frac{\Gamma(1+p + (2p+j)g)}{\Gamma(1+g)}.
\end{eqnarray}
This finally gives
\begin{eqnarray}
\label{modified:Selberg}
&&
\hskip -.3cm
\int\limits_{0}^{1}\!dt_1\ldots \int\limits_{0}^{1}\!dt_{2m}\
\prod\limits_{k=1}^{2m} t_k^A(1-t_k)^B
\hskip -10pt \prod\limits_{1 \leqslant k < l \leqslant 2m}\hskip -10pt |t_k-t_l|^{2g}\
\hbox{\boldmath $P$\!}_{2m}(t_1,\ldots,t_{2m})
\\
\nonumber
& = &
\frac{(2m)!}{2^m}
\prod\limits_{p=0}^{m-1}\prod\limits_{j=1}^2
\frac{\Gamma(1+p + (2p+j)g)\Gamma(1+A+(2p+j-1)g + p)\Gamma(1+B+(2p+j-1)g + p)
}{
\Gamma(1+g)\Gamma\big(2+A+B+(m+p-1) +(2(m+p) +j-2)g\big)}.
\end{eqnarray}
Formula (\ref{modified:Selberg}) has been derived for $g \in {\mathbb N}.$ One can however resort to the Carlson
theorem \cite{Carlson}  to check its validity also for complex $g,$ for which the integral is convergent.

Similar methods can be used to calculate the ``odd'' integral
\begin{equation}
\label{Selberg:odd:integral}
I^{(1)}_{2m-1}(A,B;g)
=
\int\limits_0^1\!dt_1\ldots\int\limits_0^1\!dt_{2m-1}\,
\hskip -5pt\prod\limits_{i=1}^{2m-1} t_i^A (1-t_i)^B
\hskip -15pt\prod\limits_{1\leqslant k < l \leqslant 2m-1}\hskip -15pt |t_k-t_l|^{2g}\,
\hbox{\boldmath $P$\!}^{\,(1)}_{2m-1}(t_1,\ldots,t_{2m-1})
\end{equation}
where
\begin{eqnarray*}
\hbox{\boldmath $P$\!}^{\,(1)}_{2m-1}(t_1,\ldots,t_{2m-1})
& = &
\hskip -10pt\prod\limits_{1\leqslant k < l \leqslant 2m-1}\hskip -10pt (t_k-t_l)\,
\langle 0_f|\psi(1)\psi(t_1)\ldots \psi(t_{2m-1})|0_f\rangle.
\end{eqnarray*}
Since we do not need this integral in the general form, we skip the calculation and just quote the result:
\begin{eqnarray}
\label{odd:Selberg:final:form}
&&
\hskip -1.2cm
I_{2m-1}^{(1)}(A,B;g)
\; = \;
\frac{(2m-1)!}{2^{m-1}}
\\[4pt]
\nonumber
& \times &
\frac{1}{B}\,\prod\limits_{p,j}
\frac{\Gamma(1+p + (2p+j)g)\Gamma(1+A+(2p+j-1)g + p)\Gamma(1+B+(2p+j-1)g + p)}
{\Gamma(1+g)\Gamma\left(2+A+B + (m+p+j-3) + \big(2(m+p) + j-3\big)g\right)}
\end{eqnarray}
where for $p = 0,1,\ldots,m-2$ we have $j=1,2$ while for $p=m-1$ we have $j=1.$

\section{Gamma Barnes identities}
\setcounter{equation}{0}
\renewcommand{\theequation}{C.\arabic{equation}}

For $\Re\,x > 0$ the Barnes double gamma function $\Gamma_b(x)$ can be defined by the integral
representation \cite{barnes,Zamolodchikov:1995aa}
\[
\log\,\Gamma_b(x)
\; = \;
\int\limits_{0}^{\infty}\frac{dt}{t}
\left[
\frac{{\rm e}^{- x t} - {\rm e}^{- {Q \over 2}t}}
{\left(1-{\rm e}^{- tb}\right)\left(1-{\rm e}^{- t/b}\right)}
-
\frac{\left({\textstyle{Q\over 2}}-x\right)^2}{2{\rm e}^{t}}
-
\frac{{\textstyle{Q\over 2}}-x}{t}
\right].
\]
It can be analytically continued to the whole complex $x$ plane as a
meromorphic function with  poles located at $ x = -m b - n{1\over b},\; m, n \in
{\mathbb N}.$

From the shift formulae
\begin{eqnarray*}
\nonumber
\Gamma_b(x+b)
& = &
\sqrt{2\pi}\,b^{bx-\frac12}\,\Gamma^{-1}\left(bx\right)\,\Gamma_b(x),
\\[-6pt]
\label{Gamma:b:shift}
\\[-6pt]
\nonumber
\Gamma_b\left(x+b^{-1}\right)
& = &
\sqrt{2\pi}\,b^{-x/b+\frac12}\,\Gamma^{-1}\left(x/b\right)\,\Gamma_b(x),
\end{eqnarray*}
one easily derives the multiple shift relations
\begin{eqnarray*}
\Gamma_b(x+jb)
& = &
\,
{(2\pi)^{\frac{j}{2}}\,b^{jbx + \frac12j(j-1)b^2 - \frac12j}\over \Big(\prod\limits_{k=0}^{j-1}\Gamma\big((x+kb)b\big)\Big) }\Gamma_b(x),
\\[2pt]
\Gamma_b(x-jb)
& = &
(2\pi)^{-\frac{j}{2}}\,b^{-jbx + \frac12j(j+1)b^2 + \frac12j}\,
{ \Big(\prod\limits_{k=1}^{j}\Gamma\big((x-kb)b\big)\Big) }\Gamma_b(x),
\\[2pt]
\Gamma_b\left(x+jb^{-1}\right)
& = &
\,
{(2\pi)^{\frac{j}{2}}\,b^{-{jx\over b} - \frac12{j(j-1)\over b^{2}} + \frac12j}
\over
\Big(\prod\limits_{k=0}^{j-1}\Gamma\big({x+{k\over b}\over b}\big)\Big)} \Gamma_b(x),
\\[2pt]
\Gamma_b\left(x- jb^{-1}\right)
& = &
(2\pi)^{-\frac{j}{2}}\,b^{{jx\over b} - \frac12{j(j+1)\over b^{2}} - \frac12j}\,
{
\Big(\prod\limits_{k=1}^{j}\Gamma\big({x-{k\over b}\over b}\big)\Big)} \Gamma_b(x).
\end{eqnarray*}
In order to verify identity (\ref{ridentity1}) one first checks the behavior of its both sides  under
the shifts $\alpha \to \alpha +2b$ and $\alpha\to \alpha +2b^{-1}$.
For non-rational $b$ it  yields the proof of (\ref{ridentity1}) up to $\alpha$-independent  factor:
\begin{eqnarray*}
{\Gamma_\bli (\ali )\over \Gamma_\bgi (\agi + \bgi)}
&=& C(b)
b^{-{b^2\over 1-b^2} {\alpha(Q-\alpha)\over 4}}
\left({1-b^2\over 2}\right)^{-{\alpha(Q-\alpha)\over 8}}\textstyle\Gamma^\NS_b\left({\alpha }\right)
\end{eqnarray*}
To find $C(b)$ one can calculate both sides at $\alpha =Q$,
hence
\begin{eqnarray*}
C(b) &=&
{\Gamma_\bli (Q^\li)\over \Gamma_\bgi ((\bgi)^{-1})\textstyle\Gamma_b\left({Q \over 2}\right)\Gamma_b\left({Q}\right)}
\;=\;
{\sqrt{2\pi\bgi}\,\Gamma_\bli (Q^\li)\over \Gamma_\bgi (\bgi+(\bgi)^{-1})\textstyle\Gamma_b\left({Q \over 2}\right)\Gamma_b\left({Q}\right)}.
\end{eqnarray*}
Identities (\ref{ridentity2}) and (\ref{ridentity3}) are obtained from (\ref{ridentity1}) by substituting
$\alpha \to \alpha + b$ and $\alpha \to \alpha - b$, respectively.

Multiplying  (\ref{ridentity1}) for $\alpha$ and   for $Q-\alpha$ side by side yields the identity for upsilon functions
proposed in \cite{Belavin:2011sw}:
\begin{eqnarray*}
{\Upsilon_\bli (\ali )\over \Upsilon_\bgi (\agi + \bgi)}
&=&
{\Upsilon_\bgi (\bgi)
{\textstyle\Upsilon_b\left(b\right)\Upsilon_b\left({Q\over 2}\right)}
\over
\Upsilon_\bli (\bli)
}
b^{-{b^2\over 1-b^2} {\alpha(Q-\alpha)\over 2}}
\left({1-b^2\over 2}\right)^{-{\alpha(Q-\alpha)\over 4}+{1\over 2}}
\textstyle\Upsilon^\NS_b\left({\alpha }\right).
\end{eqnarray*}
The relations supporting the calculations of ratios
of the LL chiral structure constants presented in Subsection 4.1
read
\begin{eqnarray}
\label{evenj}
&&\hspace{-80pt}
\frac
{\Gamma_{b^\li}\left(\alpha^\li + {1\over 2} j b^{\li}\right)\Gamma_{b^\gi}\left(\alpha^{\gi} + b^{\gi}\right)}
{\Gamma_{b^\li}\left(\alpha^{\li} \right)\Gamma_{b^\gi}\left(\alpha^{\gi} + b^{\gi} +{1\over 2}j(b^{\gi})^{-1}\right)}
\\
\nonumber
& = &
\frac
{\Gamma_{b^\li}\left(Q^\li-\alpha^\li - {1\over 2}j b^{\li}\right)\Gamma_{b^\gi}\left(-\alpha^{\gi} + (b^{\gi})^{-1}\right)}
{\Gamma_{b^\li}\left(Q^\li-\alpha^{\li} \right)\Gamma_{b^\gi}\left(-\alpha^{\gi}  -{1\over 2}j(b^{\gi})^{-1}+ (b^{\gi})^{-1}\right)}
\\[6pt]
\nonumber
&=&
b^{{j(2\alpha b+ j-2)\over 4( 1-b^2)}  +{j\over 4}}
\left({ 2-2b^2}\right)^{-{j^2\over 8}}
l^\NS(\alpha,j)
\end{eqnarray}
for $j\in 2\mathbb{N}$ and
\begin{eqnarray}
\label{oddj}
&&\hspace{-80pt}
\frac
{\Gamma_{b^\li}\left(\alpha^\li  +\frac12 jb^{\li}\right)\Gamma_{b^\gi}\left(\alpha^{\gi} +{1\over 2}(\bgi)^{-1} + b^{\gi}\right)}
{\Gamma_{b^\li}\left(\alpha^{\li}+{1\over 2}\bli \right)\Gamma_{b^\gi}\left(\alpha^{\gi}  + b^{\gi} + \frac12 j (\bgi)^{-1}\right)}
\\
&=&
\nonumber
\frac
{\Gamma_{b^\li}\left(- \alpha^\li+ \bli+(\bli)^{-1}   -\frac12 jb^{\li}\right)\Gamma_{b^\gi}\left(-\alpha^{\gi} +{1\over 2}(\bgi)^{-1} \right)}
{\Gamma_{b^\li}\left(-\alpha^{\li}+ (\bli)^{-1}+{1\over 2}\bli \right)\Gamma_{b^\gi}\left(-\alpha^{\gi}  + (\bgi)^{-1} - \frac12 j (\bgi)^{-1}\right)}
\\
\nonumber
&=&
b^{{(j-1)(2\alpha b +j  -1)\over 4(1-b^2)}+{j-1\over 4}}\left(2-2b^2\right)^{-{j^2-1\over 8}}
l^\R(\alpha,j)
\end{eqnarray}
for $j\in 2\mathbb{N}+1$. The functions $l^\NS(\alpha,j)$ and $l^\R(\alpha,j)$ are defined by (\ref{defel}) and (\ref{defelR}), respectively.
The formulae above can be
easily derived  using the multiple shift formulae and  some simple consequences of definitions (\ref{defblibgi}), (\ref{defaliagi})
$$
\bli\bgi = b,\;\;\;(\bgi)^{-2} = 2  +(\bli)^2,\;\;\;\bli\ali= (\bgi)^{-1}\agi = {b\,\alpha\over 1-b^2}.
$$

\end{document}